\documentclass[11pt,a4paper]{article}
\pdfoutput=1
\usepackage[T1]{fontenc}
\usepackage[dvipsnames]{xcolor}
\usepackage{jheppub}
\hypersetup{hypertexnames=false}

\usepackage[font=small,labelfont=bf,labelsep=period]{caption}
\usepackage{subcaption}
\usepackage{mathrsfs,pdflscape,longtable,array,slashed,braket,upgreek}
\usepackage{subfloat,ytableau,tensor}
\usepackage{tikz}
\usetikzlibrary{calc,patterns,arrows.meta,decorations.text}
\usepackage[compat=1.0.0]{tikz-feynman}
\usepackage[makeroom]{cancel}
\usepackage{bm,booktabs,multirow,dcolumn,verbatim,textcomp}
\definecolor{SchoolColor}{rgb}{0.6471,0.1098,0.1882}
\allowdisplaybreaks[4]
\graphicspath{{figure/}}
\newcommand{\bea}{\begin{eqnarray}}
\newcommand{\eea}{\end{eqnarray}}
\newcommand{\be}{\begin{equation}}
\newcommand{\ee}{\end{equation}}
\newcommand{\bs}{\begin{subequations}}
\newcommand{\es}{\end{subequations}}
\def\nn{\nonumber}

\newcommand{\n}{\nabla }
\newcommand{\beqs}{\begin{eqnarray}}
\newcommand{\eeqs}{\end{eqnarray}}
\newcommand{\cT}{\mathcal T}

\newcommand{\Rmnum}[1]{\uppercase\expandafter{\romannumeral #1\relax}}
\def\c.c.{\mathrm{c.c.}}

\def\n{\nu}

\tikzset{
    dot/.style={circle, draw=black, fill=black, inner sep=1pt},
    n_node/.style={circle, draw=black, fill=gray!10, inner sep=1.2pt, font=\tiny},
    blue_line/.style={blue, thick},
     gray line/.style={lightgray!40, thin},
    red_line/.style={red, thick, dashed, dash pattern=on 2pt off 1.5pt},
    blue arc/.style={blue, thick},
    red curve/.style={red, thick},
    label_style/.style={font=\tiny, gray}
}

\newcommand{\BasisDiagramthree}[3]{
    \begin{tikzpicture}[baseline=(current bounding box.center), scale=0.7]
        \draw[dotted, gray!30] (0,0) circle (1cm);
        \node[n_node] (N) at (0,0) {n};
        \foreach \i in {1,...,3} {
            \node[dot] (p\i) at (90-120*\i+120:1cm) {};
            \node[label_style] at (90-120*\i+120:1.3cm) {\i};
        }
        #1 
        #2 
        #3 
    \end{tikzpicture}
}

\newcommand{\BasisDiagramfour}[3]{
    \begin{tikzpicture}[baseline=(current bounding box.center), scale=0.7]
        \draw[dotted, gray!30] (0,0) circle (1cm);
        \node[n_node] (N) at (0,0) {n};
        \foreach \i in {1,...,4} {
            \node[dot] (p\i) at (90-90*\i+90:1cm) {};
            \node[label_style] at (90-90*\i+90:1.3cm) {\i};
        }
        #1 
        #2 
        #3 
    \end{tikzpicture}
}

\newcommand{\NNPointDiagram}[3]{
    \begin{tikzpicture}[baseline=(current bounding box.center), scale=0.7]
        \draw[dotted, gray!30] (0,0) circle (1.5cm);
        
        \node[n_node] (N) at (0,0) {$n$};
        
        \node[dot] (p1) at (90:1.5cm) {};
        \node[label_style] at (90:1.8cm) {1};
        
        \node[dot] (p2) at (40:1.5cm) {};
        \node[label_style] at (40:1.8cm) {2};
        
        \node[dot] (p3) at (-10:1.5cm) {};
        \node[label_style] at (-10:1.8cm) {3};
        
        \node[dot] (pn) at (140:1.5cm) {};
        \node[label_style] at (140:1.8cm) {$n$};
        
        \node[dot] (pnm) at (190:1.5cm) {};
        \node[label_style] at (190:2.3cm) {$n-1$};
        
        \node[font=\small] at (270:1.5cm) {$\dots$};

        #1 
        #2 
        #3 
    \end{tikzpicture}
}

\newcommand{\NLineDiagram}[3]{
    \begin{tikzpicture}[baseline=(current bounding box.center), scale=0.8]
        \def\gap{0.8}      
        \def\vdist{1.5}    
        \def\egap{1.2}     
        
        \def\xone{1*\gap} \def\xtwo{2*\gap} \def\xthree{3*\gap}
        \def\xnmm{3*\gap + \egap + 0.5*\gap}
        \def\xnm{3*\gap + \egap + 2*\gap}
        \def\xn{3*\gap + \egap + 3*\gap}
        \def\xmid{3*\gap + 0.5*\egap + 0.5*\gap} 

        \draw[gray line] (0.5*\gap, \vdist) -- (\xn + 0.5*\gap, \vdist);
        \draw[gray line] (0.5*\gap, 0) -- (\xn + 0.5*\gap, 0);

        \node[dot] (U1) at (\xone, \vdist) {}; \node[above=2pt, label_style] at (U1) {1};
        \node[dot] (U2) at (\xtwo, \vdist) {}; \node[above=2pt, label_style] at (U2) {2};
        \node[dot] (U3) at (\xthree, \vdist) {}; \node[above=2pt, label_style] at (U3) {3};
        \node[label_style] at (\xmid, \vdist) {$\dots$};
        \node[dot] (Unmm) at (\xnmm, \vdist) {}; \node[above=2pt, label_style] at (Unmm) {$n-2$};
        \node[dot] (Unm) at (\xnm, \vdist) {}; \node[above=2pt, label_style] at (Unm) {$n-1$};
        \node[dot] (Un) at (\xn, \vdist) {}; \node[above=2pt, label_style] at (Un) {$n$};

        \node[dot] (L1) at (\xone, 0) {}; \node[below=2pt, label_style] at (L1) {1$'$};
        \node[dot] (L2) at (\xtwo, 0) {}; \node[below=2pt, label_style] at (L2) {2$'$};
        \node[dot] (L3) at (\xthree, 0) {}; \node[below=2pt, label_style] at (L3) {3$'$};
        \node[label_style] at (\xmid, 0) {$\dots$};
        \node[dot] (Lnmm) at (\xnmm, 0) {}; \node[below=2pt, label_style] at (Lnmm) {$(n-2)'$};
        \node[dot] (Lnm) at (\xnm, 0) {}; \node[below=2pt, label_style] at (Lnm) {$(n-1)'$};
        \node[dot] (Ln) at (\xn, 0) {}; \node[below=2pt, label_style] at (Ln) {$n'$};

        #1 #2 #3
    \end{tikzpicture}
}

\newcommand{\ThreePointDiagram}[3]{
    \begin{tikzpicture}[baseline=(current bounding box.center), scale=0.8]
        \def\gap{0.8}      
        \def\vdist{1.2}    

        \draw[gray line] (0.7*\gap, \vdist) -- (5.3*\gap, \vdist);
        \draw[gray line] (0.7*\gap, 0) -- (5.3*\gap, 0);

        \foreach \i in {1,...,3} {
            \node[dot] (U\i) at (1.5*\i*\gap, \vdist) {};
            \node[above=2pt, label_style] at (U\i) {\i};
            
            \node[dot] (L\i) at (1.5*\i*\gap, 0) {};
            \node[below=2pt, label_style] at (L\i) {\i$'$};
        }

        #1 
        #2 
        #3 
    \end{tikzpicture}
}

\newcommand{\FourPointDiagram}[3]{
    \begin{tikzpicture}[baseline=(current bounding box.center), scale=0.8]
        \def\gap{0.8}      
        \def\vdist{1.2}    

        \draw[gray line] (0.7*\gap, \vdist) -- (5.3*\gap, \vdist);
        \draw[gray line] (0.7*\gap, 0) -- (5.3*\gap, 0);

        \foreach \i in {1,...,4} {
            \node[dot] (U\i) at (4*\i*\gap/3, \vdist) {};
            \node[above=2pt, label_style] at (U\i) {\i};
            
            \node[dot] (L\i) at (4*\i*\gap/3, 0) {};
            \node[below=2pt, label_style] at (L\i) {\i$'$};
        }

        #1 
        #2 
        #3 
    \end{tikzpicture}
}

\newcommand{\TwoPointMidDots}[3]{
    \begin{tikzpicture}[baseline=(current bounding box.center), scale=0.8]
        \def\gap{1.2} \def\vdist{1.2}
        \draw[gray line] (-0.5*\gap, \vdist) -- (2.5*\gap, \vdist);
        \draw[gray line] (-0.5*\gap, 0) -- (2.5*\gap, 0);
        
        \node[label_style] at (0, \vdist) {$\dots$};
        \node[dot] (Ui) at (0.5*\gap, \vdist) {}; \node[above=2pt, label_style] at (Ui) {$i$};
        \node[label_style] at (1*\gap, \vdist) {$\dots$};
        \node[dot] (Uj) at (1.5*\gap, \vdist) {}; \node[above=2pt, label_style] at (Uj) {$j$};
        \node[label_style] at (2*\gap, \vdist) {$\dots$};

        \node[label_style] at (0, 0) {$\dots$};
        \node[dot] (Li) at (0.5*\gap, 0) {};      \node[below=2pt, label_style] at (Li) {$k'$};
        \node[label_style] at (1*\gap, 0) {$\dots$};
        \node[dot] (Lj) at (1.5*\gap, 0) {};      \node[below=2pt, label_style] at (Lj) {$l'$};
        \node[label_style] at (2*\gap, 0) {$\dots$};

        #1 #2 #3
    \end{tikzpicture}
}

\newcommand{\ThreeToOneMidDots}[3]{
    \begin{tikzpicture}[baseline=(current bounding box.center), scale=0.8]
        \def\gap{1.1} \def\vdist{1.2}
        \draw[gray line] (-0.5*\gap, \vdist) -- (3.5*\gap, \vdist);
        \draw[gray line] (-0.5*\gap, 0) -- (3.5*\gap, 0);
        
        \node[label_style] at (0, \vdist) {$\dots$};
        \node[dot] (Ui) at (0.7*\gap, \vdist) {}; \node[above=2pt, label_style] at (Ui) {$i$};
        \node[label_style] at (1.5*\gap, \vdist) {$\dots$};
        \node[dot] (Uj) at (2.3*\gap, \vdist) {}; \node[above=2pt, label_style] at (Uj) {$j$};
        \node[label_style] at (3.1*\gap, \vdist) {$\dots$}; 
        \node[dot] (Uk) at (3.9*\gap, \vdist) {}; \node[above=2pt, label_style] at (Uk) {$k$};
        \node[label_style] at (4.6*\gap, \vdist) {$\dots$};

        \node[label_style] at (0.7*\gap, 0) {$\dots$};
        \node[dot] (Li) at (2.3*\gap, 0) {};      \node[below=2pt, label_style] at (Li) {$l'$};
        \node[label_style] at (3.9*\gap, 0) {$\dots$};

        #1 #2 #3
    \end{tikzpicture}
}

\newcommand{\FourFourMidDots}[3]{
    \begin{tikzpicture}[baseline=(current bounding box.center), scale=0.6]
        \def\gap{1.0} \def\vdist{1.2}
        \draw[gray line] (-0.5*\gap, \vdist) -- (8.5*\gap, \vdist);
        \draw[gray line] (-0.5*\gap, 0) -- (8.5*\gap, 0);
        
        \node[label_style] at (0*\gap, \vdist) {$\dots$};
        \node[dot] (Ui) at (1*\gap, \vdist) {}; \node[above=2pt, label_style] at (Ui) {$i$};
        \node[label_style] at (2*\gap, \vdist) {$\dots$};
        \node[dot] (Uj) at (3*\gap, \vdist) {}; \node[above=2pt, label_style] at (Uj) {$j$};
        \node[label_style] at (4*\gap, \vdist) {$\dots$};
        \node[dot] (Uk) at (5*\gap, \vdist) {}; \node[above=2pt, label_style] at (Uk) {$k$};
        \node[label_style] at (6*\gap, \vdist) {$\dots$};
        \node[dot] (Ul) at (7*\gap, \vdist) {}; \node[above=2pt, label_style] at (Ul) {$l$};
        \node[label_style] at (8*\gap, \vdist) {$\dots$};

        \node[label_style] at (0*\gap, 0) {$\dots$};
        \node[dot] (Li) at (1*\gap, 0) {}; \node[below=2pt, label_style] at (Li) {$i'$};
        \node[label_style] at (2*\gap, 0) {$\dots$};
        \node[dot] (Lj) at (3*\gap, 0) {}; \node[below=2pt, label_style] at (Lj) {$j'$};
        \node[label_style] at (4*\gap, 0) {$\dots$};
        \node[dot] (Lk) at (5*\gap, 0) {}; \node[below=2pt, label_style] at (Lk) {$k'$};
        \node[label_style] at (6*\gap, 0) {$\dots$};
        \node[dot] (Ll) at (7*\gap, 0) {}; \node[below=2pt, label_style] at (Ll) {$l'$};
        \node[label_style] at (8*\gap, 0) {$\dots$};

        #1 #2 #3
    \end{tikzpicture}
}

\title{Spinning bulk-to-boundary correlators in massless theories with Poincar\'e symmetry}
\author{Jiang Long,}
\author{Yu-Xuan Wei}
\author{and Xin-Hao Zhou}
\affiliation{School of Physics, Huazhong University of Science and Technology,\\
Luoyu Road 1037, Wuhan, Hubei 430074, China}
\emailAdd{longjiang@hust.edu.cn}
\emailAdd{u202310191@hust.edu.cn}
\emailAdd{zhouxinhao01@hust.edu.cn}

\abstract{We classify the bulk-to-boundary correlators for general integer-spin $s$ operators in a Poincar\'e-invariant theory by imposing suitable fall-off conditions near future/past null infinity. Any bulk-to-boundary correlator is a linear superposition of a set of basic tensor structures fixed by the little group ISO(2) of massless particles. We map the independent tensor structures to all possible non-crossing double-line diagrams. A further mapping from the double-line diagrams to circular diagrams shows that all independent tensor structures can be expressed as tensor products of loop diagrams. By extrapolating the bulk-to-boundary correlators to boundary-to-boundary correlators, we find a rich structure for general spin-$s$ operators. Furthermore, we  show that the extrapolated operator belongs to a Type Ib spin-$s$ multiplet representation of Carrollian conformal field theory (CCFT).  This is a net representation  generated by the Wigner translation generators.
}

\begin{document}
\maketitle
\flushbottom

\section{Introduction}
In ordinary relativistic quantum field theory (QFT), Poincar\'e symmetry   plays a fundamental role  in classifying particles with different masses and spins \cite{weinberg_1995}. 
A relativistic particle transforms in an irreducible representation of the Poincar\'e group \cite{1939AnMat..40..149W}. For a massless representation, the spin structure and the wave equation are determined by the massless little group \text{ISO}(2) \cite{1948PNAS...34..211B}. However, Poincar\'e symmetry does not by itself determine general two-point functions. Additional dynamical input is encoded, for instance, in spectral densities of the {K\"{a}ll\'{e}n-Lehmann (KL)} representation \cite{Kallen:1952zz,Lehmann1954berEV}.
This is in sharp contrast with CFT, where global conformal symmetry already imposes strong restrictions on lower-point correlators \cite{Polyakov:1970xd,1974Non,Ferrara:1973eg}. {The} tensor structures of spinning correlators in CFTs have been explored in \cite{Osborn:1993cr,Costa:2011mg} and the conformal blocks have been investigated in \cite{Dolan:2000ut,Dolan:2003hv,Costa:2011dw}.

The recent observation \cite{Long:2026cpq} that  a suitable fall-off index $\Delta$ at null infinity makes Poincar\'e bulk-to-boundary correlators highly constrained suggests that  CFT methods could be adapted to null infinity. Interestingly, the null infinity of an asymptotically flat spacetime is at the core of the program of flat holography \cite{Polchinski:1999ry,Susskind:1998vk,Giddings:1999qu,Balasubramanian:1999ri,deBoer:2003vf,Gary:2009ae}, and also plays an important role in gravitational waves \cite{Bondi:1962px,Sachs:1962wk,PhysRev.128.2851,Barnich:2010eb} and soft theorems \cite{Strominger:2013jfa,He:2014laa}. In recent years, these various developments concerning null infinity have been  unified in the framework of Carrollian holography \cite{Bagchi:2025vri,Ruzziconi:2026bix}, which claims that the dual field theory is located on the Carrollian manifold.\footnote{The geometric foundation of the Carrollian manifold is reviewed in \cite{Ciambelli:2025unn}.} 
One of the central problems is the construction of CCFTs directly at null infinity \cite{Bagchi:2016bcd,Bagchi:2019xfx,Hao:2021urq,Banerjee:2020qjj,Opreij:2026bdx}. In this language, massless particles are represented by Carrollian conformal fields on $\mathbb R\times S^2$, and the Carrollian correlators obey Ward identities inherited from the Poincar\'e/BMS invariant action. In this direction, representations of the conformal Carrollian group are studied in \cite{Chen:2021xkw}, while embedding-space methods \cite{Salzer:2023jqv} and Carrollian OPEs \cite{Nguyen:2025sqk} are developed to determine Carrollian correlators. {Composite Carrollian primaries and their relation to celestial
operators were studied in the context of multiparticle states
in \cite{Kulp:2024scx}.} More discussion on Carrollian correlators and Ward identities can be found in \cite{Bagchi:2023fbj,Saha:2023hsl,2023Univ....9..385N,Nguyen:2023miw,Bagchi:2023cen,Ruzziconi:2024kzo}.
The Carrollian correlators are related to the S-matrix in the bulk via Fourier transforms \cite{Bagchi:2022emh,Donnay:2022aba,Donnay:2022wvx,Mason:2023mti,Liu:2024nfc,Stieberger:2024shv,Liu:2024llk,Alday:2024yyj,Li:2024kbo,Kraus:2024gso,Long:2025bfi,Surubaru:2025fmg,Lipstein:2025jfj,Kulkarni:2025qcx,Adamo:2025bfr,Long:2026rpq,Nenmeli:2026ket}. In the holographic Carrollian Feynman rules \cite{Liu:2024nfc}, one of the key ingredients is the bulk-to-boundary propagator. All these developments make the classification of bulk-to-boundary correlators a natural next step.

In this work, we classify the spinning bulk-to-boundary correlators in theories with Poincar\'e symmetry. Remarkably, {when appropriate fall-off conditions are imposed near future/past null infinity}, these bulk-to-boundary correlators are fixed to a set of basic tensor structures that are in one-to-one correspondence with non-crossing double-line diagrams. For fixed rank $n$, the number of independent tensor structures is given by the Catalan number. Furthermore, to read off the tensor structures directly from the spinor diagrams, we recast the double-line diagrams as equivalent circular diagrams with petals. A circular diagram factorizes  into a tensor product of elementary building blocks, which are completely determined by traces of Dirac gamma matrices.

The boundary-to-boundary correlators are obtained by  extrapolating the bulk points to the boundary. 
In general, the extrapolated operator contains multiple components. We use the stabilizer of the Poincar\'e group at  null infinity to classify these operators. Remarkably, the boundary operator belongs to a Type Ib  multiplet representation of the boundary CCFT. 
For a general spin-$s$ operator, the representation can be decomposed into a set of irreducible representations according to the symmetry of the indices. In particular, the structure of the symmetric traceless representation can be represented by a net diagram with $2s+1$ layers. At the top layer of the diagram, there is a unique component  with Carrollian conformal weight $\bar{\Delta}=\Delta+s$. By applying the Wigner translation generators $\mathbb K_I \ (I=1,2)$ to it iteratively,  one produces all the other components at the lower levels. 
There are similar net diagrams for the other representations, although the number of components at the top layer is not necessarily one.

The layout of this work is as follows. In Section~\ref{ward}, we introduce the Ward identities for the spinning bulk-to-boundary correlators. We solve the Ward identities using a tensorial formalism in Section~\ref{tensorial} and reformulate them in the two-spinor formalism in Section~\ref{spinor}. In the following section, we discuss various properties and applications of the bulk-to-boundary correlators.  {We also examine a formal second extrapolation, in which the remaining bulk point is taken to the boundary, and determine when the resulting distribution vanishes, remains finite, or becomes divergent. Finally, we organize the components of the leading asymptotic tensor coefficient into representations of the stabilizer algebra at null infinity. These component spaces furnish Type Ib multiplets in the algebraic sense explained in Section~\ref{bcca}.} We conclude with a summary and a discussion of future directions in the final section. Technical details are relegated to  five appendices. 
\section{Ward identities}\label{ward}
\subsection{\texorpdfstring{Conventions and  notation}{Conventions and notation}}
In this section, we  derive the Ward identities for the bulk-to-boundary correlators in a Poincar\'e invariant theory for general tensor fields in the bulk.  We follow the conventions and notation of \cite{Long:2026cpq}. 
The signature of the Minkowski spacetime is $(-,+,+,+)$. {The convention for  the Levi-Civita tensor is $ \epsilon_{0123}=1,\ \epsilon^{0123}=-1$.} The Cartesian coordinates are denoted by $x^\mu=(t,x^i)$ and they are related to the retarded coordinates via the relation 
\bea 
x^\mu=u \bar m^\mu+r n^\mu\label{xcoordr}
\eea where $r$ is the spatial radius in spherical coordinates $(r,\theta,\phi)$ and $u=t-r$ is the retarded time. The $\bar m^\mu=(1,0,0,0)$ is a timelike unit vector while $n^\mu=(1,n^i)$ is a null vector with $n^i=(\sin\theta\cos\phi,\sin\theta\sin\phi,\cos\theta)$. A dual null vector $\bar n^\mu=(-1,n^i)$ is chosen such that the  inner product $n\cdot\bar n=2$. We may also construct a  purely spatial vector $m^\mu=\frac{1}{2}(n^\mu+\bar n^\mu)=(0,n^i)$. The spacetime translation vector $P_\mu=\partial_\mu$ is 
\be 
P_\mu =-n_\mu \partial_u+m_\mu \partial_r-\frac{1}{r}Y_\mu^I\partial_I,
\ee where $Y_\mu^I=-\partial^I n_\mu$. The capital Latin indices \(I,J,\ldots\) take the values \(\theta\) and \(\phi\) and are raised with the inverse metric \(\gamma^{IJ}\) of \(S^2\) with 
\be 
\gamma_{IJ}=\left(\begin{array}{cc}1&0\\ 0&\sin^2\theta\end{array}\right).
\ee
We may introduce two vectors $\zeta^I$ and $\bar\zeta^I$ on the unit sphere 
\be 
\zeta^I=\left(\begin{array}{c}1\\ -\frac{i}{\sin\theta}\end{array}\right),\quad \bar{\zeta}^{I}=\left(\begin{array}{c}1\\ \frac{i}{\sin\theta}\end{array}\right)\label{zeta}
\ee
that satisfy the following identities \bea 
&&\zeta^I\zeta_I=0,\quad \zeta^I\bar{\zeta}_I=2,\quad \gamma^{IJ}=\frac{1}{2}(\zeta^I\bar{\zeta}^J+\zeta^J\bar{\zeta}^I). \label{zetaidentity}
\eea 
It follows that any vector field $X^I$ on the sphere can be decomposed as 
\bea 
X^I=\frac{1}{2}(\zeta^I\bar x +\bar{\zeta}^Ix)
\eea with 
\be 
x=X^I\zeta_I,\quad \bar x=X^I\bar\zeta_I.
\ee In particular, for the field $Y_\mu^I$, we find 
\be 
Y_\mu^I=\frac{1}{2}(y_\mu \bar\zeta^I+\bar y_\mu\zeta^I)
\ee where $y_\mu$ and $\bar y_\mu$ obey the identities
\be 
y\cdot \bar y=2,\quad y^2=\bar y^2=0,\quad n\cdot y=n\cdot\bar y=\bar n\cdot y=\bar n\cdot\bar y=0.
\ee Thus, we may define the vielbein field $e_\mu^a$ with 
\be 
e_\mu^1=n_\mu,\quad e_\mu^2=\bar n_\mu,\quad e_\mu^3 =y_\mu,\quad e_\mu^4=\bar y_\mu.
\ee They satisfy the  orthogonality and completeness relations
\bea 
e_\mu^a e_\nu^b \eta^{\mu\nu}=g^{ab},\quad e_\mu^a e_\nu^b g_{ab}=\eta_{\mu\nu}
\eea where 
\bea 
g_{ab}=\frac{1}{2}\left(\begin{array}{cccc}0&1&0&0\\ 1&0&0&0\\ 0&0&0&1\\ 0&0&1&0\end{array}\right),\quad g^{ab}=2\left(\begin{array}{cccc}0&1&0&0\\ 1&0&0&0\\ 0&0&0&1\\ 0&0&1&0\end{array}\right).
\eea 
According to the work of  \cite{1984ssv..book.....P}, we may choose a set of commuting spinors $o_A, \iota_A$ and their conjugates $\overline o_{\dot A},\overline\iota_{\dot A}$ such that the null vectors $n_\mu,\bar n_\mu, y_\mu,\bar y_\mu$  map to
\bea 
n_\mu\sigma^\mu_{A\dot A}=-o_A\overline o_{\dot A},\quad \bar n_\mu\sigma^\mu_{A\dot A}=\iota_A\overline \iota_{\dot A},\quad y_\mu\sigma^\mu_{A\dot A}=o_A\overline\iota_{\dot A},\quad \bar y_\mu\sigma^\mu_{A\dot A}=\iota_A\overline o_{\dot A}.
\eea Here the $2\times 2$ matrices $\sigma^\mu=(\sigma^0,\sigma^1,\sigma^2,\sigma^3)$ are chosen as 
\bea 
\sigma^0=\left(\begin{array}{cc}1&0\\ 0&1\end{array}\right),\quad \sigma^1=\left(\begin{array}{cc}0&1\\ 1&0\end{array}\right),\quad \sigma^2=\left(\begin{array}{cc}0&-i\\ i&0\end{array}\right),\quad\sigma^3=\left(\begin{array}{cc}1&0\\ 0&-1\end{array}\right).
\eea  The undotted and dotted spinor indices are also denoted by capital letters, taken from the beginning of the alphabet. They are  raised  and lowered by the symbols
\bea 
\epsilon^{AB}=\epsilon^{\dot A\dot B}=\left(\begin{array}{cc}0&1\\ -1&0\end{array}\right),\quad \epsilon_{AB}=\epsilon_{\dot A\dot B}=\left(\begin{array}{cc}0&-1\\ 1&0\end{array}\right)\eea 
under the convention
\bea 
\chi^A=\epsilon^{AB}\chi_B,\quad \chi_A=\epsilon_{AB}\chi^B,\quad \bar\chi^{\dot A}=\epsilon^{\dot A\dot B}\bar\chi_{\dot B},\quad \bar\chi_{\dot A}=\epsilon_{\dot A\dot B}\bar\chi^{\dot B}. 
\eea 
Note that the commuting spinors should satisfy the identities 
\bea 
o^A\iota_A=2,\quad \iota^Ao_A=-2,\quad \overline o^{\dot A}\overline \iota_{\dot A}=2,\quad \overline\iota^{\dot A}\overline o_{\dot A}=-2.
\eea We may also define matrices $\bar\sigma^\mu=(\sigma^0,-\sigma^1,-\sigma^2,-\sigma^3)$ such that
\be 
\bar\sigma^{\mu\dot A A}=\epsilon^{AB}\epsilon^{\dot A\dot B}\sigma^\mu_{B\dot B}.
\ee The Minkowski metric can be represented by 
\be 
\eta^{\mu\nu}=-\frac{1}{2}\sigma^\mu_{A\dot A}\bar\sigma^{\nu\dot A A}=-\frac{1}{2}\text{tr}\left(\sigma^\mu\bar\sigma^\nu\right)=-\frac{1}{2}\text{tr}\left(\bar\sigma^\mu\sigma^\nu\right).
\ee

\subsection{Constraining equations}
In this work, we consider a general massless theory in which a tensor field $t_{\mu_1\cdots\mu_s}$ satisfies the following fall-off condition near $\mathscr I^+$
\be 
t_{\mu_1\cdots\mu_s}(x)=\frac{\Sigma_{\mu_1\cdots\mu_s}(u,\Omega)}{r^\Delta}+\cdots.\label{falof}
\ee The constant $\Delta$ is called the fall-off index of the field $t_{\mu_1\cdots\mu_s}$ \footnote{At the classical level, asymptotic expansions such as \eqref{falof} are a standard method for studying gravitational waves near null infinity \cite{Bondi:1962px,Sachs:1962wk}. However, such expansions should not be interpreted as operator‑level asymptotic expansions. As shown in \cite{Long:2026cpq}, upon extrapolating both operators in a two‑point function to the boundary, the boundary‑to‑boundary correlator diverges when the fall‑off index exceeds 1. We will find a similar conclusion for the spinning case later. Fortunately, the bulk‑to‑boundary correlator remains well‑defined. We preserve this abuse of notation  in more generic cases to explore boundary correlators and representations, since it leads to reasonable results when $\Delta\le 1$.}. At this stage, we impose no additional symmetries on the indices of $t_{\mu_1\cdots\mu_s}$. The associated two-point correlator is 
\bea 
G_{\mu_1\cdots\mu_s\nu_1\cdots\nu_s}(x;y)=\langle \text{T} t_{\mu_1\cdots\mu_s}(x)t_{\nu_1\cdots\nu_s}(y)\rangle
\eea where $\text{T}$ is the time-ordering operator. The Ward identity for spacetime translations is 
\bea 
\left(\partial_\alpha^x+\partial_\alpha^y\right)G_{\mu_1\cdots\mu_s\nu_1\cdots\nu_s}(x;y)=0
\eea where  $\partial_\mu^x$ is the partial derivative with respect to $x^\mu$
\be
\partial_\mu^x=\frac{\partial}{\partial x^\mu}.
\ee
The same notation applies to the coordinate $y^\mu$.  Using the fall-off condition \eqref{falof}, we expand the two-point correlator asymptotically near $\mathscr I^+$
\be 
G_{\mu_1\cdots\mu_s\nu_1\cdots\nu_s}(x;x')=\frac{D_{\mu_1\cdots\mu_s\nu_1\cdots\nu_s}(u,\Omega;x')}{r^\Delta}+\cdots.\label{GDfal}
\ee Here \(D_{\mu_1\cdots\mu_s\nu_1\cdots\nu_s}(u,\Omega;x')\) is called the bulk-to-boundary correlator where the bulk field $t_{\nu_1\cdots\nu_s}$ is inserted at $x'$ while the boundary operator $\Sigma_{\mu_1\cdots\mu_s}(u,\Omega)$ is located at $(u,\Omega)$. At leading order, we find the Ward identity associated with spacetime translation for the bulk-to-boundary correlator 
\be 
\left(\partial_\mu' -n_\mu\partial_u\right)D_{\mu_1\cdots\mu_s\nu_1\cdots\nu_s}=0
\ee whose solution is 
\be 
D_{\mu_1\cdots\mu_s\nu_1\cdots\nu_s}=f_{\mu_1\cdots\mu_s\nu_1\cdots\nu_s}(\widehat u;\Omega).
\ee We have defined a symbol \(\widehat u=u+n\cdot x'\) that is invariant under spacetime translations.  Similarly, the Ward identity of the bulk-to-boundary correlator  arising from the Lorentz transformation is 
\begin{equation}
    \left( -\frac{1}{2} n_{\alpha\beta} \widehat u \partial_u - \frac{1}{2} \Delta n_{\alpha\beta} + \frac{1}{2} Y_{\alpha\beta}^I \delta_I \right) f_{\mu_1 \cdots \mu_{n}}- \frac{1}{2} \sum_{i=1}^{n} \left( \delta^\kappa_\alpha \eta_{\mu_i \beta} - \delta^\kappa_\beta \eta_{\mu_i \alpha} \right) f_{\mu_1 \cdots\mu_{i-1}\kappa\mu_{i+1} \mu_{n}}=0\label{wa}
\end{equation} where $n=2s$. Here the antisymmetric tensors are defined as 

\be 
n^{\mu\nu}=n^{[\mu}\bar n^{\nu]},\quad Y_{\mu\nu}^I=Y_\mu^I n_\nu-Y_\nu^I n_\mu
\ee
and 
\be 
\delta_I f_{\mu_1\cdots\mu_{n}}=\frac{\partial}{\partial\theta^I}f_{\mu_1\cdots\mu_{n}}(\widehat u;\Omega)\Big|_{\widehat u}.
\ee 
The terms \(\left(\delta^\kappa_\alpha \eta_{\mu \beta} - \delta^\kappa_\beta \eta_{\mu\alpha}\right)\) in the brackets in the summation are the generators of  Lorentz transformations.

Before closing this section, we should mention that one can also consider bulk-to-boundary correlators involving two operators of different spins, e.g. \(\langle \text{T} \Sigma_{\mu_1\cdots\mu_{s_1}} t_{\nu_1\cdots\nu_{s_2}}\rangle\). We still obtain the same equation \eqref{wa} with \(n = s_1 + s_2\). In what follows, we will solve this equation for arbitrary integer \(n\).

\section{Bulk-to-boundary correlators in tensor formalism}\label{tensorial}
To solve \eqref{wa}, we should project the equations onto all possible independent  bases. Note that the  antisymmetric tensors $n_{\mu\nu}$ and $Y_{\mu\nu}^I$ can be  expanded in the bases of the tensor product of the vielbeins
\bea 
n_{\mu\nu}&=&\frac{1}{2}(e_\mu^1 e_\nu^2-e_\nu^1 e_\mu^2),\\
Y_{\mu\nu}^I&=&\frac{1}{2}(y_\mu\bar\zeta^I+\bar y_\mu\zeta^I)n_\nu-(\mu\leftrightarrow\nu)=\frac{1}{2}(e_\mu^3 e_\nu^1-e_\nu^3 e_\mu^1)\bar\zeta^I+\frac{1}{2}(e_\mu^4 e_\nu^1-e_\nu^4 e_\mu^1)\zeta^I
\eea and the function $f_{\mu_1\cdots\mu_n}$ can be written as 
\be 
f_{\mu_1\cdots\mu_n}=f_{a_1\cdots a_n}e_{\mu_1}^{a_1}\cdots e_{\mu_n}^{a_n}.
\ee The Ward identity becomes
\bea 
&&-\frac{1}{4}(e_\mu^1 e_\nu^2-e_\nu^1 e_\mu^2)\left((u+n\cdot x')\partial_u+\Delta\right) f_{a_1\cdots a_n}(u+n\cdot x';\Omega) e_{\alpha_1}^{a_1}\cdots e_{\alpha_n}^{a_n}\nn\\&&+\frac{1}{4}\left((e_\mu^3 e_\nu^1-e_\nu^3 e_\mu^1)\bar\zeta^I\delta_I+(e_\mu^4 e_\nu^1-e_\nu^4 e_\mu^1)\zeta^I\delta_I\right)f_{a_1\cdots a_n}(u+n\cdot x';\Omega)e_{\alpha_1}^{a_1}\cdots e_{\alpha_n}^{a_n}\nn\\&&- \frac{1}{2} \sum_{i=1}^n \left( \delta^\kappa_\mu \eta_{\alpha_i \nu} - \delta^\kappa_\nu \eta_{\alpha_i \mu} \right) f_{a_1\cdots a_n}e_{\alpha_1}^{a_1} \cdots e_\kappa^{a_i} \cdots e_{\alpha_{n}}^{a_n}=0.
\eea 
 Using the projectors 
\( e^\mu_{c_1}e^\nu_{c_2}e^{\alpha_1}_{b_1}\cdots e_{b_n}^{\alpha_n}\), we find the equation 
\bea 
&&-\frac{1}{4}(\delta_{c_1}^1\delta_{c_2}^2-\delta_{c_2}^1\delta_{c_1}^2)\left(\widehat u\partial_u+\Delta\right)f_{b_1\cdots b_n}\nn\\&&+\frac{1}{4}(\delta_{c_1}^3\delta_{c_2}^1-\delta_{c_2}^3 \delta_{c_1}^1)\bar\zeta^I\delta_If_{b_1\cdots b_n}+\frac{1}{4}(\delta_{c_1}^4\delta_{c_2}^1-\delta_{c_2}^4 \delta_{c_1}^1)\zeta^I\delta_I f_{b_1\cdots b_n}\nn\\&&+\frac{1}{4}(\delta_{c_1}^3\delta_{c_2}^1-\delta_{c_2}^3 \delta_{c_1}^1)\sum_{i=1}^n f_{b_1\cdots a_i\cdots b_n} e^{\alpha_i}_{b_i}\bar\zeta^I\delta_I e_{\alpha_i}^{a_i}+\frac{1}{4}(\delta_{c_1}^4\delta_{c_2}^1-\delta_{c_2}^4 \delta_{c_1}^1)\sum_{i=1}^n f_{b_1\cdots a_i\cdots b_n} e_{b_i}^{\alpha_i}\zeta^I\delta_I e_{\alpha_i}^{a_i}\nn\\&&-\frac{1}{2}\sum_{i=1}^n (\delta^{a_i}_{c_1}g_{c_2 b_i}-\delta^{a_i}_{c_2}g_{c_1 b_i})f_{b_1\cdots a_i\cdots b_n}=0.\label{3.5}
\eea 
We have used the 
inverse of the vielbein 
\[e^\mu_1=\frac{1}{2}\bar n^\mu,\quad e^\mu_2=\frac{1}{2}n^\mu,\quad e^\mu_3=\frac{1}{2}\bar y^\mu,\quad e^\mu_4=\frac{1}{2}y^\mu.\] 
Note that the following combinations appear frequently in the equations
\bea 
\Upsilon^{a_i}_{b_i}&=&e_{b_i}^{\alpha}\zeta^I \delta_I e_{\alpha}^{a_i}=-\delta^{a_i}_1\delta^3_{b_i}-\delta^{a_i}_2\delta^3_{b_i}+\delta^{a_i}_4 \delta^1_{b_i}+\delta^{a_i}_4 \delta^2_{b_i}+\cot\theta (\delta^{a_i}_3\delta^3_{b_i}-\delta^{a_i}_4\delta^4_{b_i}),\\
\overline\Upsilon^{a_i}_{b_i}&=&e_{b_i}^{\alpha}\bar\zeta^I \delta_I e_{\alpha}^{a_i}=-\delta^4_{b_i}(\delta^{a_i}_1+\delta^{a_i}_2)+\delta^{a_i}_3(\delta_{b_i}^1+\delta_{b_i}^2)+\cot\theta (\delta^{a_i}_4\delta_{b_i}^4-\delta^{a_i}_3\delta_{b_i}^3).
\eea 
We have used the identities \bs\begin{align}
    \zeta^I\delta_I e_\mu^1&=-y_\mu=-e_\mu^3,\\
    \bar\zeta^I\delta_I e_\mu^1&=-\bar y_\mu=-e_\mu^4,\\
    \zeta^I \delta_I e_\mu^2&=-y_\mu=-e_\mu^3,\\
    \bar\zeta^I\delta_I e_\mu^2&=-\bar y_\mu=-e_\mu^4,\\
    \zeta^I\delta_I e_\mu^3&=\zeta^I\partial_I y_\mu=\cot\theta y_\mu=\cot\theta e_\mu^3,\\
    \bar\zeta^I\delta_I e_\mu^3&=\bar\zeta^I\partial_I y_\mu=n_\mu+\bar n_\mu-\cot\theta  y_\mu=e_\mu^1+e_\mu^2-\cot\theta e_\mu^3,\\
    \zeta^I\delta_I e_\mu^4&=\zeta^I\partial_I \bar y_\mu=n_\mu+\bar n_\mu-\cot\theta \bar y_\mu=e_\mu^1+e_\mu^2-\cot\theta e_\mu^4,\\
    \bar\zeta^I\delta_I e_\mu^4&=\bar\zeta^I\partial_I \bar y_\mu=\cot\theta \bar y_\mu=\cot\theta e_\mu^4.
\end{align}\es
The equation \eqref{3.5} is antisymmetric under \(c_1\leftrightarrow c_2\) and then the independent equations correspond to the following choices of $(c_1,c_2)$
\bs\begin{align}
    (1,2)&\Rightarrow -\frac{1}{4}\left(\widehat u\partial_u+\Delta\right)f_{b_1\cdots b_n}-\frac{1}{2}\sum_{i=1}^n(\delta_1^{a_i}g_{2b_i}-\delta_2^{a_i}g_{1b_i})f_{b_1\cdots a_i\cdots b_n}=0\\
    (1,3)&\Rightarrow -\frac{1}{4}\bar\zeta^I\delta_I f_{b_1\cdots b_n}-\frac{1}{4}\sum_{i=1}^n f_{b_1\cdots a_i\cdots b_n}\overline\Upsilon^{a_i}_{b_i}-\frac{1}{2}\sum_{i=1}^n(\delta_1^{a_i}g_{3b_i}-\delta_3^{a_i}g_{1b_i})f_{b_1\cdots a_i\cdots b_n}=0 \\
       (1,4)&\Rightarrow -\frac{1}{4}\zeta^I\delta_If_{b_1\cdots b_n}-\frac{1}{4}\sum_{i=1}^n f_{b_1\cdots a_i\cdots b_n}\Upsilon^{a_i}_{b_i} -\frac{1}{2}\sum_{i=1}^n(\delta_1^{a_i}g_{4b_i}-\delta_4^{a_i}g_{1b_i})f_{b_1\cdots a_i\cdots b_n}=0\\ 
           (2,3)&\Rightarrow \sum_{i=1}^n(\delta_2^{a_i}g_{3b_i}-\delta_3^{a_i}g_{2b_i})f_{b_1\cdots a_i\cdots b_n}=0\\
               (2,4)&\Rightarrow \sum_{i=1}^n(\delta_2^{a_i}g_{4b_i}-\delta_4^{a_i}g_{2b_i})f_{b_1\cdots a_i\cdots b_n}=0 \\
                   (3,4)&\Rightarrow \sum_{i=1}^n(\delta_3^{a_i}g_{4b_i}-\delta_4^{a_i}g_{3b_i})f_{b_1\cdots a_i\cdots b_n}=0.
\end{align}\es 
Substituting the last three equations into the second and third equations, we find 
\be 
\zeta^I\delta_I f_{b_1\cdots b_n}=\bar\zeta^I\delta_If_{b_1\cdots b_n}=0\quad\Rightarrow\quad f_{b_1\cdots b_n}=f_{b_1\cdots b_n}(\widehat u).
\ee Then the Ward identities become 
\bs\begin{align}
    \left(\widehat u\partial_u+\Delta+J_{12}\right)f_{b_1\cdots b_n}&=0,\label{ward12}\\
    J_{23}f_{b_1\cdots b_n}&=0,\\
    J_{24}f_{b_1\cdots b_n}&=0,\\
    J_{34}f_{b_1\cdots b_n}&=0
\end{align}\es
where $J_{12}, J_{23},J_{24},J_{34}$ are four of the Lorentz transformation  generators and their actions on the function $f_{b_1\cdots b_n}$ are 
\bea 
J_{cd}f_{b_1\cdots b_n}=2\sum_i\left(\delta_c^{a_i}g_{b_id}-\delta_d^{a_i}g_{b_i c}\right)f_{b_1\cdots a_i\cdots b_n}.
\eea
We rewrite these generators as
\bs\begin{align}
L_0f_{b_1\cdots b_n}&=J_{12}f_{b_1\cdots b_n}=\sum_{i=1}^n\left( \delta^{a_i}_1 \delta_{b_i}^1-\delta^{a_i}_2 \delta_{b_i}^2\right)f_{b_1\cdots a_i\cdots b_n},\\
L_1f_{b_1\cdots b_n}&=J_{23}f_{b_1\cdots b_n}=\sum_{i=1}^n \left( \delta^{a_i}_2 \delta_{b_i}^4-\delta^{a_i}_3 \delta_{b_i}^1\right)f_{b_1\cdots a_i\cdots b_n},\\
L_2f_{b_1\cdots b_n}&=J_{24}f_{b_1\cdots b_n}=\sum_{i=1}^n \left( \delta^{a_i}_2 \delta_{b_i}^3-\delta^{a_i}_4 \delta_{b_i}^1\right)f_{b_1\cdots a_i\cdots b_n},\\
L_3 f_{b_1\cdots b_n}&=J_{34}f_{b_1\cdots b_n}=\sum_{i=1}^n \left( \delta^{a_i}_3 \delta_{b_i}^3-\delta^{a_i}_4 \delta_{b_i}^4\right)f_{b_1\cdots a_i\cdots b_n}.
\end{align}\es  To simplify the notation, we define the action of the operator $D_c^d$  on the  rank-$n$ tensor $f_{b_1\cdots b_n}$:
\begin{equation}
    (D_c^d f)_{b_1\cdots b_n} = \sum_{j=1}^{n} \delta_{b_j}^d f_{b_1\cdots b_{j-1} c b_{j+1} \cdots b_n} = \sum_{j=1}^{n} \delta_{b_j}^d \delta_c^{a_j} f_{b_1\cdots a_j\cdots b_n}
\end{equation}
The physical meaning of this operator is to replace the index $d$ in the tensor with the index $c$. Therefore, 
\bea 
[D_a^b,D_c^d]=D_a^bD_c^d-D_c^dD_a^b={\delta_a^d D_c^b-\delta_c^b D_a^d.}
\eea Interestingly, the operators $L_0,L_1,L_2,L_3$ are linear superpositions of these $D_a^b$'s
\bea 
L_0=(D_1^1-D_2^2),\quad L_1=(D_2^4-D_3^1),\quad L_2=(D_2^3-D^1_4),\quad L_3=(D_3^3-D_4^4).
\eea 
Therefore, we find the following commutators 
\bs\begin{align}
    &[L_0,L_{1}]=L_1,\quad [L_{0},L_{2}]=L_2,\quad [L_{0},L_{3}]=0,\label{L0}\\
    &[L_{1},L_{2}]=0,\quad
    [L_{1},L_{3}]=L_1,\quad 
    [L_{2},L_{3}]=-L_2.\label{L123}
\end{align}\es 
From the second line, we conclude that $L_1,L_2,L_3$ form the Lie algebra $\mathfrak{iso} (2)$ and generate the little group $\text{ISO}(2)$ for a massless particle. The equations $L_1 f=L_2f=L_3f=0$ indicate that the tensor $f_{b_1\cdots b_n}$ is invariant under the little group.
The commutators in the first line indicate that $L_0$ is an extra dilatation operator. 

To solve the equations, notice that  the vector $f_{a}$ can be expanded in the basis $\{{\delta_a^1,\delta_a^2,\delta_a^3,\delta_a^4\}}$. Using the 
definition of $L_0,L_1,L_2,L_3$, we find 
\bs\begin{align}
    L_0\delta_a^1&=\delta_a^1,\quad L_0 \delta_a^2=-\delta_a^2,\quad L_0\delta_a^3=0,\quad L_0\delta_a^4=0,\\
    L_1\delta_a^1&=0,\quad L_1 \delta_a^2=\delta_a^4,\quad L_1\delta_a^3=-\delta_a^1,\quad L_1\delta_a^4=0,\\
    L_2\delta_a^1&=0,\quad L_2 \delta_a^2=\delta_a^3,\quad L_2\delta_a^3=0,\quad L_2\delta_a^4=-\delta_a^1,\\
    L_3\delta_a^1&=0,\quad \quad L_3 \delta_a^2=0,\quad L_3\delta_a^3=\delta_a^3,\quad L_3\delta_a^4=-\delta_a^4.
\end{align}\es 
The result indicates that $v_a=\delta_a^1$ is invariant under $L_1,L_2,L_3$ while all other vectors $\delta_a^2,\delta_a^3,\delta_a^4$ are not. Therefore, the invariant vector $f_a$ has the form 
\be 
f_a=f_1 \delta_a^1
\ee and its corresponding eigenvalue is 1 under $L_0$ 
\be 
L_0 f_a=f_1 L_0 \delta_a^1=f_a.
\ee Substituting into the equation \eqref{ward12}, we find 
\be 
f_a=C\delta_a^1 \widehat u^{-\Delta-1}
\ee where $C$ is a constant. Switching back to the Lorentz indices, we find the unique solution 
\be 
f_\mu=f_ae_\mu^a=C n_\mu \widehat u^{-\Delta-1}.
\ee 
The method can be extended to higher rank tensor fields. According to the theorem of Weyl \cite{Weyl1946ClassicalGroups}, all the tensors invariant under the little group are constructed from the following structures 
\be 
\delta_a^1,\quad g_{ab},\quad \epsilon_{abcd}\label{invariants}
\ee where $\epsilon_{abcd}$ is the Levi-Civita symbol. The $L_
0$ eigenvalues of both $g_{ab}$ and $\epsilon_{abcd}$ are zero 
\be 
L_0 g_{ab}=L_0\epsilon_{abcd}=0.
\ee 
For a general tensor with $n$ indices, we should construct a set of linearly independent invariant tensor bases $\{ \mathbf{T}^{(N)}_{a_1\cdots a_n} \}$ of rank $n$. Here, $N$ enumerates the independent tensor structures. We assume  the eigenvalue of a basis tensor $ \mathbf{T}^{(N)}_{a_1\cdots a_n}$  is $q_N$ \footnote{$q_N$ is determined by the number of $\delta_a^1$ in the structure.},
\be 
L_0  \mathbf{T}^{(N)}_{a_1\cdots a_n}=q_N \mathbf{T}^{(N)}_{a_1\cdots a_n}.
\ee 
Solving the Ward identity \eqref{ward12}, we find 
\bea 
 \mathbf{T}^{(N)}_{a_1\cdots a_n}=\mathbb{T}^{(N)}_{a_1\cdots a_n} \widehat u^{-\Delta-q_N}.
\eea 
{With} the vielbein field, the solution $\{ \mathbf{T}^{(N)}_{a_1\cdots a_n} \}$ is transformed into the basis with Lorentz indices $\{ \mathbf{T}^{(N)}_{\mu_1\cdots \mu_n} \}$
\bea 
 \mathbf{T}^{(N)}_{\mu_1\cdots \mu_n}= \mathbb{T}^{(N)}_{\mu_1\cdots \mu_n} \widehat u^{-\Delta-q_N}
\eea where $\mathbb{T}^{(N)}_{\mu_1\cdots \mu_n}$ is constructed from the three invariant structures 
\be 
n_\mu,\quad \eta_{\mu\nu},\quad \epsilon_{\mu\nu\rho\sigma}.\label{tensors}
\ee
We conclude that the solution of the bulk-to-boundary correlator $f_{\mu_1\cdots\mu_n}$ is a linear combination of $\{ \mathbf{T}^{(N)}_{\mu_1\cdots \mu_n} \}$
\be 
f_{\mu_1\cdots\mu_n}=\sum_{N}^{\#_n}C_N\mathbf{T}^{(N)}_{\mu_1\cdots \mu_n} 
\ee 
where $C_N$ are constants. The summation is over all possible independent rank-$n$ tensor structures that could be constructed via \(n_\mu, \eta_{\mu\nu}, \epsilon_{\mu\nu\rho\sigma}\). We use the symbol $\#_n$ to count the independent tensor structures of rank-$n$. In this case, $q_N$ counts the number of $n_\mu$'s in the tensor basis.

We list the independent structures in Table \ref{Is} for small  $n$. For small values of $ n$, these solutions can be enumerated by brute force.
\begin{table}[h]
    \centering
    \begin{tabular}{|c|c|c|}
\hline
  $n$ &  independent structures& $\#_n$ \\
\hline
0 & 1&1\\ \hline
1 & $n_\mu$&1\\ \hline
2 & $\eta_{\mu\nu},\ n_\mu n_\nu$&2\\ \hline
3&$\eta_{\mu\nu}n_\rho,\ n_\mu n_\nu n_\rho,\omega_{\mu\nu\rho}$&5\\\hline
4&$\eta_{\mu\nu}\eta_{\rho\sigma},\ \eta_{\mu\nu}n_\rho n_\sigma,\ n_\mu n_\nu n_\rho n_\sigma ,\epsilon_{\mu\nu\rho\sigma}, \omega_{\mu\nu\rho}n_\sigma$&14\\\hline
\end{tabular}
    \caption{\centering Independent structures for $n=0,1,2,3,4$.}
    \label{Is}
\end{table}
Note that for $n=3$, every basis tensor contains at least one \(n_\mu\). When the number of $n_\mu$'s is 3, there is a unique structure 
\be 
\mathbb T^{(1)}_{\mu\nu\rho}=n_\mu n_\nu n_\rho.
\ee When the number of $n_\mu$'s is 1, there are four independent tensor structures 
\bea 
\mathbb T^{(2)}_{\mu\nu\rho}=\eta_{\mu\nu}n_\rho,\quad \mathbb T^{(3)}_{\mu\nu\rho}=\eta_{\nu\rho}n_\mu,\quad \mathbb T^{(4)}_{\mu\nu\rho}=\eta_{\mu\rho}n_\nu,\quad \mathbb T^{(5)}_{\mu\nu\rho}=\epsilon_{\mu\nu\rho\sigma}n^\sigma.
\eea The first three structures are built by one $\eta$ and the null vector $n^\mu$. We just list $\eta_{\mu\nu}n_\rho$ in the table since the other two can be found by permutation. The last structure is built by the contraction of the Levi-Civita tensor $\epsilon_{\mu\nu\rho\sigma}$ and the null vector $n^\sigma$. We have used the notation 
\be 
\omega_{\mu\nu\rho}=\epsilon_{\mu\nu\rho\sigma}n^\sigma
\ee in the table. 

For $n=4$, we have listed all possible structures in the last row. They can be written schematically as
\be 
\eta \eta,\quad \eta n n,\quad nnnn,\quad \epsilon,\quad \omega n.
\ee 
Naively, the numbers of the structures are 
\(3,6,1,1,4\), respectively. However, the
identity 
\be 
\omega_{[\mu\nu\rho}n_{\sigma]}=0\label{antisym}
\ee  reduces the number of independent structures of $\omega n$ by 1. Therefore, we have $\#_4=3+6+1+1+3=14$ in the last column.

For $n\ge 5$, the number of independent solutions increases quickly. There are also various identities among the naive solutions. To illustrate the problem, we set $n=5$ and list the solutions. First, since two Levi-Civita tensors can always reduce to a set of products of $\eta$'s, we can decompose the solutions into parity-even and parity-odd parts. The parity-even part contains no Levi-Civita tensor and {its terms} are among the following types
\be 
\eta \eta n,\quad \eta n n n,\quad nnnnn.
\ee The number of independent terms is 
\bea 
\frac{1}{2}\left(\begin{array}{c}5\\ 2\end{array}\right)\left(\begin{array}{c}3\\ 2\end{array}\right)+\left(\begin{array}{c}5\\ 2\end{array}\right)+1=26.
\eea The parity-odd sector contains one Levi-Civita tensor and consists of structures of the following types:
\be 
\epsilon n,\quad \omega n n,\quad \omega\eta.
\ee 
Now we consider the type $\epsilon n$. Naively, there are 5 independent
terms. However, the antisymmetrization of any five indices in four dimensions always leads to zero. Therefore, there is an identity 
\be 
\epsilon_{[\mu\nu\rho\sigma}n_{\alpha]}=0.
\ee We conclude that
\be 
\text{number of independent structures of type}\ \epsilon n=5-1=4.
\ee 
Next, we consider the type $\omega n n$. There are $\binom{5}{3}=10$ naive combinations. However, \eqref{antisym} provides five constraints. Unfortunately, the five constraints are not completely independent. More precisely, the five constraints are 
\be 
n_\kappa n_{[\sigma}\omega_{\mu\nu\rho]}=0.
\ee There is a trivial identity for the five constraints 
\be 
n_{[\kappa}n_{\sigma}\omega_{\mu\nu\rho]}=0.
\ee Thus, we conclude that
\be 
\text{number of independent structures of type}\ \omega n n=10-(5-1)=6.
\ee 
We still need to count the number of {structures of} type $\omega \eta$. Naively, there are 10 independent structures. However, we note that there are four constraints
\be 
\omega_{[\mu\nu\rho}\eta_{\sigma]\alpha}=\frac{1}{4}\epsilon_{\mu\nu\rho\sigma}n_\alpha.
\ee We find
\be 
\text{number of independent structures of type}\ \omega \eta =10-4=6.
\ee Therefore, the parity-odd sector has $4+6+6=16$ independent structures. In total, we find $26+16=42$ independent structures for rank-5 correlators. 

Now we consider the case $n=6$. The parity-even solutions take the forms
\bea 
\eta\eta\eta,\quad \eta \eta nn,\quad \eta nnnn,\quad nnnnnn.
\eea The number of independent tensor structures is
\bea 
\frac{1}{3!}\left(\begin{array}{c}6\\2\end{array}\right)\left(\begin{array}{c}4\\2\end{array}\right)+\frac{1}{2}\left(\begin{array}{c}6\\2\end{array}\right)\left(\begin{array}{c}4\\2\end{array}\right)+\left(\begin{array}{c}6\\2\end{array}\right)+1=15+45+15+1=76.
\eea The parity-odd solutions are in the forms  of 
\bea 
\epsilon \eta,\quad \epsilon nn,\quad \omega \eta n,\quad \omega nnn.
\eea For the type $\epsilon \eta$, we have 6 constraints
\be 
\epsilon_{[\mu\nu\rho\sigma}\eta_{\alpha]\beta}=0.
\ee The above constraints themselves obey one constraint
\be 
\epsilon_{[\mu\nu\rho\sigma}\eta_{\alpha\beta]}=0.
\ee Therefore, we conclude that
\be 
\text{number of independent structures of type}\ \epsilon \eta =15-(6-1)=10.
\ee For the type $\epsilon n n$, we still have $6$ constraints 
\be 
\epsilon_{[\mu\nu\rho\sigma}n_{\alpha]}n_{\beta}=0.
\ee These constraints obey 1 additional constraint 
\be 
\epsilon_{[\mu\nu\rho\sigma}n_{\alpha}n_{\beta]}=0.
\ee It follows that 
\be 
\text{number of independent structures of type}\ \epsilon n n =15-(6-1)=10.
\ee For the type $\omega \eta n$, naively there are 60 independent terms. However, there are 15 constraints in the form of  
\bea 
\omega_{[\mu\nu\rho}n_{\sigma]}\eta_{\alpha\beta}=0.
\eea 
The above constraint equations obey $6-1=5$ constraints 
\be 
\omega_{[\mu\nu\rho}n_{\sigma}\eta_{\alpha]\beta}=0.\label{6omega}
\ee 

Besides, there are {30 additional} constraints 
\bea 
\omega_{\mu[\nu\rho}n_\sigma\eta_{\alpha]\beta}= \frac{1}{12}\epsilon_{\nu\rho\sigma\alpha}n_\mu n_\beta,
\eea five of which are not independent. We should mention that the condition $\omega_{[\mu\nu\rho}n_{\sigma}\eta_{\alpha\beta]}=0$ is subtracted twice. 
We conclude 
\be 
\text{number of independent structures of type}\ \omega \eta n =60-(15-5)-(30-5)+1=26.
\ee Finally, there are 20 tensors of the form $\omega nnn$ which obey fifteen constraints 
\be 
\omega_{[\mu\nu\rho}n_{\sigma]}n_{\alpha}n_{\beta}=0.
\ee These constraints obey six new constraints 
\be 
\omega_{[\mu\nu\rho}n_{\sigma}n_{\alpha]}n_{\beta}=0.
\ee 
The above constraints still satisfy one additional condition 
\be 
\omega_{[\mu\nu\rho}n_{\sigma}n_{\alpha}n_{\beta]}=0.
\ee 
We conclude that 
\be 
\text{number of independent structures of type}\ \omega nn n =20-(15-(6-1))=10.
\ee Therefore, there are $10+10+26+10=56$ independent rank-6 tensor structures in the parity-odd sector. In total, we find $76+56=132$ independent structures for rank-6 correlators.

To study the general solutions for arbitrary $n$, it would be better to switch to the  two-spinor formalism.

\section{Bulk-to-boundary correlators in two-spinor formalism}\label{spinor}
In the two-spinor formalism, a rank-$n$ tensor $f_{\mu_1\cdots\mu_n}$ is transformed into 
\be 
f_{A_1A_2\cdots A_n\dot A_1\dot A_2\cdots\dot A_n}=f_{\mu_1\mu_2\cdots\mu_n}\sigma^{\mu_1}_{A_1\dot A_1}\cdots\sigma^{\mu_n}_{A_n\dot A_n}.
\ee Conversely, we have
\bea 
f_{\mu_1\cdots\mu_n}=\left(-\frac{1}{2}\right)^n f_{A_1A_2\cdots A_n\dot A_1\dot A_2\cdots \dot A_n}\bar\sigma_{\mu_1}^{\dot A_1A_1}\cdots\bar\sigma_{\mu_n}^{\dot A_nA_n}.\label{invtrans}
\eea In this notation, the basic building blocks $n_\mu,\eta_{\mu\nu},\epsilon_{\mu\nu\rho\sigma}$ are transformed  as
\bs\begin{align}
    n_{A\dot A}&=n_\mu\sigma^\mu_{A\dot A}=-o_A\overline o_{\dot A},\\
    \eta_{AB\dot A\dot B}&=\eta_{\mu\nu}\sigma^\mu_{A\dot A}\sigma^\nu_{B\dot B}=-2\epsilon_{AB}\epsilon_{\dot A\dot B},\\
   \epsilon_{ABCD\dot A\dot B\dot C\dot D}&=\epsilon_{\mu\nu\rho\sigma}\sigma^\mu_{A\dot A}\sigma^\nu_{B\dot B}\sigma^\rho_{C\dot C}\sigma^\sigma_{D\dot D}=-4i\left( \epsilon_{AB}\epsilon_{\dot B\dot C}\epsilon_{CD}\epsilon_{\dot D\dot A} - \epsilon_{\dot{A}\dot{B}}\epsilon_{B C}\epsilon_{\dot{C}\dot{D}}\epsilon_{DA} \right).
\end{align}\es Therefore, we conclude that the building blocks of the independent structures 
\be \mathbb T^{(N)}_{A_1\cdots A_n\dot A_1\cdots \dot A_n}=T^{(N)}_{\mu_1\cdots\mu_n}\sigma^{\mu_1}_{A_1\dot A_1}\cdots\sigma^{\mu_n}_{A_n\dot A_n} \ee  are $\epsilon_{AB},\epsilon_{\dot A\dot B}, o_A,\overline o_{\dot A}$.
Furthermore, the number of $\epsilon_{AB}$'s should be equal to the number of $\epsilon_{\dot A\dot B}$'s. In addition, the number of $o_A$'s should be the same as the number of $\overline o_{\dot A}$'s. One could also solve the Ward identities in the two-spinor formalism and obtain this result explicitly.

\begin{figure}
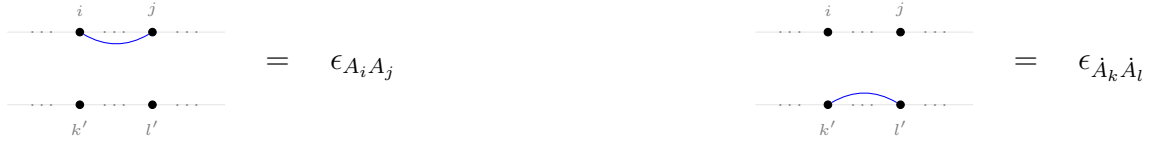

\begin{center}
\begin{minipage}{0.45\textwidth}
\centering
\[
    \TwoPointMidDots{\draw[blue] (Ui) to[out=-30, in=-150] (Uj);}{;}{;} \quad = \quad \epsilon_{A_i A_j}
\]
\end{minipage}\hfill
\begin{minipage}{0.45\textwidth}
\centering
\[
    \TwoPointMidDots{\draw[blue] (Li) to[out=30, in=150] (Lj);}{;}{;} \quad = \quad \epsilon_{\dot{A}_k\dot{A}_l}
\]
\end{minipage}
\end{center}\caption{\centering The diagrammatic representation of $\epsilon_{A_iA_j}$ and $\epsilon_{\dot A_i\dot A_j}$.}\label{ep}\end{figure}

There is {a }diagrammatic way to 
represent a quantity such as $\mathbb T^{(N)}_{A_1\cdots A_n\dot A_1\cdots \dot A_n}$. We draw two  horizontal straight lines, one above the other, each with $n$ points. The $n$ indices are placed on the upper line in the sequence $A_1,A_2,\cdots,A_n$, and the $n$ indices $\dot A_1,\dot A_2,\cdots,\dot A_n$ on the lower line. We represent $\epsilon_{A_iA_j}$ by a blue string joining $A_i$ and $A_j$, and represent $\epsilon_{\dot A_i\dot A_j}$ by a blue string joining $\dot A_i$ and $\dot A_j$. This is shown in Figure \ref{ep}. Note that in the diagram we omit the capital letter $A$ and keep only the numerical subscripts. Thus an index $A_i$ is simplified to $i$, and a dotted index $\dot A_k$ becomes $k'$. Thus, the following figure represents $\epsilon_{A_1A_2}\epsilon_{\dot A_1\dot A_2}$ which is proportional to $\eta_{\mu_1\mu_2}$
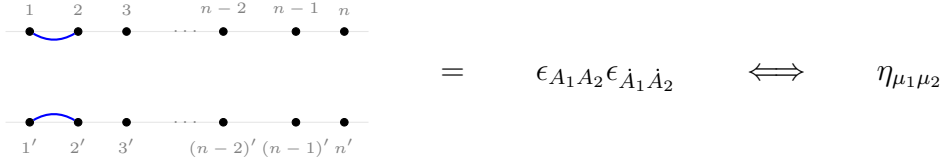
\begin{equation}
        \NLineDiagram{  
        \draw[blue arc] (U1) to[out=-30, in=-150] (U2);
        \draw[blue arc] (L1) to[out=30, in=150] (L2);
    ;}{
    ;}
    {
    }\qquad =\qquad \epsilon_{A_1A_2}\epsilon_{\dot A_1\dot A_2}\qquad\Longleftrightarrow \qquad \eta_{\mu_1\mu_2}\nn
    \end{equation}

Furthermore, a red string joining an undotted index $A_i$ with a dotted index $\dot A_j$ is used to represent the pair $o_{A_i}\overline o_{\dot A_j}$. For example, the following figure is $o_{A_1}\overline o_{\dot A_1}$ and can be transformed into $n_{\mu_1}$ 
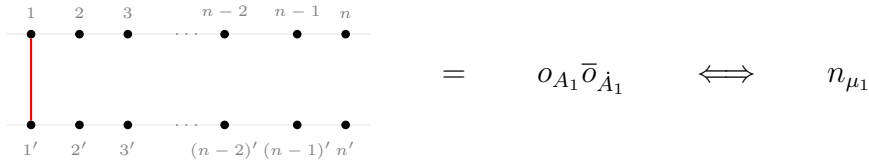
\begin{equation}
        \NLineDiagram{  
    ;}{
    \draw[red curve] (U1) to[out=-90, in=90] (L1);}
    {
    }\qquad =\qquad o_{A_1}\overline o_{\dot A_1}\qquad \Longleftrightarrow \qquad n_{\mu_1}\nn
    \end{equation}

In principle, we can use $n$ strings to connect $2n$ points freely and use \eqref{invtrans} to transform it back to the corresponding tensor. As an example, the following diagram is translated as
\bea \begin{tikzpicture}[
    baseline=(current bounding box.center), 
    dot/.style={circle, fill=black, inner sep=1.2pt}, 
    gray line/.style={lightgray!40, thin},            
    blue arc/.style={blue, thick},                
    red curve/.style={red, thick},                
    node distance=0.6cm                               
]

    \def\n{6}         
    \def\gap{0.8}     
    \def\vdist{1.2}   

    \draw[gray line] (0.7*\gap, \vdist) -- (\n*\gap + 0.3*\gap, \vdist);
    \draw[gray line] (0.7*\gap, 0) -- (\n*\gap + 0.3*\gap, 0);

    \foreach \i in {1,...,\n} {
        \node[dot] (U\i) at (\i*\gap, \vdist) {};
        \node[above=2pt, font=\scriptsize] at (U\i) {\i};
        
        \node[dot] (L\i) at (\i*\gap, 0) {};
        \node[below=2pt, font=\scriptsize] at (L\i) {\i$'$};
    }

    \draw[blue arc] (U1) to[bend right=50] (U3);
    \draw[blue arc] (L2) to[bend left=50] (L4);

    \draw[red curve] (U2) to[out=-90, in=90] (L1);
    \draw[red curve] (U5) to[out=-120, in=60] (L3);
    \draw[red curve] (U6) to[out=-90, in=90] (L6);
    \draw[red curve] (U4) to[out=-90, in=90] (L5);
\end{tikzpicture}\quad = \quad \epsilon_{A_1A_3}\epsilon_{\dot A_2\dot A_4}o_{A_2}\overline o_{\dot A_1}o_{A_4}\overline o_{\dot A_5}o_{A_5}\overline o_{\dot A_3}o_{A_6}\overline o_{\dot A_6}\label{mixetype}\eea and its corresponding tensor is  \[\left(-\frac{1}{2}\right)^6 \left(\prod_{j=1}^6 \bar \sigma^{\mu_j\dot A_j A_j}\right)\epsilon_{A_1A_3}\epsilon_{\dot A_2\dot A_4}o_{A_2}\overline o_{\dot A_1}o_{A_4}\overline o_{\dot A_5}o_{A_5}\overline o_{\dot A_3}o_{A_6}\overline o_{\dot A_6}.\] To find all independent structures for a fixed $n$, we 
should take into account the Schouten identity
\be
\epsilon_{AB}\epsilon_{CD}+\epsilon_{BC}\epsilon_{AD}+\epsilon_{CA}\epsilon_{BD}=0\label{rule21}
\ee which can be represented by the following diagrams
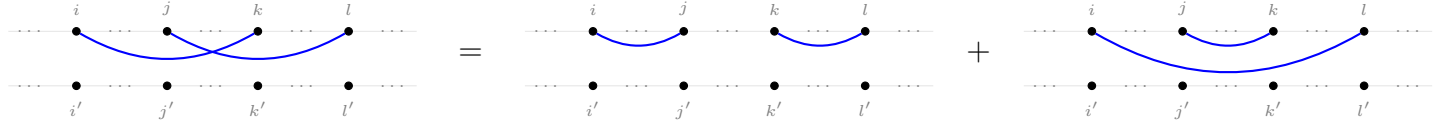
\begin{equation}
    \begin{aligned}
\FourFourMidDots{\draw[blue arc] (Ui) to[out=-30, in=-150] (Uk);\draw[blue arc] (Uj) to[out=-30, in=-150] (Ul);}{;}{;} &=\quad \FourFourMidDots{\draw[blue arc] (Ui) to[out=-30, in=-150] (Uj);\draw[blue arc] (Uk) to[out=-30, in=-150] (Ul);}{;}{;}\\ &\quad +\quad \FourFourMidDots{\draw[blue arc] (Ui) to[out=-30, in=-150] (Ul);\draw[blue arc] (Uj) to[out=-30, in=-150] (Uk);}{;}{;}
    \end{aligned}\nn
\end{equation} Similarly, we have 
\be 
\epsilon_{\dot A\dot B}\epsilon_{\dot C\dot D}+\epsilon_{\dot B\dot C}\epsilon_{\dot A\dot D}+\epsilon_{\dot C\dot A}\epsilon_{\dot B\dot D}=0\label{rule222}
\ee and the corresponding diagrams are
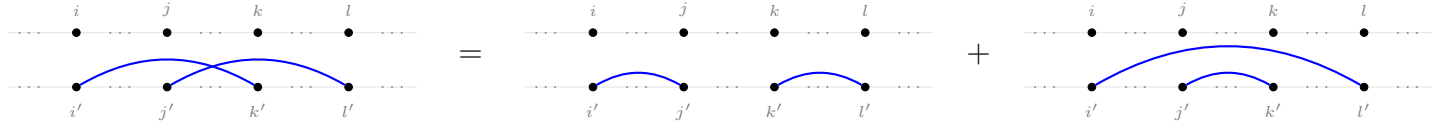
\begin{equation}
    \begin{aligned}
\FourFourMidDots{\draw[blue arc] (Li) to[out=30, in=150] (Lk);\draw[blue arc] (Lj) to[out=30, in=150] (Ll);}{;}{;} &=\quad \FourFourMidDots{\draw[blue arc] (Li) to[out=30, in=150] (Lj);\draw[blue arc] (Lk) to[out=30, in=150] (Ll);}{;}{;}\\ &\quad +\quad \FourFourMidDots{\draw[blue arc] (Li) to[out=30, in=150] (Ll);\draw[blue arc] (Lj) to[out=30, in=150] (Lk);}{;}{;}
    \end{aligned}\nn
\end{equation} The strings on the left-hand side cross over each other, while those on the right-hand side do not. Therefore, the Schouten identity can be used to transform a plain crossing diagram to two  plain non-crossing diagrams. A similar Schouten identity is 
\be 
\epsilon_{AB}o_C \overline o_{\dot D}+\epsilon_{BC}o_A\overline o_{\dot D}+\epsilon_{CA}o_B\overline o_{\dot D}=0.    \label{rule3}
\ee The diagrammatic representation is 

\begin{equation}
    \begin{aligned}
\ThreeToOneMidDots{\draw[red curve] (Uj) to[out=-90, in=90] (Li);\draw[blue arc] (Ui) to[out=-30, in=-150] (Uk);}{;}{;} &=\quad             \ThreeToOneMidDots{\draw[red curve] (Uk) to[out=-90, in=90] (Li);\draw[blue arc] (Ui) to[out=-30, in=-150] (Uj);}{;}{;}\\ &\quad +\quad             \ThreeToOneMidDots{\draw[red curve] (Ui) to[out=-90, in=90] (Li);\draw[blue arc] (Uj) to[out=-30, in=-150] (Uk);}{;}{;}
    \end{aligned}\nn
\end{equation} This shows that the Schouten identity can untie plain crossing diagrams with one red  string and one blue  string. In addition, there is a trivial identity for four commuting spinors \be 
o_{A}\overline o_{\dot B}o_C\overline o_{\dot D}=o_A\overline o_{\dot D}o_C\overline o_{\dot B}. \label{rule1}
\ee The diagrammatic representation is 
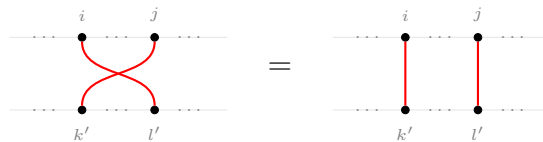
\begin{equation}
    \TwoPointMidDots{\draw[red curve] (Ui) to[out=-90, in=90] (Lj);\draw[red curve] (Uj) to[out=-90, in=90] (Li);}{;}{;} \quad  =\quad     \TwoPointMidDots{\draw[red curve] (Ui) to[out=-90, in=90] (Li);\draw[red curve] (Uj) to[out=-90, in=90] (Lj);}{;}{;}\nn
\end{equation} Again, a  plain crossing diagram can be untied to a  plain non-crossing diagram. 

We have identified three independent elementary crossing-diagrams above. We will call the crossing diagram 
\begingroup\postdisplaypenalty=10000 $$ \TwoPointMidDots{\draw[red curve] (Ui) to[out=-90, in=90] (Lj);\draw[red curve] (Uj) to[out=-90, in=90] (Li);}{;}{;}$$\endgroup
 the red X-shape, the diagram 
\begingroup\postdisplaypenalty=10000 $$\FourFourMidDots{\draw[blue arc] (Li) to[out=30, in=150] (Lk);\draw[blue arc] (Lj) to[out=30, in=150] (Ll);}{;}{;}$$\endgroup
 {the blue X-shape} and the diagram 
\begingroup\postdisplaypenalty=10000 $$\ThreeToOneMidDots{\draw[red curve] (Uj) to[out=-90, in=90] (Li);\draw[blue arc] (Ui) to[out=-30, in=-150] (Uk);}{;}{;}$$\endgroup
 the $\Psi$-shape. Note that there are more complicated crossing diagrams such as \eqref{mixetype}. However, all crossing diagrams are woven from these three basic crossing diagrams. Consider a crossing diagram containing $N_1$ red X-shapes, $N_2$ blue X-shapes and $N_3$ $\Psi$-shapes. Locally, a blue X-shape crossing is untied by \eqref{rule21} or \eqref{rule222} while keeping all the remaining X-shape crossings unchanged. Repeating this process, we conclude that all the blue X-shape crossings can be  untied. Now we use \eqref{rule3} or its complex conjugate  to untie a $\Psi$-shape crossing. Clearly, this process does not create any new blue X-shape crossings. Repeating the process, we untie all the $\Psi$-shape crossings without increasing any blue X-shape crossings. Now the remaining crossings are red  X-shape crossings. Assuming that there are $m$ red strings, possibly with red X-shape crossings, there must be $m$ undotted points  on the upper line and $m$ dotted points on the lower line. We could label them as $C_1,C_2,\cdots,C_m$ and $\dot C_1,\dot C_2,\cdots,\dot C_m$, respectively. We can remove the original red strings and replace them with strings connecting $C_i$ and $\dot C_i$ respectively. Then all the red X-shape crossings are untied. Therefore,  
we conclude that all the independent structures can be represented by non-crossing diagrams. The rules to draw these independent diagrams are as follows: 
\begin{enumerate}
    \item There are $n$ upper points (corresponding to $A_i$) and $n$  lower points (corresponding to $\dot{A}_i$), arranged evenly on two horizontal straight lines.
    \item  Rules for adding strings.
    \begin{enumerate}
            \item Each blue string links two upper points, or two lower points. The number of blue strings connected to the upper line equals the number connected to the lower line.
            \item Each red string links an upper point and a lower point.
            \item Each point can be linked by a string only once, regardless of color.
            \item  No two strings cross.
    \end{enumerate}
\end{enumerate}

In the following, we list all the independent diagrams for  $n=0,1,2,3,4$.
\begin{enumerate}
\item $n=0$. The diagram is empty and the tensor structure is $1$.
    \item $n=1$. There is only one independent diagram:
    \begin{equation}
 \begin{tikzpicture}[baseline=(current bounding box.center), scale=0.8]
        \def\gap{0.8}     
        \def\vdist{1.2}   
        \draw[gray line] (0.7*\gap, \vdist) -- (1.7*\gap, \vdist);
        \draw[gray line] (0.7*\gap, 0) -- (1.7*\gap, 0);
            \node[dot] (U1) at (1.2, \vdist) {};
            \node[above=2pt, label_style] at (U1) {1};
            \node[dot] (L1) at (1.2, 0) {};
            \node[below=2pt, label_style] at (L1) {1$'$};
            \draw[red curve] (U1) to[out=-90, in=90] (L1);
           \end{tikzpicture} \quad \Leftrightarrow \quad n_{\mu_1}.   \end{equation}
    \item $n=2$. There are two independent diagrams.  One has no blue strings
       \begin{equation}
\begin{tikzpicture}[baseline=(current bounding box.center), scale=0.8]
        \def\gap{1.2} \def\vdist{1.2}
        \draw[gray line] (-0.5*\gap, \vdist) -- (2.5*\gap, \vdist);
        \draw[gray line] (-0.5*\gap, 0) -- (2.5*\gap, 0);
        \node[dot] (U1) at (0.5*\gap, \vdist) {}; \node[above=2pt, label_style] at (U1) {$1$};
        \node[dot] (U2) at (1.5*\gap, \vdist) {}; \node[above=2pt, label_style] at (U2) {$2$};
        \node[dot] (L1) at (0.5*\gap, 0) {};      \node[below=2pt, label_style] at (L1) {$1'$};
        \node[dot] (L2) at (1.5*\gap, 0) {};      \node[below=2pt, label_style] at (L2) {$2'$};
\draw[red curve] (U1) to[out=-90, in=90] (L1);
\draw[red curve] (U2) to[out=-90, in=90] (L2);
    \end{tikzpicture} \quad \Leftrightarrow \quad n_{\mu_1}  n_{\mu_2}.    
\end{equation}
{The other} contains only blue strings
    \begin{equation}
\begin{tikzpicture}[baseline=(current bounding box.center), scale=0.8]
        \def\gap{1.2} \def\vdist{1.2}
        \draw[gray line] (-0.5*\gap, \vdist) -- (2.5*\gap, \vdist);
        \draw[gray line] (-0.5*\gap, 0) -- (2.5*\gap, 0);
        \node[dot] (U1) at (0.5*\gap, \vdist) {}; \node[above=2pt, label_style] at (U1) {$1$};
        \node[dot] (U2) at (1.5*\gap, \vdist) {}; \node[above=2pt, label_style] at (U2) {$2$};
        \node[dot] (L1) at (0.5*\gap, 0) {};      \node[below=2pt, label_style] at (L1) {$1'$};
        \node[dot] (L2) at (1.5*\gap, 0) {};      \node[below=2pt, label_style] at (L2) {$2'$};
\draw[blue arc] (U1) to[out=-30, in=-150] (U2);
\draw[blue arc] (L1) to[out=30, in=150] (L2);
    \end{tikzpicture} \quad \Leftrightarrow \quad \eta_{\mu_1\mu_2}.     
\end{equation}
    \item $n=3$. There are five independent diagrams. The diagram without blue strings is 
    \be 
    \ThreePointDiagram{
}{
    \draw[red curve] (U1) to[out=-90, in=90] (L1);
      \draw[red curve] (U2) to[out=-90, in=90] (L2);    \draw[red curve] (U3) to[out=-90, in=90] (L3);  
}{
   
} \quad=\quad o_{A_1}\overline o_{\dot A_1}o_{A_2}\overline o_{\dot A_2}o_{A_3}\overline o_{\dot A_3}\quad\Leftrightarrow\quad n_{\mu_1}n_{\mu_2}n_{\mu_3}.
    \ee In addition, there are four diagrams with two blue strings
    \bs\begin{align}
    \ThreePointDiagram{   \draw[blue arc] (U1) to[out=-30, in=-150] (U2);  \draw[blue arc] (L1) to[out=30, in=150] (L2);
}{
   \draw[red curve] (U3) to[out=-90, in=90] (L3);
}{
} \quad& =\quad \epsilon_{A_1A_2}\epsilon_{\dot A_1\dot A_2}o_{A_3}\overline o_{\dot A_3}\quad\Leftrightarrow\quad \eta_{\mu_1\mu_2}n_{\mu_3},\\
\ThreePointDiagram{   \draw[blue arc] (U2) to[out=-30, in=-150] (U3);  \draw[blue arc] (L2) to[out=30, in=150] (L3);
}{
   \draw[red curve] (U1) to[out=-90, in=90] (L1);
}{
} \quad& =\quad \epsilon_{A_2A_3}\epsilon_{\dot A_2\dot A_3}o_{A_1}\overline o_{\dot A_1}\quad\Leftrightarrow\quad \eta_{\mu_2\mu_3}n_{\mu_1}
    \end{align}\es and
    \bs\begin{align}
     \ThreePointDiagram{   \draw[blue arc] (U1) to[out=-30, in=-150] (U2);  \draw[blue arc] (L2) to[out=30, in=150] (L3);
}{
   \draw[red curve] (U3) to[out=-90, in=90] (L1);
}{
} \quad &=\quad \epsilon_{A_1A_2}\epsilon_{\dot A_2\dot A_3}o_{A_3}\overline o_{\dot A_1}\quad\nonumber\\*
&\Leftrightarrow\quad {\eta_{\mu_1\mu_2}n_{\mu_3}+\eta_{\mu_2\mu_3}n_{\mu_1}-\eta_{\mu_1\mu_3}n_{\mu_2}+i\omega_{\mu_1\mu_2\mu_3}},\\
\ThreePointDiagram{   \draw[blue arc] (U2) to[out=-30, in=-150] (U3);  \draw[blue arc] (L1) to[out=30, in=150] (L2);
}{
   \draw[red curve] (U1) to[out=-90, in=90] (L3);
}{
} \quad &=\quad \epsilon_{A_2A_3}\epsilon_{\dot A_1\dot A_2}o_{A_1}\overline o_{\dot A_3}\quad\nonumber\\*
&\Leftrightarrow\quad { \eta_{\mu_1\mu_2}n_{\mu_3}+\eta_{\mu_2\mu_3}n_{\mu_1}-\eta_{\mu_1\mu_3}n_{\mu_2}-i\omega_{\mu_1\mu_2\mu_3}}.
    \end{align}\es  
    Note that there is no diagram in which the blue string connects points 1 and 3 as there would be a $\Psi$-shape crossing. Interestingly, the last two diagrams are related via complex conjugation 
    \be 
    \left( \ThreePointDiagram{   \draw[blue arc] (U1) to[out=-30, in=-150] (U2);  \draw[blue arc] (L2) to[out=30, in=150] (L3);
}{
   \draw[red curve] (U3) to[out=-90, in=90] (L1);
}{
} \right)^*=\ThreePointDiagram{   \draw[blue arc] (U2) to[out=-30, in=-150] (U3);  \draw[blue arc] (L1) to[out=30, in=150] (L2);
}{
   \draw[red curve] (U1) to[out=-90, in=90] (L3);
}{
} .
    \ee 
    The other diagrams correspond to real tensors since the diagrams are invariant under complex conjugation. For example, 
    \be 
   \left( \ThreePointDiagram{   \draw[blue arc] (U2) to[out=-30, in=-150] (U3);  \draw[blue arc] (L2) to[out=30, in=150] (L3);
}{
   \draw[red curve] (U1) to[out=-90, in=90] (L1);
}{
} \right)^*=\ThreePointDiagram{   \draw[blue arc] (U2) to[out=-30, in=-150] (U3);  \draw[blue arc] (L2) to[out=30, in=150] (L3);
}{
   \draw[red curve] (U1) to[out=-90, in=90] (L1);
}{
} .
    \ee 
    
{A general diagrammatic rule for complex conjugation can be stated as
follows. Given a double-line diagram \(\mathfrak D\), its complex conjugate is
obtained by reflecting \(\mathfrak D\) about the horizontal axis midway between
the two lines. 
Equivalently, one interchanges \(i\leftrightarrow i'\) for every
\(i=1,\ldots,n\). Thus, a blue string on the upper (lower) line is
mapped to the corresponding blue string on the lower (upper) line,
whereas a red string connecting \(i\) to \(j'\) is mapped to a red
string connecting \(j\) to \(i'\).

More formally, suppose that a diagram contains \(N_{\rm U}\) blue strings on the
upper line, \(N_{\rm L}\) blue strings on the lower line, and
\(N_{\rm R}\) red strings connecting the two lines. Denote their
endpoints on the diagram by
\[
(i_a,j_a),\qquad
(i'_b,j'_b),\qquad
(k_c,\ell'_c),
\]
respectively, where
\[
a=1,\ldots,N_{\rm U},\qquad
b=1,\ldots,N_{\rm L},\qquad
c=1,\ldots,N_{\rm R}.
\]
The two-spinor structure represented by the diagram \(\mathfrak D\) is then
\begin{equation}
\mathbb{T}(\mathfrak D)
=
\prod_{a=1}^{N_{\rm U}}
\epsilon_{A_{i_a}A_{j_a}}
\prod_{b=1}^{N_{\rm L}}
\epsilon_{\dot A_{i_b}\dot A_{j_b}}
\prod_{c=1}^{N_{\rm R}}
 o_{A_{k_c}}\bar{o}_{\dot A_{\ell_c}}.
\label{eq:diagram-spinor-structure}
\end{equation}
Here the endpoints of each blue string are ordered from left to
right. Under complex conjugation, {the representations} $(\frac{1}{2},0)$ and $(0,\frac{1}{2})$ exchange roles. Thus one should exchange the dotted and undotted indices as follows \cite{Dreiner:2008tw}
\begin{equation}
\begin{aligned}
\epsilon_{A_iA_j}
&\longleftrightarrow
\epsilon_{\dot A_{i}\dot A_{j}},\\
o_{A_i}\bar{o}_{\dot A_{j}}
&\longleftrightarrow
o_{A_j}\bar{o}_{\dot A_{i}}.
\end{aligned}
\label{eq:elementary-conjugation-rules}
\end{equation}
Therefore,
\begin{equation}
\mathbb{T}(\mathfrak D)^*
=
\prod_{b=1}^{N_{\rm L}}
\epsilon_{A_{i_b}A_{j_b}}
\prod_{a=1}^{N_{\rm U}}
\epsilon_{\dot A_{i_a}\dot A_{j_a}}
\prod_{c=1}^{N_{\rm R}}
o_{A_{\ell_c}}\bar{o}_{\dot A_{k_c}} .
\label{eq:conjugate-diagram-spinor-structure}
\end{equation}
This is precisely the spinor structure obtained by interchanging the
upper and lower lines. Equivalently, on the diagram,
\[
i\longleftrightarrow i'
\]
for every pair of vertically aligned points.

}

    \item $n=4$. There are fourteen independent diagrams in total. The unique diagram without any blue string is 
    \be 
    \FourPointDiagram{
}{
    \draw[red curve] (U1) to[out=-90, in=90] (L1);
      \draw[red curve] (U2) to[out=-90, in=90] (L2);    \draw[red curve] (U3) to[out=-90, in=90] (L3);  \draw[red curve] (U4) to[out=-90, in=90] (L4); 
}{
   
} \quad=\quad o_{A_1}\overline o_{\dot A_1}o_{A_2}\overline o_{\dot A_2}o_{A_3}\overline o_{\dot A_3}o_{A_4}\overline o_{\dot A_4}\quad\Leftrightarrow\quad n_{\mu_1}n_{\mu_2}n_{\mu_3}n_{\mu_4}.
    \ee There are nine independent diagrams with two blue strings:
    \bs\begin{align}
    \FourPointDiagram{\draw[blue arc] (U1) to[out=-30, in=-150] (U2);  \draw[blue arc] (L1) to[out=30, in=150] (L2);
}{
    \draw[red curve] (U3) to[out=-90, in=90] (L3);    \draw[red curve] (U4) to[out=-90, in=90] (L4); 
}{
} &= \epsilon_{A_1A_2}\epsilon_{\dot A_1\dot A_2}o_{A_3}\overline o_{\dot A_3}o_{A_4}\overline o_{\dot A_4}\nonumber\\*
&\Leftrightarrow  \eta_{\mu_1\mu_2}n_{\mu_3}n_{\mu_4}\label{per12},\\
\FourPointDiagram{\draw[blue arc] (U1) to[out=-30, in=-150] (U2);  \draw[blue arc] (L2) to[out=30, in=150] (L3);
}{
    \draw[red curve] (U3) to[out=-90, in=90] (L1);    \draw[red curve] (U4) to[out=-90, in=90] (L4); 
}{
} &=\epsilon_{A_1A_2}\epsilon_{\dot A_2\dot A_3}o_{A_3}\overline o_{\dot A_1}o_{A_4}\overline o_{\dot A_4}\nonumber\\*
&\Leftrightarrow \left(\eta_{\mu_1\mu_2}n_{\mu_3}+\eta_{\mu_2\mu_3}n_{\mu_1}-\eta_{\mu_1\mu_3}n_{\mu_2}+i\omega_{\mu_1\mu_2\mu_3}\right) n_{\mu_4},\label{per13} \\ 
\FourPointDiagram{\draw[blue arc] (U1) to[out=-30, in=-150] (U2);  \draw[blue arc] (L3) to[out=30, in=150] (L4);
}{
    \draw[red curve] (U3) to[out=-90, in=90] (L1);    \draw[red curve] (U4) to[out=-90, in=90] (L2); 
}{
} &=\epsilon_{A_1A_2}\epsilon_{\dot A_3\dot A_4}o_{A_3}\overline o_{\dot A_1}o_{A_4}\overline o_{\dot A_2}\nonumber\\*
&\Leftrightarrow 
{ n_{[\mu_1}\eta_{\mu_2][\mu_3}n_{\mu_4]}+\frac{i}{4}(\omega_{\mu_1\mu_2[\mu_3}n_{\mu_4]}+n_{[\mu_1}\omega_{\mu_2]\mu_3\mu_4})}.
\label{per14}
    \end{align}\es   There are three diagrams of type \eqref{per12}, four of type \eqref{per13} \footnote{Including the complex conjugation.}, and two of type \eqref{per14}. Furthermore, there are four independent diagrams with four blue strings.
    \bs\begin{align}
        \FourPointDiagram{\draw[blue arc] (U1) to[out=-30, in=-150] (U2);  \draw[blue arc] (L1) to[out=30, in=150] (L2);\draw[blue arc] (U3) to[out=-30, in=-150] (U4);  \draw[blue arc] (L3) to[out=30, in=150] (L4);
}{
}{
}&=\epsilon_{A_1A_2}\epsilon_{A_3A_4}\epsilon_{\dot A_1\dot A_2}\epsilon_{\dot A_3\dot A_4}\nonumber\\*
&\Leftrightarrow \eta_{\mu_1\mu_2}\eta_{\mu_3\mu_4},\\
\FourPointDiagram{\draw[blue arc] (U1) to[out=-30, in=-150] (U2);  \draw[blue arc] (L1) to[out=30, in=150] (L4);\draw[blue arc] (U3) to[out=-30, in=-150] (U4);  \draw[blue arc] (L2) to[out=30, in=150] (L3);
}{
}{
}&=\epsilon_{A_1A_2}\epsilon_{A_3A_4}\epsilon_{\dot A_1\dot A_4}\epsilon_{\dot A_2\dot A_3}\nonumber\\*
&\Leftrightarrow  -\eta_{\mu_1\mu_2}\eta_{\mu_3\mu_4} - \eta_{\mu_2\mu_3}\eta_{\mu_1\mu_4} + \eta_{\mu_1\mu_3}\eta_{\mu_2\mu_4} - \mathrm{i}\,\epsilon_{\mu_1\mu_2\mu_3\mu_4},\\
\FourPointDiagram{\draw[blue arc] (U1) to[out=-30, in=-150] (U4);  \draw[blue arc] (L1) to[out=30, in=150] (L4);\draw[blue arc] (U2) to[out=-30, in=-150] (U3);  \draw[blue arc] (L2) to[out=30, in=150] (L3);
}{
}{
}&=\epsilon_{A_1A_4}\epsilon_{A_2A_3}\epsilon_{\dot A_1\dot A_4}\epsilon_{\dot A_2\dot A_3}\nonumber\\*
&\Leftrightarrow \eta_{\mu_1\mu_4}\eta_{\mu_2\mu_3}.
    \end{align}\es Only the second diagram has an independent complex conjugation.
\end{enumerate}
Notice that the number of independent tensor structures is the same as { in} the previous section. It is easy to extend the  diagrammatic method to higher rank bulk-to-boundary correlators. To count the number of independent non-crossing diagrams for general $n$, we first deform the two horizontal lines to a circle by joining the endpoints of the lines, as is shown in the following diagram:
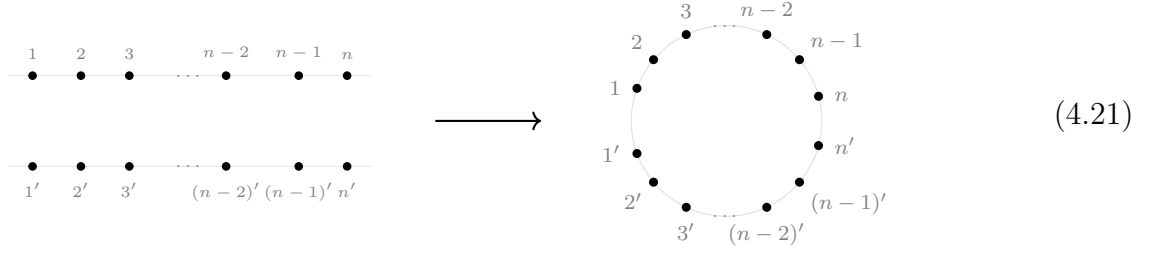
\begin{equation}
\NLineDiagram{;}{;}{;} \qquad 
\begin{tikzpicture}[baseline=(current bounding box.center), scale=0.7]
    \tikzset{
        dot/.style={circle, fill=black, inner sep=1.2pt}, 
        gray circle/.style={lightgray!50, thin},          
        blue line/.style={blue, thick},                  
        red line/.style={red, thick},                   
        label_style/.style={font=\scriptsize, text=gray},
        dots_style/.style={text=gray, font=\tiny}             
    }

    \quad \quad \draw[thick, ->] (-5.5, 0) -- (-3.5, 0); \quad \quad

    \draw[gray circle] (0,0) circle (1.8cm);

    \node[dot] (p1) at (160:1.8) {};  \node[label_style, left=2pt] at (160:1.8) {1};
    \node[dot] (p2) at (140:1.8) {};  \node[label_style, above left] at (140:1.8) {2};
    \node[dot] (p3) at (115:1.8) {};  \node[label_style, above=2pt] at (115:1.8) {3};
    
    \node[dots_style] at (90:1.8) {$\dots$};

    \node[dot] (pnm2) at (65:1.8) {}; \node[label_style, above=2pt] at (65:1.8) {$n-2$};
    \node[dot] (pnm1) at (40:1.8) {}; \node[label_style, above right] at (40:1.8) {$n-1$};
    \node[dot] (pn) at (15:1.8) {};   \node[label_style, right=2pt] at (15:1.8) {$n$};

    \node[dot] (p1p) at (200:1.8) {};  \node[label_style, left=2pt] at (200:1.8) {$1'$};
    \node[dot] (p2p) at (220:1.8) {};  \node[label_style, below left] at (220:1.8) {$2'$};
    \node[dot] (p3p) at (245:1.8) {};  \node[label_style, below=2pt] at (245:1.8) {$3'$};

    \node[dots_style] at (270:1.8) {$\dots$};

    \node[dot] (pnm2p) at (295:1.8) {}; \node[label_style, below=2pt] at (295:1.8) {$(n-2)'$};
    \node[dot] (pnm1p) at (320:1.8) {}; \node[label_style, below right] at (320:1.8) {$(n-1)'$};
    \node[dot] (pnp) at (345:1.8) {};   \node[label_style, right=2pt] at (345:1.8) {$n'$};

\end{tikzpicture}
\end{equation}
Note that the diagram is still non-crossing. The number of ways to pair 
$2n$ points on a circle into 
$n$ non-intersecting strings is the Catalan number 
\cite{RomanSteven2015AItC}
\be 
C_n=\frac{1}{n+1}\left(\begin{array}{cc}2n\\n\end{array}\right).
\ee Therefore, the number of independent tensor structures for any fixed $n$ is exactly the Catalan number
\be 
\#_n=C_n.
\ee For small values of $n$, the Catalan numbers are 
\bea 
1,1,2,5,14,42,132,\cdots.
\eea Note that this series matches the counting numbers in the tensor formalism. 

 For a symmetric traceless tensor in the irreducible representation of the Lorentz group, all the parity-odd terms vanish and 
 the candidate structure can only contain the following three types of metric
factors:
\[
\eta_{\mu_i\mu_j},\qquad
\eta_{\nu_i\nu_j},\qquad
\eta_{\mu_i\nu_j}.
\]
After projecting onto the symmetric traceless part in the \(\mu\)-indices,
all terms containing \(\eta_{\mu_i\mu_j}\) vanish, since they are traces within
the \(\mu\)-indices. Similarly, all terms containing \(\eta_{\nu_i\nu_j}\)
vanish after the symmetric traceless projection in the \(\nu\)-indices.
Thus only cross metrics \(\eta_{\mu_i\nu_j}\) can survive.
If there are \(J\) cross metrics, the remaining indices must be filled with 
\(n_\mu\). Since the \(\mu\)-indices and \(\nu\)-indices are separately
symmetrized, only \(J\) matters. The surviving structures are therefore
\[
\mathcal T^{(J)}_{\mu_1\cdots\mu_s,\nu_1\cdots\nu_s}
=
\left(
\eta_{\mu_1\nu_1}\cdots \eta_{\mu_J\nu_J}
n_{\mu_{J+1}}\cdots n_{\mu_s}
n_{\nu_{J+1}}\cdots n_{\nu_s}
\right)^{\text{STT}},
\qquad
J=0,1,\ldots,s .
\]
They are symmetric and traceless in each index group by construction and we  accordingly use a superscript $\text{STT}$ in the expression. Moreover, for a fixed 
\(J\), \(\mathcal T^{(J)}_{\mu_1\cdots\mu_s,\nu_1\cdots\nu_s}\) contains  \(2(s-J)\) factors of \(n_\mu\), so under
the rule that only terms with the same number of \(n_\mu\)'s may be linearly
combined, each belongs to a distinct sector. Hence the number of independent
structures for a rank-$n$ symmetric traceless tensor is
\be \#_s^{\text{STT}}=s+1.\label{numberSTT}\ee

\paragraph{An equivalent diagrammatic representation} Interestingly, we may also transform a double-line diagram to a circular diagram. To be more precise, we transform a pair of upper and lower points $i$ and $i'$ in the double-line diagram into a single point on a circle, which simultaneously represents the spinor indices $A_i$ and $\dot A_i$
as well as {the Lorentz index} $\mu_i$. When converting from the double-line diagram to the circular diagram, a blue string on the upper part becomes a solid blue line connecting two points on the circle, while a blue string on the lower part becomes a red dashed line also connecting two points on the circle. The red strings connecting the upper and lower parts become: 1) The end starting from the upper line becomes a solid blue line on the circular diagram connecting the boundary point to the center. 2)
The end starting from the lower line corresponds to a red dashed line on the circular diagram connecting the boundary point to the center. Now it is easy to observe that:
 \begin{enumerate}
        \item Each point on the circle can only be and must be linked by one dashed line and one solid line.
         \item The total number of solid lines linking the center must equal  that of the dashed lines.
         \item Each  diagram can be divided into several loops.
        \end{enumerate}
In the following, we explain the circular diagrams  for $n=0,1,2,3,4$. 
\begin{enumerate}
    \item $n=0$. The circular diagram  is empty.
    \item $n=1$. There is only one   circular diagram:
     \begin{equation}
         \begin{tikzpicture}[baseline=(current bounding box.center), scale=0.7]
        \draw[dotted, gray!30] (0,0) circle (1cm);
        \node[n_node] (N) at (0,0) {n};
        \foreach \i in {1,...,1} {
            \node[dot] (p\i) at (90-90*\i+90:1cm) {};
            \node[label_style] at (90-90*\i+90:1.3cm) {\i};
        }
        \draw[red_line](p1)to[bend left](N);
        \draw[blue_line](p1)to[bend right](N);
    \end{tikzpicture} \quad \Leftrightarrow \quad n_{\mu_1}
    \end{equation} Note that we have omitted the circle and added a label \begin{tikzpicture}[baseline=-0.6ex, scale=0.23]
        \draw[dotted, gray!30] (0,0) circle (1cm);
        \node[n_node] (N) at (0,0) {n};
    \end{tikzpicture} at the center of the circle. Once a pair of blue and red lines connects to the center point, this implies that there should be a null vector $n_\mu$ in the corresponding tensor.
    
    \item $n=2$. There are two inequivalent circular diagrams: 
    \begin{equation}
        \begin{tikzpicture}[baseline=(current bounding box.center), scale=0.7]
        \draw[dotted, gray!30] (0,0) circle (1cm);
        \node[n_node] (N) at (0,0) {n};
        \foreach \i in {1,...,2} {
            \node[dot] (p\i) at (90-180*\i+90:1cm) {};
            \node[label_style] at (90-180*\i+90:1.3cm) {\i};
        }
        \draw[red_line](p1)to[bend left](N);
        \draw[blue_line](p1)to[bend right](N);
        \draw[red_line](p2)to[bend left](N);
        \draw[blue_line](p2)to[bend right](N);
    \end{tikzpicture} \quad \Leftrightarrow \quad n_{\mu_1}n_{\mu_2}
    \end{equation}

    and 
    \begin{equation}
    \begin{tikzpicture}[baseline=(current bounding box.center), scale=0.7]
        \draw[dotted, gray!30] (0,0) circle (1cm);
        \node[n_node] (N) at (0,0) {n};
        \foreach \i in {1,...,2} {
            \node[dot] (p\i) at (90-180*\i+90:1cm) {};
            \node[label_style] at (90-180*\i+90:1.3cm) {\i};
        }
        \draw[red_line](p1)to[bend left](p2);
        \draw[blue_line](p1)to[bend right](p2);
    \end{tikzpicture}
     \quad \Leftrightarrow \quad \eta_{\mu_1\mu_2}
    \end{equation}

    \item $n=3$. There are five independent circular diagrams. One diagram {has} three loops{:}
    \be 
    \BasisDiagramthree{;}{;}{\draw[blue_line](p1)to[bend left](N);\draw[red_line](p1)to[bend right](N);\draw[blue_line](p2)to[bend left](N);\draw[red_line](p2)to[bend right](N);\draw[blue_line](p3)to[bend left](N);\draw[red_line](p3)to[bend right](N);}  \quad\Leftrightarrow\quad
n_{\mu_1} n_{\mu_2} n_{\mu_3}{.}
\ee {Two} diagrams {have} two loops:
\bs\begin{align} 
\BasisDiagramthree{\draw[blue_line](p1)to[bend left](p2);}{\draw[red_line](p1)--(p2);}{\draw[blue_line](p3)to[bend left](N);\draw[red_line](p3)to[bend right](N);} \quad&\Leftrightarrow\quad \eta_{\mu_1\mu_2}n_{\mu_3}\\
\BasisDiagramthree{\draw[blue_line](p2)to[bend left](p3);}{\draw[red_line](p2)--(p3);}{\draw[blue_line](p1)to[bend left](N);\draw[red_line](p1)to[bend right](N);} \quad&\Leftrightarrow\quad \eta_{\mu_2\mu_3}n_{\mu_1}{.}
\end{align}\es  {Two} diagrams {have} only one loop:
\bs\begin{align}
    \BasisDiagramthree{\draw[blue_line](p1)to[bend left](p2);\draw[red_line](p2)to[bend left](p3);}{;}{\draw[red_line](p1)--(N);\draw[blue_line](p3)--(N);} \quad&\Leftrightarrow\quad 
\eta_{\mu_1\mu_2}n_{\mu_3}+\eta_{\mu_2\mu_3}n_{\mu_1}-\eta_{\mu_1\mu_3}n_{\mu_2}+i\omega_{\mu_1\mu_2\mu_3}\label{dia31}\\
\BasisDiagramthree{\draw[red_line](p1)to[bend left](p2);\draw[blue_line](p2)to[bend left](p3);}{;}{\draw[blue_line](p1)--(N);\draw[red_line](p3)--(N);} \quad&\Leftrightarrow\quad 
\eta_{\mu_1\mu_2}n_{\mu_3}+\eta_{\mu_2\mu_3}n_{\mu_1}-\eta_{\mu_1\mu_3}n_{\mu_2}-i\omega_{\mu_1\mu_2\mu_3}\label{dia32}{.}
\end{align}\es Note that the complex conjugate is equivalent to exchanging the solid blue line and the dashed red line. Therefore, we have 
\be 
\left(\BasisDiagramthree{\draw[blue_line](p1)to[bend left](p2);\draw[red_line](p2)to[bend left](p3);}{;}{\draw[red_line](p1)--(N);\draw[blue_line](p3)--(N);} \right)^*=\BasisDiagramthree{\draw[red_line](p1)to[bend left](p2);\draw[blue_line](p2)to[bend left](p3);}{;}{\draw[blue_line](p1)--(N);\draw[red_line](p3)--(N);}.
\ee We should  remind the reader that the following diagram is not independent since this is a crossing diagram in the double-line diagram
\be 
\BasisDiagramthree{\draw[red_line](p1)to[bend right](p3);}{\draw[blue_line](p1)--(p3);}{\draw[blue_line](p2)to[bend left](N);\draw[red_line](p2)to[bend right](N);} \quad\Leftrightarrow\quad \eta_{\mu_1\mu_3}n_{\mu_2}.
\ee 
    \item $n=4$. There are fourteen independent circular diagrams. The diagram with four loops is 
    \be 
    \BasisDiagramfour{;}{;}{\draw[blue_line](p1)to[bend left](N);\draw[red_line](p1)to[bend right](N);\draw[blue_line](p2)to[bend left](N);\draw[red_line](p2)to[bend right](N);\draw[blue_line](p3)to[bend left](N);\draw[red_line](p3)to[bend right](N);\draw[blue_line](p4)to[bend left](N);\draw[red_line](p4)to[bend right](N);} \quad\Leftrightarrow\quad  
n_{\mu_1} n_{\mu_2} n_{\mu_3} n_{\mu_4}.
    \ee The diagrams with three loops are 
    \bs\begin{align}
        \BasisDiagramfour{\draw[blue_line](p1)to[bend left](p2);}{\draw[red_line](p1)--(p2);}{\draw[blue_line](p3)to[bend left](N);\draw[red_line](p3)to[bend right](N);\draw[blue_line](p4)to[bend left](N);\draw[red_line](p4)to[bend right](N);} \quad& \Leftrightarrow\quad \eta_{\mu_1\mu_2} n_{\mu_3}n_{\mu_4}\\
         \BasisDiagramfour{\draw[blue_line](p2)to[bend left](p3);}{\draw[red_line](p2)--(p3);}{\draw[blue_line](p1)to[bend left](N);\draw[red_line](p1)to[bend right](N);\draw[blue_line](p4)to[bend left](N);\draw[red_line](p4)to[bend right](N);} \quad& \Leftrightarrow\quad \eta_{\mu_2\mu_3} n_{\mu_1}n_{\mu_4}\\
          \BasisDiagramfour{\draw[blue_line](p3)to[bend left](p4);}{\draw[red_line](p3)--(p4);}{\draw[blue_line](p1)to[bend left](N);\draw[red_line](p1)to[bend right](N);\draw[blue_line](p2)to[bend left](N);\draw[red_line](p2)to[bend right](N);} \quad& \Leftrightarrow\quad \eta_{\mu_3\mu_4} n_{\mu_1}n_{\mu_2}.
    \end{align}\es Diagrams with two loops are 
    \bs\begin{align}
        \BasisDiagramfour{\draw[blue_line](p1)to[bend left](p2);\draw[blue_line](p3)to[bend left](p4);}{\draw[red_line](p1)--(p2);\draw[red_line](p3)--(p4);}{;} \quad& \Leftrightarrow\quad 
\eta_{\mu_1\mu_2}\eta_{\mu_3\mu_4}\\
\BasisDiagramfour{\draw[blue_line](p1)to[bend right](p4);\draw[blue_line](p2)to[bend left](p3);}{\draw[red_line](p1)--(p4);\draw[red_line](p2)--(p3);}{;} \quad& \Leftrightarrow\quad 
\eta_{\mu_1\mu_4}\eta_{\mu_2\mu_3}
    \end{align}\es  and
    \bs
    \begin{align}
        \BasisDiagramfour{\draw[blue_line](p1)to[bend left](p2);\draw[red_line](p2)to[bend left](p3);}{;}{\draw[red_line](p1)--(N);\draw[blue_line](p3)--(N);\draw[blue_line](p4)to[bend left](N);\draw[red_line](p4)to[bend right](N);}  \quad& \Leftrightarrow\quad
\left(\eta_{\mu_1\mu_2}n_{\mu_3}+\eta_{\mu_2\mu_3}n_{\mu_1}-\eta_{\mu_1\mu_3}n_{\mu_2}+i\omega_{\mu_1\mu_2\mu_3}\right)n_{\mu_4},\\
   \BasisDiagramfour{\draw[blue_line](p2)to[bend left](p3);\draw[red_line](p3)to[bend left](p4);}{;}{\draw[red_line](p2)--(N);\draw[blue_line](p4)--(N);\draw[blue_line](p1)to[bend left](N);\draw[red_line](p1)to[bend right](N);}  \quad& \Leftrightarrow\quad
\left(\eta_{\mu_2\mu_3}n_{\mu_4}+\eta_{\mu_3\mu_4}n_{\mu_2}-\eta_{\mu_2\mu_4}n_{\mu_3}+i\omega_{\mu_2\mu_3\mu_4}\right)n_{\mu_1},\\
\left( \BasisDiagramfour{\draw[blue_line](p1)to[bend left](p2);\draw[red_line](p2)to[bend left](p3);}{;}{\draw[red_line](p1)--(N);\draw[blue_line](p3)--(N);\draw[blue_line](p4)to[bend left](N);\draw[red_line](p4)to[bend right](N);} \right)^*=\BasisDiagramfour{\draw[red_line](p1)to[bend left](p2);\draw[blue_line](p2)to[bend left](p3);}{;}{\draw[blue_line](p1)--(N);\draw[red_line](p3)--(N);\draw[red_line](p4)to[bend left](N);\draw[blue_line](p4)to[bend right](N);} \quad& \Leftrightarrow\quad
\left(\eta_{\mu_1\mu_2}n_{\mu_3}+\eta_{\mu_2\mu_3}n_{\mu_1}-\eta_{\mu_1\mu_3}n_{\mu_2}-i\omega_{\mu_1\mu_2\mu_3}\right)n_{\mu_4},\\
\left(\BasisDiagramfour{\draw[blue_line](p2)to[bend left](p3);\draw[red_line](p3)to[bend left](p4);}{;}{\draw[red_line](p2)--(N);\draw[blue_line](p4)--(N);\draw[blue_line](p1)to[bend left](N);\draw[red_line](p1)to[bend right](N);}  \right)^*=\BasisDiagramfour{\draw[red_line](p2)to[bend left](p3);\draw[blue_line](p3)to[bend left](p4);}{;}{\draw[blue_line](p2)--(N);\draw[red_line](p4)--(N);\draw[red_line](p1)to[bend left](N);\draw[blue_line](p1)to[bend right](N);}\quad& \Leftrightarrow\quad
\left(\eta_{\mu_2\mu_3}n_{\mu_4}+\eta_{\mu_3\mu_4}n_{\mu_2}-\eta_{\mu_2\mu_4}n_{\mu_3}-i\omega_{\mu_2\mu_3\mu_4}\right)n_{\mu_1}.
    \end{align}\es The diagrams with only one loop are 
    \bs\begin{align}
        \BasisDiagramfour{\draw[blue_line](p1)to[bend left](p2);\draw[red_line](p2)to[bend left](p3);\draw[blue_line](p3)to[bend left](p4);\draw[red_line](p4)to[bend left](p1);}{;}{;}  \quad& \Leftrightarrow\quad 
\eta_{\mu_1\mu_2}\eta_{\mu_3\mu_4} - \eta_{\mu_1\mu_3}\eta_{\mu_2\mu_4} + \eta_{\mu_1\mu_4}\eta_{\mu_2\mu_3} + i\epsilon_{\mu_1\mu_2\mu_3\mu_4}\\
\left(\BasisDiagramfour{\draw[blue_line](p1)to[bend left](p2);\draw[red_line](p2)to[bend left](p3);\draw[blue_line](p3)to[bend left](p4);\draw[red_line](p4)to[bend left](p1);}{;}{;}\right)^* =\BasisDiagramfour{\draw[red_line](p1)to[bend left](p2);\draw[blue_line](p2)to[bend left](p3);\draw[red_line](p3)to[bend left](p4);\draw[blue_line](p4)to[bend left](p1);}{;}{;} \quad& \Leftrightarrow\quad 
\eta_{\mu_1\mu_2}\eta_{\mu_3\mu_4} - \eta_{\mu_1\mu_3}\eta_{\mu_2\mu_4} + \eta_{\mu_1\mu_4}\eta_{\mu_2\mu_3} - i\epsilon_{\mu_1\mu_2\mu_3\mu_4}
    \end{align}\es 
\end{enumerate}
and \footnote{The two diagrams in the following cannot be decomposed into two independent loops since the solid blue lines and the red dashed lines in a loop should appear alternately. }
    \bs\begin{align}
        \BasisDiagramfour{\draw[blue_line](p1)to[bend left](p2);\draw[red_line](p1)--(N);\draw[red_line](p2)--(N);}{\draw[red_line](p3)to[bend left](p4);\draw[blue_line](p3)--(N);\draw[blue_line](p4)--(N);}{;} \quad& \Leftrightarrow\quad \begin{aligned}
&{ n_{[\mu_1}\eta_{\mu_2][\mu_3}n_{\mu_4]}+\frac{i}{4}(\omega_{\mu_1\mu_2[\mu_3}n_{\mu_4]}+n_{[\mu_1}\omega_{\mu_2]\mu_3\mu_4})},\\
\end{aligned}
\\
      \left(\BasisDiagramfour{\draw[blue_line](p1)to[bend left](p2);\draw[red_line](p1)--(N);\draw[red_line](p2)--(N);}{\draw[red_line](p3)to[bend left](p4);\draw[blue_line](p3)--(N);\draw[blue_line](p4)--(N);}{;}\right)^*=  \BasisDiagramfour{\draw[red_line](p1)to[bend left](p2);\draw[blue_line](p1)--(N);\draw[blue_line](p2)--(N);}{\draw[blue_line](p3)to[bend left](p4);\draw[red_line](p3)--(N);\draw[red_line](p4)--(N);}{;} \quad& \Leftrightarrow\quad \begin{aligned}
&{n_{[\mu_1}\eta_{\mu_2][\mu_3}n_{\mu_4]}-\frac{i}{4}(\omega_{\mu_1\mu_2[\mu_3}n_{\mu_4]}+n_{[\mu_1}\omega_{\mu_2]\mu_3\mu_4})}.\\
\end{aligned}
    \end{align}\es
A circular diagram has the advantage that when one divides it into several loops , the corresponding tensor structure is exactly the tensor product of each individual loop. For example, 
\bs
\begin{align}
    \BasisDiagramthree{\draw[blue_line](p1)to[bend left](p2);}{\draw[red_line](p1)--(p2);}{\draw[blue_line](p3)to[bend left](N);\draw[red_line](p3)to[bend right](N);}&= \begin{tikzpicture}[baseline=(current bounding box.center), scale=0.7]
        \draw[dotted, gray!30] (0,0) circle (1cm);
       \node[n_node] (N) at (0,0) {n};
        \foreach \i in {1,...,2} {
            \node[dot] (p\i) at (90-180*\i+90:1cm) {};
            \node[label_style] at (90-180*\i+90:1.3cm) {\i};
        }
        \draw[red_line](p1)to[bend left](p2);
        \draw[blue_line](p1)to[bend right](p2);
    \end{tikzpicture}
     \quad \otimes \begin{tikzpicture}[baseline=(current bounding box.center), scale=0.7]
        \draw[dotted, gray!30] (0,0) circle (1cm);
        \node[n_node] (N) at (0,0) {n};
        \foreach \i in {1,...,1} {
            \node[dot] (p\i) at (90-90*\i+90:1cm) {};
            \node[label_style] at (90-90*\i+90:1.3cm) {3};
        }
        \draw[red_line](p1)to[bend left](N);
        \draw[blue_line](p1)to[bend right](N);
    \end{tikzpicture}\\
    \BasisDiagramfour{\draw[blue_line](p1)to[bend left](p2);\draw[red_line](p2)to[bend left](p3);}{;}{\draw[red_line](p1)--(N);\draw[blue_line](p3)--(N);\draw[blue_line](p4)to[bend left](N);\draw[red_line](p4)to[bend right](N);}&=\begin{tikzpicture}[baseline=(current bounding box.center), scale=0.7]
        \draw[dotted, gray!30] (0,0) circle (1cm);
        \node[n_node] (N) at (0,0) {n};
        \foreach \i in {1,...,1} {
            \node[dot] (p\i) at (90-90*\i+90:1cm) {};
            \node[label_style] at (90-90*\i+90:1.3cm) {4};
        }
        \draw[red_line](p1)to[bend left](N);
        \draw[blue_line](p1)to[bend right](N);
    \end{tikzpicture}\quad\otimes\quad \BasisDiagramthree{\draw[blue_line](p1)to[bend left](p2);\draw[red_line](p2)to[bend left](p3);}{;}{\draw[red_line](p1)--(N);\draw[blue_line](p3)--(N);}.\end{align}\es Therefore, the building blocks of the  circular diagrams are the loop diagrams.\footnote{By a loop diagram we mean a diagram with only one loop.} To show this explicitly, consider a loop diagram with $m$ points on the circle, and the basic diagrams for small values of $m$ are shown  as follows:
\begin{enumerate}
    \item $m=1$. 
    \be 
    \begin{tikzpicture}[baseline=(current bounding box.center), scale=0.7]
        \draw[dotted, gray!30] (0,0) circle (1cm);
        \node[n_node] (N) at (0,0) {n};
        \foreach \i in {1,...,1} {
            \node[dot] (p\i) at (90-90*\i+90:1cm) {};
            \node[label_style] at (90-90*\i+90:1.3cm) {1};
        }
        \draw[red_line](p1)to[bend left](N);
        \draw[blue_line](p1)to[bend right](N);
    \end{tikzpicture}
    \ee 
    \item $m=2$. 
    \be 
    \begin{tikzpicture}[baseline=(current bounding box.center), scale=0.7]
        \draw[dotted, gray!30] (0,0) circle (1cm);
       \node[n_node] (N) at (0,0) {n};
        \foreach \i in {1,...,2} {
            \node[dot] (p\i) at (90-180*\i+90:1cm) {};
            \node[label_style] at (90-180*\i+90:1.3cm) {\i};
        }
        \draw[red_line](p1)to[bend left](p2);
        \draw[blue_line](p1)to[bend right](p2);
    \end{tikzpicture}
    \ee 
    \item $m=3$.
    \be 
     \BasisDiagramthree{\draw[blue_line](p1)to[bend left](p2);\draw[red_line](p2)to[bend left](p3);}{;}{\draw[red_line](p1)--(N);\draw[blue_line](p3)--(N);}
    \ee 
    \item $m=4$. 
    \be 
     \BasisDiagramfour{\draw[blue_line](p1)to[bend left](p2);\draw[red_line](p2)to[bend left](p3);\draw[blue_line](p3)to[bend left](p4);\draw[red_line](p4)to[bend left](p1);}{;}{;} 
    \ee 
\end{enumerate}
When $m\ge 3$, the tensor of the $m$-circle is not real. Therefore, one should take into account its complex conjugation. Interestingly, the tensor corresponding to the above diagram with $m=3$ is equal to {the} contraction of the tensor associated to the diagram with $m=4$ and the null vector associated with the $m=1$ {diagram}:
\be 
\BasisDiagramthree{\draw[blue_line](p1)to[bend left](p2);\draw[red_line](p2)to[bend left](p3);}{;}{\draw[red_line](p1)--(N);\draw[blue_line](p3)--(N);}=\BasisDiagramfour{\draw[blue_line](p1)to[bend left](p2);\draw[red_line](p2)to[bend left](p3);\draw[blue_line](p3)to[bend left](p4);\draw[red_line](p4)to[bend left](p1);}{;}{;} \cdot \begin{tikzpicture}[baseline=(current bounding box.center), scale=0.7]
        \draw[dotted, gray!30] (0,0) circle (1cm);
        \node[n_node] (N) at (0,0) {n};
        \foreach \i in {1,...,1} {
            \node[dot] (p\i) at (90-90*\i+90:1cm) {};
            \node[label_style] at (90-90*\i+90:1.3cm) {4};
        }
        \draw[red_line](p1)to[bend left](N);
        \draw[blue_line](p1)to[bend right](N);
    \end{tikzpicture}.
    \label{eq:diagcontract4->3}
\ee To check it, we find
\bea 
\text{r.h.s.}&=&\left(-\frac{1}{2}\right)^4 \epsilon_{A_1A_2}\epsilon_{\dot A_2\dot A_3}\epsilon_{A_3A_4}\epsilon_{\dot A_1\dot A_4}\prod_{j=1}^4 \bar\sigma^{\mu_j \dot A_j A_j} \times \left(-\frac{1}{2}\right)o_{A}\overline o_{\dot A}\bar\sigma_{\mu_4}^{\dot A A}\nn\\&=&\left(-\frac{1}{2}\right)^5 \epsilon_{A_1A_2}\epsilon_{\dot A_2\dot A_3}\epsilon_{A_3A_4}\epsilon_{\dot A_1\dot A_4}\prod_{j=1}^4 \bar\sigma^{\mu_j\dot A_j A_j}o_{A}\overline o_{\dot A}\sigma_{\mu_4C\dot C}\epsilon^{CA}\epsilon^{\dot C\dot A}\nn\\&=&\left(-\frac{1}{2}\right)^4\epsilon_{A_1A_2}\epsilon_{\dot A_2\dot A_3}o_{A_3}\overline o_{\dot A_1}\prod_{j=1}^3 \bar\sigma^{\mu_j \dot A_j A_j}\nn\\&=&-\frac{1}{2}\text{l.h.s.}.
\eea 
A more general identity of this type is 
\begin{equation}
   \begin{tikzpicture}[baseline=(current bounding box.center), scale=0.7]
        \draw[dotted, gray!30] (0,0) circle (1.5cm);
        \node[n_node] (N) at (0,0) {$n$};
        \node[dot] (p1) at (90:1.5cm) {};
        \node[label_style] at (90:1.8cm) {1};
        
        \node[dot] (p2) at (40:1.5cm) {};
        \node[label_style] at (40:1.8cm) {2};
        
        \node[dot] (p3) at (-10:1.5cm) {};
        \node[label_style] at (-10:1.8cm) {3};
        
        \node[dot] (pn) at (140:1.5cm) {};
        \node[label_style] at (140:1.8cm) {$n-1$};
        
        \node[dot] (pnm) at (190:1.5cm) {};
        \node[label_style] at (190:2.3cm) {$n-2$};
        \node[font=\small] at (270:1.5cm) {$\dots$};
        \draw[blue, thick](p1)to[bend left](p2);\draw[red, dashed](pnm)to[bend left](pn);
        \draw[red, dashed](p2)to[bend left](p3);\draw[red, dashed](p1)--(N);\draw[blue, thick](pn)--(N);
    \end{tikzpicture} = \NNPointDiagram{\draw[blue, thick](p1)to[bend left](p2);\draw[blue, thick](pnm)to[bend left](pn);}{\draw[red, dashed](p2)to[bend left](p3);\draw[red, dashed](p1)to[bend right](pn);}{;} \cdot \begin{tikzpicture}[baseline=(current bounding box.center), scale=0.7]
        \draw[dotted, gray!30] (0,0) circle (1cm);
        \node[n_node] (N) at (0,0) {n};
        \foreach \i in {1,...,1} {
            \node[dot] (p\i) at (90-90*\i+90:1cm) {};
            \node[label_style] at (90-90*\i+90:1.3cm) {n};
        }
        \draw[red_line](p1)to[bend left](N);
        \draw[blue_line](p1)to[bend right](N);
    \end{tikzpicture}.
    \label{eq:diagramcontract}
     \end{equation}

It follows that the building blocks are the indecomposable loop diagrams with $2m$ points on the circle, none of which connects to the center\begin{tikzpicture}[baseline=-0.6ex, scale=0.23]
        \draw[dotted, gray!30] (0,0) circle (1cm);
        \node[n_node] (N) at (0,0) {n};
    \end{tikzpicture}. One reduces all basic diagrams with $2m-1$ points to the building blocks by applying \eqref{eq:diagramcontract},  e.g., as in \eqref{eq:diagcontract4->3}.
     
The tensor of the indecomposable loop diagram with $2m$ points is 
\bea 
e_{\mu_1\mu_2\cdots\mu_{2m}}&=&\left(-\frac{1}{2}\right)^{2m}\epsilon_{A_1A_2}\epsilon_{\dot A_2\dot A_3}\cdots\epsilon_{A_{2m-1}A_{2m}}\epsilon_{\dot A_1\dot A_{2m}}\prod_{j=1}^{2m}\bar\sigma_{\mu_j}^{\dot A_j A_j}\nn\\&=&\frac{(-1)^{m+1}}{2^{2m}}\text{Tr}\left( \bar\sigma_{\mu_1}\sigma_{\mu_2}\bar\sigma_{\mu_3}\sigma_{\mu_4}\cdots \bar\sigma_{\mu_{2m-1}}\sigma_{\mu_{2m}}\right)
\eea or its complex conjugate
\bea 
e^*_{\mu_1\mu_2\cdots\mu_{2m}}&=&{ \frac{(-1)^{m+1}}{2^{2m}}}\text{Tr}\left(\sigma_{\mu_1}\bar\sigma_{\mu_2}\sigma_{\mu_3}\bar\sigma_{\mu_4}\cdots \sigma_{\mu_{2m-1}}\bar\sigma_{\mu_{2m}}\right)
\eea where $\sigma$ and $\bar \sigma$ appear alternately in the trace. Using the Dirac gamma matrices 
\bea
\gamma^{\mu}=\left(\begin{array}{cc}0&\sigma^\mu\\ \bar\sigma^\mu&0\end{array}\right)
\eea and 
\bea 
\gamma^5=\left(\begin{array}{cc}1&0\\0&-1\end{array}\right),
\eea we find 
\bs\begin{align}
e_{\mu_1\mu_2\cdots\mu_{2m}}&={\frac{(-1)^{m+1}}{2^{2m}}}\text{Tr}\left(\frac{1-\gamma^5}{2}\gamma_{\mu_1}\gamma_{\mu_2}\cdots\gamma_{\mu_{2m}}\right),\\ e^*_{\mu_1\mu_2\cdots\mu_{2m}}&={\frac{(-1)^{m+1}}{2^{2m}}}\text{Tr}\left(\frac{1+\gamma^5}{2}\gamma_{\mu_1}\gamma_{\mu_2}\cdots\gamma_{\mu_{2m}}\right).
\end{align}\es  
They can be written down explicitly using the trace of Dirac gamma {matrices~}\cite{ERCaianiello:1952rwm}
\bs\begin{align}
    \text{Tr}\left(\gamma_{\mu_1}\gamma_{\mu_2}\cdots\gamma_{\mu_{2m}}\right)&=\frac{4(-1)^m}{2^m m!}\sum_{\pi \in S_{2m}} \text{sgn}(\pi) \prod_{j=1}^m \left( \eta_{\mu_{\pi(2j-1)} \mu_{\pi(2j)}} \cdot \kappa_{\pi(2j-1), \pi(2j)} \right),\\
    \text{Tr}\left(\gamma^5\gamma_{\mu_1}\gamma_{\mu_2}\cdots\gamma_{\mu_{2m}}\right)&=\frac{-4i (-1)^{m-2}}{4! \cdot 2^{m-2} \cdot (m-2)!} \sum_{\sigma \in S_{2m}} \text{sgn}(\sigma)\nonumber\\* &\quad \epsilon_{\mu_{\sigma(1)} \mu_{\sigma(2)} \mu_{\sigma(3)} \mu_{\sigma(4)}} \prod_{k=3}^m \left( \eta_{\mu_{\sigma(2k-1)} \mu_{\sigma(2k)}} \cdot \kappa_{\sigma(2k-1), \sigma(2k)} \right)
\end{align}\es where $\pi,\sigma$ are permutation elements of the symmetry group $S_{2m}$ and $\kappa_{h,k}$ is defined as 
\begin{equation}
    \kappa_{h,k} = \begin{cases} +1 & \text{for } h < k \\ 0 & \text{for } h = k \\ -1 & \text{for } h > k \end{cases}.
\end{equation}
For a permutation $\pi$ in the symmetric group $S_{2m}$, $\operatorname{sgn}(\pi)$ denotes its signature. Namely,
$\operatorname{sgn}(\pi)=1$ if $\pi$ is even and $\operatorname{sgn}(\pi)=-1$ if $\pi$ is odd. 
\section{Applications}\label{app}
In this section, we use examples to verify the bulk-to-boundary correlators obtained in the previous sections. The relation to {the} KL representation is discussed in detail. 
\subsection{Bulk-to-boundary correlator for spin-1 operators}
 The spin-1 field is denoted by $t_\mu$ and its asymptotic expansion near $\mathscr I^+$ is 
\be 
t_\mu(x)=\frac{\Sigma_\mu(u,\Omega)}{r^\Delta}+\cdots.\label{fall-offtmu}
\ee The bulk-to-boundary correlator is 
\be 
D_{\mu\nu}(u,\Omega;x')=\langle \Sigma_\mu(u,\Omega)t_\nu(x')\rangle.
\ee Poincar\'e symmetry fixes the general solution to be 
\bea 
D_{\mu\nu}(u,\Omega;x')=C^{(1)}_1\frac{\eta_{\mu\nu}}{(u+n\cdot x')^\Delta}+C^{(1)}_2\frac{n_\mu n_\nu}{(u+n\cdot x')^{\Delta+2}}\label{twostr}
\eea where $\Delta$ is the fall-off index of the corresponding spin-1 field in the bulk.  $C^{(1)}_1$ and $C^{(1)}_2$ are constants.
We discuss several examples below. 
\begin{enumerate}

 \item For a conserved current $j_\mu$ 
    \be 
    \partial^\mu j_\mu=0,
    \ee there should be an extra Ward identity
    \be 
    n^\mu \dot D_{\mu\nu}=0.
    \ee Comparing it with \eqref{twostr}, we conclude that $C^{(1)}_1=0$ for the bulk-to-boundary correlator and then  
    \be 
    D_{\mu\nu}(u,\Omega;x')=C^{(1)}_2\frac{n_\mu n_\nu}{(u+n\cdot x')^{\Delta+2}}.\label{consbtob}
    \ee 
    \item We may also consider the spin-1 operator 
    \be 
    t_\mu=\partial_\mu\Phi
    \ee where $\Phi$ is a bulk scalar field whose corresponding bulk-to-boundary correlator is fixed to be 
    \be 
    D(u,\Omega;x')=\frac{C^{(0)}}{(u+n\cdot x')^\Delta}.
    \ee It follows that 
    \bea 
    D_{\mu\nu}(u,\Omega;x')=-n_\mu \partial_u \partial'_\nu D(u,\Omega;x')=-\frac{C^{(0)} \Delta(\Delta+1)n_\mu n_\nu}{(u+n\cdot x')^{\Delta+2}}.
    \eea In this case, we should identify 
    \bea 
    C^{(1)}_1=0,\quad C^{(1)}_2=-C^{(0)}\Delta(\Delta+1).
    \eea 
\item Note that the previous derivation depends on the assumption that the constants $C^{(1)}_1$ and $C^{(1)}_2$ are finite,  i.e., independent of the IR cutoff. As an illustration, consider an electromagnetic field $a_\mu$.  The Feynman propagator in the $R_\xi$ gauge is \cite{Moga:2025ads}
\begin{align}
    G_{\mu\nu}(x;x')&=\int \frac{d^4 p}{(2\pi)^4}e^{i p\cdot (x-x')}G_{\mu\nu}(p)\nonumber\\* &=\frac{1}{4\pi^2} \frac{1}{(x-x')^2 + i\epsilon} \left[ { \frac{1+\xi}{2}} \eta_{\mu\nu} + (1-\xi) \frac{(x-x')_\mu (x-x')_\nu}{(x-x')^2 + i\epsilon} \right]. \end{align}
{The propagator can be expanded near $\mathscr I^+$ as 
\be 
G_{\mu\nu}(x;x')=\sum_{a\ge 0}\frac{D^{(a)}_{\mu\nu}(u,\Omega;x')}{r^{a}}
\ee whose leading term is 
\be 
D_{\mu\nu}(u,\Omega;x')=D^{(0)}_{\mu\nu}(u,\Omega;x')=\lim_{r\to\mathscr I^+} G_{\mu\nu}(u,\Omega;x')=\frac{1-\xi}{16\pi^2}\frac{n_\mu n_\nu}{\widehat u^2}.\label{D0}
\ee Therefore, we may identify 
\be 
\Delta=0,\quad C_2^{(1)}=\frac{1-\xi}{16\pi^2},\quad C_1^{(1)}=0\quad \text{for}\ \xi\not=1.\label{514}
\ee  In  Fourier space, this bulk-to-boundary correlator is proportional to the momentum and thus corresponds to the longitudinal mode \cite{weinberg_1995}. Thus, this correlator does not contain the contribution from the transverse radiative mode. In this case, one should expand the bulk-to-bulk correlator to  next-to-leading order  
\begin{align}
    &D_{\mu\nu}^{(1)}(u,\Omega;x')\nn\\=&-\frac{1+\xi}{16\pi^2}
		\frac{\eta_{\mu\nu}}{\hat{u}}
+
		\frac{1-\xi}{16\pi^2}\left[
		\frac{
			u(n_\mu\bar m_\nu+\bar m_\mu n_\nu)
			-(n_\mu x'_\nu+x'_\mu n_\nu)
		}{\hat{u}^2}
+
		\frac{
			\big(x'^2-u^2-2u\,\bar m\cdot x'\big)n_\mu n_\nu
		}{\hat{u}^3}\right].\nn\label{Dmunu1}\\
\end{align}
The expression does not have the general form \eqref{twostr} since it is not the leading-order bulk-to-boundary correlator.  
However, one can show that the form of the  next-to-leading-order bulk-to-boundary correlator can also be fixed to several structures. Further details are given in Appendix \ref{nextL}.

In Feynman gauge, $\xi=1$, the  correlator \eqref{D0} vanishes and the leading term becomes 
\be 
D_{\mu\nu}(u,\Omega;x')=D_{\mu\nu}^{(1)}(u,\Omega;x')=\lim_{r\to\mathscr I^+}\ r\ G_{\mu\nu}(x;x')=-\frac{1}{8\pi^2}\frac{\eta_{\mu\nu}}{\widehat u}.\label{leadingD-photon}
\ee Thus one should identify 
\be 
\Delta=1,\quad C_1^{(1)}=-\frac{1}{8\pi^2},\quad C_2^{(1)}=0\quad \text{for}\ \xi=1.\label{xi1}
\ee 
Comparing \eqref{514} with \eqref{xi1}, we conclude that the fall-off index $\Delta$ depends on the gauge choice for general massless spinning fields. Feynman gauge is rather special in the sense that the corresponding fall-off index is distinguished from the one in the general $R_\xi$ gauge. In Feynman gauge, we find the boundary-to-boundary correlator in the usual way.  For the advanced coordinates $(v',\Omega')$ and the associated vectors, we follow the conventions of \cite{Long:2026cpq}.
\begin{align}
 B_{\mu\nu}(u,\Omega;v',\Omega')
 =-\frac{1}{8\pi^2}
 \frac{\eta_{\mu\nu}}{1+\cos\gamma}+\frac{1}{4\pi}
 \eta_{\mu\nu}\ln (u-v')\delta^{(2)}(\Omega'-\Omega^P).
\end{align}
Here $\Omega$ denotes the spherical angles $\Omega=(\theta,\phi)$ whose antipodal point is $\Omega^{P}=(\pi-\theta,\pi+\phi)$.

}

  \item For a CFT, the two-point function of a spin-1 primary operator is fixed by conformal symmetry 
    \be 
    \langle \text{T}\left(j_\mu(x)j_\nu(x')\right)\rangle=\frac{C_j I_{\mu\nu}(x-x')}{(x-x')^{2\bar\Delta}}
    \ee where the function $I_{\mu\nu}(x)$ is 
    \be 
    I_{\mu\nu}(x)=\eta_{\mu\nu}-\frac{2x_\mu x_\nu}{x^2}.\label{Imunu}
    \ee The constant $C_j$ denotes the normalization and $\bar\Delta$ is the conformal dimension of the spin-1 operator. Note that  the conformal dimension $\bar\Delta$ is not necessarily  equal to the fall-off index. Extrapolating $x$ to the null boundary while keeping $x'$ finite, we obtain 
    \begin{equation}
        \begin{aligned}
I_{\mu\nu}(x-x')&\to \eta_{\mu\nu}+\frac{rn_\mu n_\nu}{\widehat u}\\
&\quad+\frac{u\left(n_\mu\bar m_\nu+\bar m_\mu n_\nu\right)-\left(n_\mu x'_\nu+x'_\mu n_\nu\right)}{\widehat u}+\frac{\left(x'^2-u^2-2u\,\bar m\cdot x'\right)n_\mu n_\nu}{2\widehat u^2}.
        \end{aligned}
    \end{equation}
    The second term dominates and the leading fall-off behavior of the two-point correlator is 
    \bea 
    \langle \text{T}\left(j_\mu(x)j_\nu(x')\right)\rangle=\frac{C_j n_\mu n_\nu}{(-2)^{\bar\Delta} (u+n\cdot x')^{\bar{\Delta}+1}}\frac{1}{r^{\bar\Delta-1}}+\cdots.
    \eea Therefore, we may identify 
    \be 
    \Delta=\bar\Delta-1,\quad C^{(1)}_1=0,\quad C^{(1)}_2=\frac{C_j}{(-2)^{\bar\Delta}}.
    \ee We conclude that the fall-off index of a spin-1 operator is smaller than the conformal dimension by 1 for a CFT. An analogous relation holds in the Dirac theory where the fall-off index of a spin-$1/2$ operator is smaller than the conformal dimension by $1/2$ \cite{Long:2026cpq}. To be more precise, we consider a spin-1 operator in a free complex scalar theory 
    \be 
    j_\mu=-i(\Phi\partial_\mu\overline\Phi-\overline\Phi\partial_\mu\Phi).\label{jmuo}
    \ee The conformal dimension of $j_\mu$ is $\bar\Delta=3$ and the fall-off index of $\Phi$ is 1
    \be 
    \Phi(x)=\frac{\Sigma(u,\Omega)}{r}+\cdots.\label{fall-offPhi}
    \ee A straightforward computation leads to 
    \bea 
    j_\mu=\frac{in_\mu (\Sigma\dot{\overline\Sigma}-\overline\Sigma\dot\Sigma)}{r^2}+\cdots.
    \eea It follows that the fall-off index of $j_\mu$ is $\Delta=2${,} which matches exactly the formula \be \Delta=\bar\Delta-1.\label{bardelta}\ee 
    
    We may extrapolate the remaining bulk point $x'$ to null infinity using the same fall-off index 
    \bea 
    B_{\mu\nu}(u,\Omega;v',\Omega')=\lim_{r'\to\mathscr I^-}r'^{\Delta} D_{\mu\nu}(u,\Omega;x').
    \eea When $\Omega\not=\Omega'^{\text{P}}$, the boundary-to-boundary correlator $B_{\mu\nu}(u,\Omega;v',\Omega')$ vanishes since the leading behavior of the denominator is $r'^{\Delta+2}$ while the numerator only increases as $r'^{\Delta}$ for large $r'$. Therefore, as in \cite{Long:2026cpq}, we obtain 
    \be 
    B_{\mu\nu}(u,\Omega;v',\Omega')=\kappa(u-v') n_\mu n_\nu \delta^{(2)}(\Omega-\Omega'^{\text{P}}) \label{assumpB}
    \ee where 
    \bea 
    \kappa(u-v')=\left\{\begin{array}{cc} 0& 0<\Delta<1,\\
    \frac{\pi C^{(1)}_2}{(u-v')^2}&\Delta=1,\\
\text{divergent}&\Delta>1.\end{array}\right.
    \eea 
    When $0<\Delta<1$, there is no magnetic  or electric branch. When $\Delta=1$, only the electric branch exists in this case. When $\Delta>1$, the electric branch is IR divergent with 
    \be 
    \kappa(u-v')=\frac{2\pi {C^{(1)}_2} }{\Delta+1}\frac{r'^{\Delta-1}}{(u-v')^{\Delta+1}}.
    \ee In this case, the assumption \eqref{assumpB} is not valid and the coefficient needs regularization. In what follows, we always find similar IR  divergences for $\Delta>1$. We postpone the discussion of the regularization of the IR divergences  to the last section.
    \item The previous discussion can be extended to general  nonconformal theories. Given a general solution \eqref{twostr}, 
    the extrapolation limit is classified according to {the values of} $C^{(1)}_1$ and $C^{(1)}_2$.
      \begin{align}
        B_{\mu\nu}(u,\Omega; v',\Omega ')&=\left\{\begin{array}{cc} \frac{C^{(1)}_1\eta_{\mu\nu}}{(1+\cos\gamma)^{\Delta}}&0<\Delta<1,\\
\begin{aligned}&-2\pi C^{(1)}_1\eta_{\mu\nu}\ln (u-v')\delta^{(2)}(\Omega-\Omega'^{\text{P}})+\frac{C^{(1)}_1\eta_{\mu\nu}}{1+\cos\gamma}\\ &+ \frac{\pi C^{(1)}_2 n_\mu n_\nu}{(u-v')^2}\delta^{(2)}(\Omega-\Omega'^{\text{P}})\end{aligned}&\Delta=1,\\
\text{divergent}&\Delta>1.\end{array}\right.\label{caseI} 
        \end{align}
         \item Our result should be valid non-perturbatively. In an interacting  theory with a spin-1 operator $t_\mu$, the two-point correlator $G_{\mu\nu}(x;x')=\langle \text{T}\left(t_{\mu}(x)t_{\nu}(x')\right)\rangle $ is fixed in {the KL representation~}\cite{Karateev:2020axc}
\begin{equation}
G^{\mu\nu}(x;x') = -i\int_{0}^{\infty}ds\left(-\rho_{v}^{0}(s)\Delta_{F,0}^{\mu\nu}(x-x';s) + \rho_{v}^{1}(s)\Delta_{F,1}^{\mu\nu}(x-x';s)\right)\label{Gmunu}
\end{equation}
where  $\Delta_{F,0}^{\mu\nu}(x;s),\Delta_{F,1}^{\mu\nu}(x;s)$ are 
\begin{equation}
\Delta_{F,0}^{\mu\nu}(x;s) = {\partial^\mu\partial^\nu}\,\Delta_{F}(x;s), \quad \Delta_{F,1}^{\mu\nu}(x;s) = \bigl(s\,\eta^{\mu\nu} - \partial^{\mu}\partial^{\nu}\bigr)\,\Delta_{F}(x;s)
\end{equation}
with $\Delta_{F}(x;s)$ the scalar Feynman propagator with mass $m=\sqrt{s}$
\begin{equation}
    \begin{aligned}
          -i\Delta_{F}(x;s) 
&= (2\pi)^{-d/2}\,s^{\frac{d-2}{2}} \times \left(\theta(+x^{2})\,\frac{K_{\frac{d-2}{2}}(\sqrt{sx^{2}})}{(\sqrt{sx^{2}})^{\frac{d-2}{2}}}- \frac{i\pi}{2}\theta(-x^{2})\,\frac{H_{\frac{2-d}{2}}^{(2)}(\sqrt{-sx^{2}})}{(\sqrt{-sx^{2}})^{\frac{d-2}{2}}}\right).
    \end{aligned}
\end{equation}
Here, $K_\nu(x)$ is the modified Bessel function of the second kind and $H^{(2)}_\nu(x)$ is the Hankel function of the second kind. The functions $\rho_{v}^0(s)$ and $\rho_{v}^1(s)$ are the spin 0 and spin-1 components of the spectral density. 
{For an elementary field that satisfies the canonical equal-time commutation relations, the spectral density $\rho_v^1(s)$ is non-negative \cite{Karateev:2020axc}, i.e., $\rho_v^1(s)\ge 0$, and it satisfies the sum rule~\cite{Zimmermann:1969BFb0045925}
\begin{equation}
\int_0^\infty ds\, s\,\rho_{v}^1(s)=1.\label{uni}
\end{equation}
The factor of $s$ in the integrand originates from factoring out an $s$ from the spectral function. Consequently, both $\rho_v^0$ and $\rho_v^1$ have mass dimension $-4$; the insertion of the factor $s$ is thus required by dimensional analysis.  
For a generic vector operator (e.g., a composite operator), however, the canonical commutation relations are not necessarily satisfied, and hence the sum rule for the spectral density need not hold.}

We set the dimension $d=4$ in this work. For spacelike distance, we define $h=-(x-x')^2$ and 
\be 
G(x-x';s)=\frac{\sqrt{s}}{4\pi^2}\frac{K_1(\sqrt{-sh})}{\sqrt{-h}}.
\ee The two-point function reduces to 
\begin{align}
    G_{\mu\nu}(x;x')&=\int_0^\infty ds \Bigl[\rho_v^1(s)s \eta_{\mu\nu}G(x-x';s) -(\rho_v^0(s)+\rho_v^1(s))\partial_\mu\partial_\nu G(x-x';s)\Bigr].
\end{align}
 Using the identities 
\bs\begin{align}
    \partial_\mu G(x-x';s)&=-2(x-x')_\mu\frac{d}{dh}G(x-x';s),\\
    \partial_\mu\partial_\nu G(x-x';s)&=-2\eta_{\mu\nu}\frac{d}{dh}G(x-x';s)+4(x-x')_\mu (x-x')_\nu \frac{d^2}{dh^2}G(x-x';s),
\end{align}\es we find 
\begin{align}
G_{\mu\nu}(x;x') = \int_0^\infty ds\biggl[ 
&\eta_{\mu\nu}\Bigl(\rho_v^1(s)sG(x-x';s) + 2(\rho_v^0(s)+\rho_v^1(s))\frac{d}{dh}G(x-x';s)\Bigr) \nonumber \\
&-4(\rho_v^0(s)+\rho_v^1(s))(x-x')_\mu(x-x')_\nu \frac{d^2}{dh^2}G(x-x';s) \biggr] \label{eq:KL-forspin-1}
\end{align}
The first line is proportional to $\eta_{\mu\nu}$ and the second line is proportional to $n_\mu n_\nu$ after extrapolating $x$ to $\mathscr I^+$. 

{Keeping the spectral densities general, the coefficients in
\eqref{twostr} satisfy
\begin{align}
\frac{C_1^{(1)}}{\widehat u^\Delta}
&=\lim_{r\to\infty}r^\Delta
\int_0^\infty ds\,
\left[
s\rho_v^1(s)G
+2\bigl(\rho_v^0(s)+\rho_v^1(s)\bigr)\frac{d}{dh}G
\right],
\label{spin1-KL-general-C1}\\
\frac{C_2^{(1)}}{\widehat u^{\Delta+2}}
&=-4\lim_{r\to\infty}r^{\Delta+2}
\int_0^\infty ds\,
\bigl(\rho_v^0(s)+\rho_v^1(s)\bigr)\frac{d^2}{dh^2}G.
\label{spin1-KL-general-C2}
\end{align}
The limits are taken after the spectral integrations.}

{%

We set $\rho_v=\rho_v^0+\rho_v^1$ and assume that the effective spectral combinations $\rho_v$ and
$s\rho_v^1$ consist of functions that are locally integrable on $s\in (0,\infty)$, possibly
supplemented by $\delta(s)$ terms and satisfy\footnote{For example, suppose that the function parts of
$\rho_v(s)$ and $s\rho_v^1(s)$ each obey a bound of the form
$O(e^{cs^\kappa})$, with $c>0$ and $0\leq\kappa<1/2$.
Such bounds satisfy our UV assumption. This growth is also
compatible with Jaffe's criterion for strict localizability
\cite{Jaffe1967}: the corresponding UV integral is required {to converge}
\[
\int_0^{\infty}
\frac{c s^\kappa}{2\sqrt{s}(1+s)}\,ds<\infty.
\] Since the Bessel function has the asymptotic behavior
\(
  x^{\nu}H_{\nu}(x)\to \mathrm{const}\) as \(x\to 0^{+}\),
the IR condition imposed here is a sufficient condition for the convergence of the KL spectral representation. This does not exclude the possibility of QFTs whose IR behavior does not satisfy this condition.}
\bs\label{UVIR} \begin{align}
\exists \ \alpha,\alpha_1&>0,\qquad  |\rho_v(s)|=o\!\left(e^{\alpha\sqrt{s}}\right),&
|s\rho_v^1(s)|&=o\!\left(e^{\alpha_1\sqrt{s}}\right),
&&s\to\infty,
\label{KL-UV-convergence}\\
\exists \ \epsilon,\epsilon_1&>0,\qquad |\rho_v(s)|=o\!\left(s^{-1+\epsilon}\right),&
|s\rho_v^1(s)|&=o\!\left(s^{-1+\epsilon_1}\right),
&&s\to0^+.
\label{KL-IR-convergence}
\end{align}\es 
For a gauge-invariant local vector operator acting on the physical
Hilbert space of a unitary QFT, Wightman positivity, together with
Poincar\'e covariance and the spectral condition\footnote{Wightman positivity means that states created from the vacuum by
smeared local operators have non-negative norm:
$\langle 0|O(f)^\dagger O(f)|0\rangle\geq 0$.
The spectral condition requires the energy-momentum spectrum to lie
in the closed forward light cone. In other words, 
$p^0\geq 0$ and $p^2\leq 0$, so the invariant mass squared
$s=-p^2$ is non-negative.}, implies that the
physical spin-0 and spin-1 spectral components are non-negative,
locally finite  measures~\cite{Wightman1956}. Whenever these locally finite measures admit density functions
on $s>0$, the densities are non-negative almost everywhere~\cite{Karateev:2020axc} and
locally integrable on $(0,\infty)$~\cite{Hormander1990}. Consequently, the effective
combinations $\rho_v$ and $s\rho_v^1$ are also locally integrable
on $(0,\infty)$.

Under these conditions, the density function contribution to the
KL representation is $o(r^{-1})$, while a $\delta(s)$ contribution
is at most $O(r^{-1})$. Hence any nonzero leading pure-power
extrapolation satisfies
\begin{equation}
\Delta\geq1.
\label{spin1Delta>1}
\end{equation}
The equality is valid if and only if there is a massless-pole contribution. 
A detailed proof is given in Appendix~\ref{spin-1KLrepproof}.

We now analyze some examples below:

\begin{itemize}
    \item A simple power-law choice without massless poles is
\begin{equation}
\rho_v(s)=B s^{\Delta-2}+\cdots,
\qquad
\rho_v^1(s)=A s^{\Delta-3}+\cdots,
\qquad
\Delta>1.
\label{spin-1rhobehave}
\end{equation}
Here the ellipses contribute only at subleading order in the null
extrapolation and  the full spectral densities are locally integrable and satisfy the conditions in \eqref{KL-IR-convergence}. 
For this choice, the leading Bessel integrals
can be evaluated explicitly, giving
\begin{equation}
C^{(1)}_1
=
\frac{(-1)^{-\Delta}2^{\Delta-4}
\Gamma(\Delta)\Gamma(\Delta-1)}{\pi^2}\,A,
\label{C11}
\end{equation}
and
\begin{equation}
C^{(1)}_2
=
\frac{(-1)^{1-\Delta}2^{\Delta-4}
\Gamma(\Delta+2)\Gamma(\Delta-1)}{\pi^2}\,B.
\label{C12}
\end{equation}
This construction gives an explicit relation between the leading
Ward coefficients and the spectral coefficients. The positivity of the spectral density requires $A\ge 0,\ B\ge 0$. When $A>0$, the condition $0\le\rho_v^1\le\rho_v$ is violated near $s=0$. Therefore, the positivity condition implies $A=0$ in this case.
\item 
 A gauge field in a generic gauge can contain $\delta$-derivative contributions~\cite{Lowdon2017}. These contributions are not locally integrable, fall outside our assumptions, and can lead to $\Delta<1$. For example, in our conventions the free photon in the 
$R_\xi$ gauge has
\begin{equation}
s\rho_v^1(s)=\delta(s),
\qquad
\rho_v(s)=-(1-\xi)\delta'(s)
\end{equation}
and leads to 
\begin{equation}
G_{\mu\nu}(x;x')
=
\frac{1+\xi}{8\pi^2(-h)}\eta_{\mu\nu}
+
\frac{1-\xi}{4\pi^2(-h)^2}
(x-x')_\mu(x-x')_\nu
\xrightarrow{r\to\infty}
\frac{1-\xi}{16\pi^2\widehat u^2}n_\mu n_\nu.
\end{equation}
Thus, for $\xi\ne1$, $\delta'(s)$  produces a
nonzero leading $n_\mu n_\nu$ term with $\Delta=0$, reproducing
the photon example discussed above. Moreover, it violates Wightman positivity due to  longitudinal degrees of freedom.
\item For a spin-1 primary operator in a four-dimensional unitary CFT,
the unitarity bound is $\bar\Delta\geq3$, with
$\Delta=\bar\Delta-1\geq2$. Each nonvanishing continuous physical
spin density scales as~\cite{Karateev:2020axc}
\begin{equation}
\begin{split}
2\pi \rho_v^0(s) &= C_v \times \frac{\Delta- 2}{\Delta + 1}\theta(s) s^{\Delta- 2 }+ C_v \times \frac{2}{\Delta + 1} \delta(s) s^{\Delta + 1 - 2}, \\
2\pi \rho_v^1(s) &= C_v \times \frac{\Delta}{\Delta + 1}  \theta(s) s^{\Delta - 2}.
\end{split}
\end{equation}
We can see that these spectral densities, after deleting the massless pole,  match \eqref{spin-1rhobehave}, with $A=0$ and $B=C_v\frac{\Delta-1}{\pi(\Delta+1)}$.

\end{itemize}

}

    \end{enumerate}

\subsection{Bulk-to-boundary correlator for spin-2 operators}
In this case, the bulk-to-boundary correlator is fixed to be 
\bea 
D_{\mu\nu\rho\sigma}&=& C^{(2)}_1 \frac{\eta_{\mu\nu}\eta_{\rho\sigma}}{\widehat u^{\Delta}}+C^{(2)}_2\frac{\eta_{\mu\rho}\eta_{\nu\sigma}}{\widehat u^{\Delta}}+C^{(2)}_3 \frac{\eta_{\mu\sigma}\eta_{\nu\rho}}{\widehat u^{\Delta}}\nn+C^{(2)}_4\frac{\eta_{\mu\nu}n_\rho n_\sigma}{\widehat u^{\Delta+2}}\\&&+C^{(2)}_5\frac{\eta_{\mu\rho}n_\nu n_\sigma}{\widehat u^{\Delta+2}}+C^{(2)}_6\frac{\eta_{\mu\sigma}n_\nu n_\rho}{\widehat u^{\Delta+2}}\nn+C^{(2)}_7\frac{\eta_{\nu\rho}n_\mu n_\sigma}{\widehat u^{\Delta+2}}+C^{(2)}_8\frac{\eta_{\nu\sigma}n_\mu n_\rho}{\widehat u^{\Delta+2}}\\&&+C^{(2)}_9\frac{\eta_{\rho\sigma}n_\mu n_\nu}{\widehat u^{\Delta+2}}\nn+C^{(2)}_{10}\frac{n_\mu n_\nu n_\rho n_\sigma}{\widehat u^{\Delta+4}}+C^{(2)}_{11}\frac{\omega_{\mu\nu\rho}n_\sigma }{\widehat u^{\Delta+2}}+C^{(2)}_{12}\frac{\omega_{\mu\nu\sigma}n_\rho }{\widehat u^{\Delta+2}}\nn\\&&+C^{(2)}_{13}\frac{\omega_{\mu\rho\sigma}n_\nu }{\widehat u^{\Delta+2}}+C^{(2)}_{14}\frac{\epsilon_{\mu\nu\rho\sigma} }{\widehat u^{\Delta}}\label{fullD}
\eea We don't include the term proportional to  $\omega_{\nu\rho\sigma}n_\mu$ since it is not independent, as  shown in \eqref{antisym}. The last four terms are parity-odd and should vanish in a parity-invariant theory. 
\begin{itemize}
    \item For a symmetric tensor $t_{\mu\nu}$, the bulk-to-boundary correlator should satisfy the algebraic equation 
\be 
D_{\mu\nu\rho\sigma}=D_{\nu\mu\rho\sigma}=D_{\mu\nu\sigma\rho}=D_{\nu\mu\sigma\rho}.
\ee Therefore, several coefficients are related to each other 
\be 
C^{(2)}_2=C^{(2)}_3,\quad C^{(2)}_5=C^{(2)}_6=C^{(2)}_7=C^{(2)}_8
\ee and 
\be 
C^{(2)}_{11}=C^{(2)}_{12}=C^{(2)}_{13}=C^{(2)}_{14}=0.
\ee 
The bulk-to-boundary correlator is reduced to 
\bea 
D_{\mu\nu\rho\sigma}&=&C^{(2)}_1\frac{\eta_{\mu\nu}\eta_{\rho\sigma}}{\widehat u^{\Delta}}+C^{(2)}_2\frac{\eta_{\mu\rho}\eta_{\nu\sigma}+\eta_{\mu\sigma}\eta_{\nu\rho}}{\widehat u^\Delta}\nn\\&&+{ C^{(2)}_4\frac{\eta_{\mu\nu}n_\rho n_\sigma}{\widehat u^{\Delta+2}}}+C^{(2)}_9\frac{\eta_{\rho\sigma}n_\mu n_\nu}{\widehat u^{\Delta+2}}+C^{(2)}_{10}\frac{n_\mu n_\nu n_\rho n_\sigma}{\widehat u^{\Delta+4}}\nn\\&&+C^{(2)}_5\frac{\eta_{\mu\rho}n_\nu n_\sigma+\eta_{\mu\sigma}n_\nu n_\rho +\eta_{\nu\rho}n_\mu n_\sigma+\eta_{\nu\sigma}n_\mu n_\rho}{\widehat u^{\Delta+2}}.\label{symt}
\eea 
If $t_{\mu\nu}$ is symmetric traceless, we find three further relations 
\be 
4C^{(2)}_1+2C^{(2)}_{2}=0,\quad C^{(2)}_5{ +C^{(2)}_4}=0,\quad C^{(2)}_9{+C^{(2)}_5}=0
\ee Therefore, the bulk-to-boundary correlator becomes 
\begin{align}
    D_{\mu\nu\rho\sigma}={}&C^{(2)}_2\frac{\eta_{\mu\rho}\eta_{\nu\sigma}+\eta_{\mu\sigma}\eta_{\nu\rho}-\frac{1}{2}\eta_{\mu\nu}\eta_{\rho\sigma}}{\widehat u^\Delta}+C^{(2)}_{10}\frac{n_\mu n_\nu n_\rho n_\sigma}{\widehat u^{\Delta+4}}\nn\\
&{+C^{(2)}_4\frac{
\eta_{\mu\nu}n_\rho n_\sigma+\eta_{\rho\sigma}n_\mu n_\nu
-\bigl(\eta_{\mu\rho}n_\nu n_\sigma+\eta_{\mu\sigma}n_\nu n_\rho
+\eta_{\nu\rho}n_\mu n_\sigma+\eta_{\nu\sigma}n_\mu n_\rho\bigr)
}{\widehat u^{\Delta+2}} }.\label{symtraceless}
\end{align}
\item For an antisymmetric tensor $t_{\mu\nu}$, the bulk-to-boundary correlator should satisfy the equation 
\be 
D_{\mu\nu\rho\sigma}=-D_{\nu\mu\rho\sigma}=-D_{\mu\nu\sigma\rho}=D_{\nu\mu\sigma\rho}.
\ee In this case, we find 
\bea 
C^{(2)}_1=C^{(2)}_4=C^{(2)}_9=C^{(2)}_{10}=0
\eea and 
\bea 
\begin{gathered}
C^{(2)}_3=-C^{(2)}_2,\quad C^{(2)}_5=-C^{(2)}_6=-C^{(2)}_7=C^{(2)}_8,\\
C^{(2)}_{11}=-C^{(2)}_{12},\quad C^{(2)}_{13}=0.
\end{gathered}
\eea The corresponding bulk-to-boundary correlator is 
\bea 
D_{\mu\nu\rho\sigma}&=&C^{(2)}_2\frac{\eta_{\mu\rho}\eta_{\nu\sigma}-\eta_{\mu\sigma}\eta_{\nu\rho}}{\widehat u^{\Delta}}+C^{(2)}_5\frac{\eta_{\mu\rho}n_{\nu}n_\sigma-\eta_{\mu\sigma}n_\nu n_\rho-\eta_{\nu\rho}n_\mu n_\sigma +\eta_{\nu\sigma}n_\mu n_\rho}{\widehat u^{\Delta+2}}\nn\\&&+C^{(2)}_{11}\frac{\omega_{\mu\nu\rho}n_\sigma-\omega_{\mu\nu\sigma}n_\rho}{\widehat u^{\Delta+2}}+C^{(2)}_{14}\frac{\epsilon_{\mu\nu\rho\sigma}}{\widehat u^\Delta}.\label{antiD}
\eea 
\end{itemize} We will discuss several interesting examples in the following. 
\begin{enumerate}
    \item Graviton propagator. The graviton propagator in momentum space can be found in \cite{tHooft:1974toh}. One can use the Pauli-Fierz action and add the gauge-fixing term {with parameters $(\alpha,\xi)$} to the Lagrangian to obtain the quadratic terms
\begin{align}
\mathcal{L} =\ & -\frac{1}{4} \partial_\alpha h_{\mu\nu} \partial^\alpha h^{\mu\nu} + \frac{1}{4} \partial_\alpha h \partial^\alpha h -\frac{1}{2}\partial_\alpha h \partial_\beta h^{\alpha\beta} +\frac{1}{2} \partial_\alpha h_{\mu\beta} \partial^\beta h^{\mu\alpha} \nn\\
& -\frac{1}{2\xi} \left( \partial^\nu h_{\mu\nu} - \frac{\alpha}{2} \partial_\mu h \right) \left( \partial_\rho h^{\mu\rho} - \frac{\alpha}{2} \partial^\mu h \right)
\end{align} 
In momentum space, the linearized action is 
\bea 
S=\frac{1}{2}\int \frac{d^4k}{(2\pi)^4}h_{\mu\nu}(-k)O^{\mu\nu\rho\sigma}(k)h_{\rho\sigma}(k),
\eea where 
\bea 
O^{\mu\nu\rho\sigma}(k)=\beta_1 k^2P_1^{\mu\nu\rho\sigma}+\beta_2 k^2P_2^{\mu\nu\rho\sigma}+\beta_3P_3^{\mu\nu\rho\sigma}+\beta_4P_4^{\mu\nu\rho\sigma}
\eea where 
\bs\begin{align}
    P_1^{\mu\nu\rho\sigma}&=\eta^{\mu\rho}\eta^{\nu\sigma}+\eta^{\mu\sigma}\eta^{\nu\rho},\\
    P_2^{\mu\nu\rho\sigma}&=\eta^{\mu\nu}\eta^{\rho\sigma},\\
    P_3^{\mu\nu\rho\sigma}&=\eta^{\mu\nu}k^\rho k^\sigma+\eta^{\rho\sigma}k^\mu k^\nu,\\
    P_4^{\mu\nu\rho\sigma}&=\eta^{\mu\rho}k^\nu k^\sigma+\eta^{\mu\sigma}k^\nu k^\rho +\eta^{\nu\rho}k^\mu k^\sigma+\eta^{\nu\sigma}k^\mu k^\rho
\end{align}\es and 
\bea 
\beta_1=-\frac{1}{4},\quad \beta_2=\frac{1}{2}(1-\frac{\alpha^2}{2\xi}),\quad \beta_3=-\frac{1}{2}(1-\frac{\alpha}{\xi}),\quad \beta_4=\frac{1}{4}(1-\frac{1}{\xi}).
\eea 
The inverse of $O^{\mu\nu\rho\sigma}$ is the Feynman propagator in momentum space
\bea 
G^{\mu\nu\rho\sigma}=\frac{\gamma_1}{k^2}P_1^{\mu\nu\rho\sigma}+\frac{\gamma_2}{k^2}P_2^{\mu\nu\rho\sigma}+\frac{\gamma_3}{k^4}P_3^{\mu\nu\rho\sigma}+\frac{\gamma_4}{k^4}P_4^{\mu\nu\rho\sigma}+\frac{\gamma_5 k^\mu k^\nu k^\rho k^\sigma}{k^6}
\eea with 
\begin{align}
    \gamma_1=1 ,\quad \gamma_2=-1,\quad \gamma_3=\frac{2\alpha-2}{\alpha-2},\quad \gamma_4=\xi-1 , \quad
\gamma_5=-\frac{4(\alpha-1)[(\alpha-3)\xi+\alpha+1]}{(\alpha-2)^2}.
\end{align}
These parameters satisfy
\be
{O_{\mu\nu}}^{\rho\sigma}G_{\rho\sigma\alpha\beta}=-\frac{1}{2}(\eta_{\mu\alpha}\eta_{\nu\beta}+\eta_{\mu\beta}\eta_{\nu\alpha}).
\ee
 Our result could match the general conclusion of  \cite{VanNieuwenhuizen:1973fi} by restricting to the symmetric tensor Lagrangian in that paper.

{
In position space, the first two terms lead to the following form of the
Feynman propagator
\bea
G^{(1)\mu\nu\rho\sigma}(x;x') 
&=&
\frac{1}{4\pi^2}
\frac{
\eta^{\mu\rho}\eta^{\nu\sigma}
+\eta^{\mu\sigma}\eta^{\nu\rho}
-\eta^{\mu\nu}\eta^{\rho\sigma}
}{
(x-x')^2+i\epsilon
}.
\eea

The last three terms are gauge-dependent and give
\bea
G^{(2)\mu\nu\rho\sigma}(x;x')
&=&
\frac{1}{4\pi^2}
\frac{1}{(x-x')^2+i\epsilon}
\Bigg\{
\gamma_3\left(P_2^{\mu\nu\rho\sigma}-Q_3^{\mu\nu\rho\sigma}\right)
+\gamma_4\left(P_1^{\mu\nu\rho\sigma}-Q_4^{\mu\nu\rho\sigma}\right)
\nonumber\\
&&\quad
+\gamma_5\left[
\frac{P_1^{\mu\nu\rho\sigma}+P_2^{\mu\nu\rho\sigma}}{8}
-\frac{Q_3^{\mu\nu\rho\sigma}+Q_4^{\mu\nu\rho\sigma}}{4}
+Q_5^{\mu\nu\rho\sigma}
\right]
\Bigg\},
\eea
where
\bs\begin{align}
    Q_3^{\mu\nu\rho\sigma}
    &=
    \frac{(x-x')^\mu (x-x')^\nu \eta^{\rho\sigma}
    +(x-x')^\rho (x-x')^\sigma \eta^{\mu\nu}}
    {(x-x')^2+i\epsilon},
    \\
    Q_4^{\mu\nu\rho\sigma}
    &=
    \frac{\begin{gathered}
    (x-x')^\mu (x-x')^\rho \eta^{\nu\sigma}
    +(x-x')^\mu (x-x')^\sigma \eta^{\nu\rho}
    \\
    +(x-x')^\nu (x-x')^\rho \eta^{\mu\sigma}
    +(x-x')^\nu (x-x')^\sigma \eta^{\mu\rho}\end{gathered}}
    {(x-x')^2+i\epsilon},
    \\
    Q_5^{\mu\nu\rho\sigma}
    &=
    \frac{(x-x')^\mu (x-x')^\nu
    (x-x')^\rho (x-x')^\sigma}
    {\left[(x-x')^2+i\epsilon\right]^2}.
\end{align}\es
Note that \(G^{(2)}\) depends on the gauge parameters. A similar result can be found in \cite{Glavan:2019}. {When extrapolated to the boundary}, these gauge-dependent terms  in $D=\lim_+rG^{(2)}$
blow up at null infinity unless their coefficients vanish. This requires
\(\alpha=\xi=1\) and thus
\be
\gamma_3=\gamma_4=\gamma_5=0.
\ee
This is the Feynman choice, corresponding to the Lorenz-type
condition
\be
\partial^\nu \left(h_{\mu\nu}-\frac{1}{2}\eta_{\mu\nu}h\right)=0.
\ee

Therefore, in this gauge the Feynman propagator simplifies to
\be
G^{\mu\nu\rho\sigma}(x;x')
=
\frac{1}{4\pi^2}
\frac{
\eta^{\mu\rho}\eta^{\nu\sigma}
+\eta^{\mu\sigma}\eta^{\nu\rho}
-\eta^{\mu\nu}\eta^{\rho\sigma}
}{
(x-x')^2+i\epsilon
}.
\ee
From this expression we obtain the fall-off index \(\Delta=1\). Using
the same extrapolation as in the spin-1 bulk-to-boundary analysis, one obtains
\bea
D_{\mu\nu\rho\sigma}(u,\Omega;x')
&=&
-\frac{1}{8\pi^2}
\frac{
\eta_{\mu\rho}\eta_{\nu\sigma}
+\eta_{\mu\sigma}\eta_{\nu\rho}
-\eta_{\mu\nu}\eta_{\rho\sigma}
}{
u+n\cdot x'
}.
\eea
Therefore,
\be
C^{(2)}_2=C^{(2)}_3=-C^{(2)}_1
=
-\frac{1}{8\pi^2}.
\ee
}
{The bulk-to-boundary correlator in a general gauge can be found in Appendix \ref{nextL}. }
    \item In a four-dimensional CFT, the two-point function for a spin-2 operator with conformal dimension $\bar\Delta$ has the form  \cite{Costa:2011mg}
    \be 
    \langle\text{T} t_{\mu\nu}(x)t_{\rho\sigma}(x')\rangle=\frac{C_t}{(x-x')^{2\bar\Delta}}P_{\mu\nu\rho\sigma}(x-x')
    \ee where the rank-4 tensor $P$ is 
    \be 
    P_{\mu\nu\rho\sigma}(x)=\frac{1}{2}\left(I_{\mu\rho}(x)I_{\nu\sigma}(x)+I_{\mu\sigma}(x)I_{\nu\rho}(x)-\frac{1}{2}\eta_{\mu\nu}\eta_{\rho\sigma}\right)
    \ee and $I_{\mu\nu}(x)$ is defined in \eqref{Imunu}. There is only one structure in the expression and $C_t$ is the unique constant. The fall-off index of the spin-2 operator should be $\Delta=\bar\Delta-2$. The bulk-to-boundary correlator is 
    \be 
    D_{\mu\nu\rho\sigma}(u,\Omega;x')=\lim_{r\to \mathscr I^+}r^{\bar\Delta-2} \langle\text{T} t_{\mu\nu}(x)t_{\rho\sigma}(x')\rangle= 
    {\frac{C_t n_\mu n_\nu n_\rho n_\sigma}{ (-2)^{\bar\Delta}(u+n\cdot x')^{\bar\Delta+2}
    }}.\label{dmunurhosigma}
    \ee As in the spin-1 case, the fall-off index $\Delta$ is not equal to the conformal dimension $\bar\Delta$. Instead, we should identify 
    \be 
   \Delta=\bar\Delta-2\label{spin2fall}
    \ee to obtain a finite result. The constants in \eqref{symt} are 
    \bea 
    C^{(2)}_1=C^{(2)}_2=C^{(2)}_9=C^{(2)}_5=0,\quad C^{(2)}_{10}={ \frac{C_t}{(-2)^{\bar\Delta}}}.
    \eea To check the relation between the fall-off index and the conformal weight, we consider a free scalar theory with the fall-off in \eqref{fall-offPhi}. A  simple spin-2 operator is the stress tensor
    \be 
    t_{\mu\nu}=\partial_\mu\Phi \partial_\nu\Phi-\frac{1}{2}\eta_{\mu\nu}(\partial_\rho\Phi)^2-\frac{1}{6}(\partial_\mu\partial_\nu-\eta_{\mu\nu}\Box)\Phi^2\label{stressfreescalar}
    \ee with the fall-off index $\Delta=2$
    \be 
    t_{\mu\nu}=\frac{n_\mu n_\nu(2\dot\Sigma^2 -\Sigma\ddot\Sigma)}{3r^2}+\cdots.  
    \ee The corresponding boundary spin-2 operator is 
    \be 
    \Sigma_{\mu\nu}={\frac{n_\mu n_\nu(2\dot\Sigma^2 -\Sigma\ddot\Sigma)}{3}}.
    \ee Therefore, the bulk-to-boundary correlator is
    \bea 
    \langle \text{T}
\left(\Sigma_{\mu\nu}(u,\Omega)t_{\rho\sigma}(x')\right)\rangle= { \frac{4 \left(C^{(0)}\right)^2 n_\mu n_\nu n_\rho n_\sigma}{3(u+n\cdot x')^6}}.
    \eea     The above result matches  \eqref{dmunurhosigma}.
    \item Extrapolation limit. We  consider the most general bulk-to-boundary correlator \eqref{fullD}. The magnetic branch is obtained directly 
    \bea 
    B^{\text{mag}}_{\mu\nu\rho\sigma}(u,\Omega;v',\Omega')=\lim_{r'\to\mathscr I^-}r'^{\Delta} D_{\mu\nu\rho\sigma}(u,\Omega;x')=\frac{Q^{(1)}_{\mu\nu\rho\sigma}}{(n\cdot \bar n')^\Delta}
    \eea with 
    \be 
    Q^{(1)}_{\mu\nu\rho\sigma}=C^{(2)}_1 \eta_{\mu\nu}\eta_{\rho\sigma}+C^{(2)}_2\eta_{\mu\rho}\eta_{\nu\sigma}+C^{(2)}_3\eta_{\mu\sigma}\eta_{\nu\rho}+C^{(2)}_{14}\epsilon_{\mu\nu\rho\sigma}.\label{q1}
    \ee 
    There are four independent structures in the magnetic branch. The first three terms are parity-invariant while the last term is parity-odd. Note that the last term does not appear in the scalar or spin-1 case.  The electric branch is much more complicated:  
 \begin{align}
        B_{\mu\nu\rho\sigma}^{\text{ele}}(u,\Omega;v',\Omega')&=\left\{\begin{array}{cc} 0&0<\Delta<1,\\
     \begin{gathered}\pi \Bigl( -2Q^{(1)}_{\mu\nu\rho\sigma}\ln (u-v')
+ \frac{Q^{(2)}_{\mu\nu\rho\sigma} }{(u-v')^2}+\frac{1}{2} \frac{Q^{(3)}_{\mu\nu\rho\sigma} }{(u-v')^4}\Bigr)\delta^{(2)}(\Omega-\Omega'^{\text{P}})\end{gathered} &\Delta=1,\\
       \text{divergent} &\Delta>1.\end{array}\right.
    \end{align} 
    {Here,} the tensor $Q^{(1)}$ is given in \eqref{q1} and the other two tensors $Q^{(2)}$ and $Q^{(3)}$ are 
    \bs\begin{align}
Q^{(2)}_{\mu\nu\rho\sigma}=&C^{(2)}_4\eta_{\mu\nu}n_\rho n_\sigma+C^{(2)}_5 \eta_{\mu\rho}n_\nu n_\sigma +C^{(2)}_6 \eta_{\mu\sigma}n_\nu n_\rho\nn\\&+C^{(2)}_7 \eta_{\nu\rho}n_\mu n_\sigma +C^{(2)}_8 \eta_{\nu\sigma}n_\mu n_\rho+C^{(2)}_9  \eta_{\rho\sigma}n_\mu n_\nu \nn\\&+C^{(2)}_{11}\omega_{\mu\nu\rho}n_\sigma+C^{(2)}_{12}\omega_{\mu\nu\sigma}n_\rho+C^{(2)}_{13}\omega_{\mu\rho\sigma}n_\nu,\\
        Q^{(3)}_{\mu\nu\rho\sigma}=&C^{(2)}_{10}n_\mu n_\nu n_\rho n_\sigma.
    \end{align}\es 
        \item KL representation. The KL representation for a symmetric traceless spin-2 operator can be recast into three independent structures \cite{Karateev:2020axc} 
        \begin{equation}
            \begin{aligned}
                  G^{\mu\nu\rho\sigma}(x;x')&=-i\int_0^\infty ds \Bigl(\rho_t^0(s) \Delta_{F,0}^{\mu\nu\rho\sigma}(x-x') { -}\rho_t^1(s)\Delta_{F,1}^{\mu\nu\rho\sigma}(x-x')+\rho_t^2(s)\Delta_{F,2}^{\mu\nu\rho\sigma}(x-x')\Bigr)
            \end{aligned}
        \end{equation}
       with
        \bs\begin{align}
            \Delta_{F,0}^{\mu\nu\rho\sigma}&=\frac{4}{3}(\frac{s}{4}\eta^{\mu\nu}-\partial^\mu \partial^\nu)(\frac{s}{4}\eta^{\rho\sigma}-\partial^\rho \partial^\sigma) \Delta_F(x-x';s),\\
            \Delta_{F,1}^{\mu\nu\rho\sigma}&=\frac{1}{2}\Bigl[s\Bigl(\eta^{\nu\sigma}\partial^\mu\partial^\rho+\eta^{\nu\rho}\partial^\mu\partial^\sigma\nn +\eta^{\mu\rho}\partial^\nu\partial^\sigma+\eta^{\mu\sigma}\partial^\nu\partial^\rho\Bigr)-4\partial^\mu\partial^\nu\partial^\rho\partial^\sigma\Bigr]\Delta_F(x-x';s)
            \\
    \Delta_{F,2}^{\mu\nu\rho\sigma}&=\Bigl[\frac{1}{2}(s\eta^{\mu\rho}-\partial^\mu\partial^\rho)(s\eta^{\nu\sigma}-\partial^\nu\partial^\sigma)+(\rho\leftrightarrow\sigma)\nn\\&\qquad -\frac{1}{3}(s\eta^{\mu\nu}-\partial^\mu\partial^\nu)(s\eta^{\rho\sigma}-\partial^\rho\partial^\sigma)\Bigr]\Delta_F(x-x';s)
        \end{align}\es 
where $\rho_t^0,\ \rho_t^1,\ \rho_t^2$ are spectral densities associated with the spin-0, spin-1, and spin-2 representations of the little group, respectively. A straightforward calculation leads to 
\bea 
G^{\mu\nu\rho\sigma}(x;x')&=&\int_0^\infty ds s^2\left[ \frac{\rho_t^2}{2}(\eta^{\mu\rho}\eta^{\nu\sigma}+\eta^{\mu\sigma}\eta^{\nu\rho})+\left(\frac{1}{12}\rho_t^0-\frac{1}{3}\rho_t^2\right)\eta^{\mu\nu}\eta^{\rho\sigma}\right]G\nn\\&&+\int_0^\infty ds\frac{s}{3}(\rho_t^2-\rho_t^0)\left[\eta^{\mu\nu}\partial^\rho\partial^\sigma+\eta^{\rho\sigma}\partial^\mu\partial^\nu\right]G\nn\\&&+\int_0^\infty ds\frac{s}{2}(-\rho_t^2{-}\rho_t^1)\left[\eta^{\mu\rho}\partial^\nu\partial^\sigma+\eta^{\nu\sigma}\partial^\mu \partial^\rho+\eta^{\mu\sigma}\partial^\nu\partial^\rho+\eta^{\nu\rho}\partial^\mu\partial^\sigma\right]G\nn\\&&+\int_0^\infty ds\left(\frac{2}{3}\rho_t^2{+}2\rho_t^1+\frac{4}{3}\rho_t^0\right)\partial^\mu\partial^\nu\partial^\rho\partial^\sigma G.
\eea 

{We have defined} the effective combinations of the spectral densities $\rho^{\mathrm{eff}}_i$ with $i=0,1,2$ {%
\begin{align}
\rho^{\mathrm{eff}}_0
&= \frac{4}{3}\rho_t^0 + 2\rho_t^1 + \frac{2}{3}\rho_t^2,\qquad 
\rho^{\mathrm{eff}}_1= 2s\bigl(\rho_t^1+\rho_t^2\bigr), \qquad 
\rho^{\mathrm{eff}}_2= s^2\rho_t^2.
\end{align}
 Keeping the spectral densities general, the three
independent coefficients in \eqref{symtraceless} satisfy
\bs\begin{align}
\frac{C_2^{(2)}}{\widehat u^\Delta}
&=\lim_{r\to\infty}r^\Delta
\int_0^\infty ds\,
\left[
\frac{1}{2}\rho_2^{\mathrm{eff}}G
+\rho_1^{\mathrm{eff}}\partial_hG
+4\rho_0^{\mathrm{eff}}\partial_h^2G
\right],
\label{spin2-KL-general-C2}\\
\frac{C_4^{(2)}}{\widehat u^{\Delta+2}}
&=\lim_{r\to\infty}r^{\Delta+2}
\int_0^\infty ds\,
\left[
\bigl(\rho_1^{\mathrm{eff}}-s\rho_0^{\mathrm{eff}}\bigr)
\partial_h^2G
-8\rho_0^{\mathrm{eff}}\partial_h^3G
\right],
\label{spin2-KL-general-C4}\\
\frac{C_{10}^{(2)}}{\widehat u^{\Delta+4}}
&=16\lim_{r\to\infty}r^{\Delta+4}
\int_0^\infty ds\,
\rho_0^{\mathrm{eff}}\partial_h^4G.
\label{spin2-KL-general-C10}
\end{align}
\es 

{As in} the spin-1 case, 
we assume that the three effective spectral densities
$\rho^{\mathrm{eff}}_i$ consist of  functions
that are locally integrable on $s\in(0,\infty)$, possibly
supplemented by $\delta(s)$ terms. For  $i=0,1,2$, they
satisfy
\bs\label{iruv}
\begin{align}
\exists\ \alpha_i>0,\qquad
\left|\rho^{\mathrm{eff}}_i(s)\right|
&=o\!\left(e^{\alpha_i\sqrt{s}}\right),
&&s\to\infty,
\label{spin2-KL-UV-convergence}
\\
\exists\ \epsilon_i>0,\qquad
\left|\rho^{\mathrm{eff}}_i(s)\right|
&=o\!\left(s^{-1+\epsilon_i}\right),
&&s\to0^+.
\label{spin2-KL-IR-convergence}
\end{align}
\es
For a gauge-invariant local symmetric traceless tensor operator
acting on the physical Hilbert space of a unitary QFT, the spectral densities are non-negative 
\begin{equation}
\rho_t^0(s)\geq0,\qquad
\rho_t^1(s)\geq0,\qquad
\rho_t^2(s)\geq0
\quad\text{for} s>0.
\label{spin2-KL-positivity}
\end{equation}  The Wightman
positivity also ensures the local integrability of
the three effective densities on $(0,\infty)$.

Under these conditions, one can also show that any leading pure-power
extrapolation satisfies
\begin{equation}
\Delta\geq1.
\label{spin2-KL-bound}
\end{equation}
The equality holds if and only if there is a nonzero
massless-pole contribution in the effective combinations.
Furthermore, in the absence of massless poles, Wightman
positivity requires $C_4^{(2)}=C_2^{(2)}=0$.
A detailed proof is given in Appendix~\ref{expan}.

We now analyze some examples below:

\begin{itemize}

\item
A simple power-law choice without massless poles is
\begin{align}
\rho^{\mathrm{eff}}_0(s)
&=A s^{\Delta-2},
\nonumber\\
\rho^{\mathrm{eff}}_1(s)
&=B s^{\Delta-2},
\nonumber\\
\rho^{\mathrm{eff}}_2(s)
&=C s^{\Delta-2},
\qquad \Delta>1.
\label{spin2-rho-power}
\end{align}
For this choice, the leading Bessel integrals can be evaluated
explicitly, giving
\bs \begin{align}
C_{10}^{(2)}
&=
\frac{(-1)^{-\Delta}2^{\Delta-4}
\Gamma(\Delta-1)\Gamma(\Delta+4)}{\pi^2}\,A,
\\
C_4^{(2)}
&=
\frac{(-1)^{-\Delta}2^{\Delta-6}
\Gamma(\Delta-1)\Gamma(\Delta+2)}{\pi^2}\,B,
\\
C_2^{(2)}
&=
\frac{(-1)^{-\Delta}2^{\Delta-5}
\Gamma(\Delta-1)\Gamma(\Delta)}{\pi^2}\,C.
\label{spin2-KL-coefficients}
\end{align}\es 
This construction gives an explicit relation between the leading
Ward coefficients and the spectral coefficients.
Positivity of the physical spin-channel densities requires
$A\geq0$, $B\geq0$, and $C\geq0$, and also gives
\begin{equation}
0\leq\rho^{\mathrm{eff}}_1(s)
\leq3s\rho^{\mathrm{eff}}_0(s),
\qquad
0\leq\rho^{\mathrm{eff}}_2(s)
\leq\frac{3}{2}s^2\rho^{\mathrm{eff}}_0(s).
\end{equation}
For the power-law choice \eqref{spin2-rho-power}, a nonzero
$B$ or $C$ violates these inequalities near $s=0$.
Therefore, positivity implies
\begin{equation}
B=C=0,\qquad C_4^{(2)}=C_2^{(2)}=0.
\label{spin2-positive-leading}
\end{equation}
Note that the bound $\Delta\ge 1$ is a consequence of the local integrability of the spectral densities and their UV and IR asymptotic behavior \eqref{iruv}. The Wightman positivity is not necessary in the proof. Thus in a general QFT, the coefficients $B$ or $C$ need not be zero.

\item
A spectral density of the graviton in a generic covariant gauge can contain
$\delta$-derivative contributions, which fall outside the
assumed class of spectral densities
and can lead to $\Delta<1$.
For example, in our conventions the traceless projection of
the free graviton propagator \cite{Glavan:2019} has
\begin{equation}
\rho^{\mathrm{eff}}_0(s)
=\frac{\gamma_5}{2}\delta''(s),
\qquad
\rho^{\mathrm{eff}}_1(s)=\gamma_4\delta'(s),
\qquad
\rho^{\mathrm{eff}}_2(s)=2\delta(s).
\end{equation}
The $\delta''(s)$ term gives $\Delta=-1$ when $\gamma_5\ne0$.
When $\gamma_5=0$ and $\gamma_4\ne0$, the $\delta'(s)$ term
gives $\Delta=0$.
These gauge-dependent $\delta$-derivative contributions are
excluded by the assumptions leading to \eqref{spin2-KL-bound}
and are also incompatible with Wightman positivity.

\item
For a symmetric traceless spin-2 primary operator in a
four-dimensional unitary CFT, the unitarity bound is
$\bar\Delta\geq4$, with $\Delta=\bar\Delta-2\geq2$.
Each nonvanishing continuous physical spin density scales as
\cite{Karateev:2020axc}
\begin{equation}
\begin{aligned}
2\pi \rho_t^0(s) &= C_t \times \frac{(\Delta -  2)(\Delta - 1)}{(\Delta + 2)(\Delta + 3)}  \theta(s) s^{\Delta - 2} \\
2\pi \rho_t^1(s) &= C_t \times \frac{(\Delta - 2)(\Delta + 1)}{(\Delta + 2)(\Delta + 3)}  \theta(s) s^{\Delta - 2} \\
2\pi \rho_t^2(s) &= C_t \times \frac{\Delta + 1}{\Delta + 3} \theta(s) s^{\Delta- 2},
\end{aligned}
\end{equation}

with a non-negative proportionality coefficient.
These densities have polynomial UV growth and satisfy the
IR and local-integrability conditions above.
They match \eqref{spin2-rho-power} if we consider $\rho^{\mathrm{eff}}_i$ to contain additional terms that contribute only at subleading order in the null
extrapolation. In this case, we can see that $A=C_t\frac{2\Delta(\Delta-1)}{\pi(\Delta+3)(\Delta+2)}$ and $B=C=0$. 
\end{itemize}}
\item Antisymmetric field. For any bulk vector field $t_\mu$ with the fall-off behavior \eqref{fall-offtmu}, we can construct an antisymmetric field 
   \(t_{\mu\nu}=\partial_\mu t_\nu -\partial_\nu t_\mu\) whose leading coefficient is 
   \be 
   \Sigma_{\mu\nu}=-n_\mu \dot \Sigma_\nu+n_\nu \dot \Sigma_\mu
   \ee and the fall-off index of $t_{\mu\nu}$ is still $\Delta$, the same as {that of} $t_\mu$. Then the bulk-to-boundary correlator is 
   \bea 
   D_{\mu\nu\rho\sigma}&=&-n_\mu \partial'_\rho \dot D_{\nu\sigma}+n_\mu \partial'_\sigma \dot D_{\nu\rho}+n_\nu \partial'_\rho \dot D_{\mu\sigma}-n_\nu \partial'_\sigma\dot D_{\mu\rho}\nn\\&=&-C^{(1)}_1\Delta(\Delta+1)\frac{\eta_{\mu\rho}n_\nu n_\sigma-\eta_{\mu\sigma}n_\nu n_\rho-\eta_{\nu\rho}n_\mu n_{\sigma}+\eta_{\nu\sigma}n_\mu n_\rho}{\widehat u^{\Delta+2}}.
   \eea Comparing with \eqref{antiD}, {we find that} the unique non-vanishing coefficient is $C^{(2)}_5=-C^{(1)}_1\Delta(\Delta+1)$. A corollary is that the bulk-to-boundary correlator $\langle \Sigma_{\mu\nu} t_{\rho\sigma}\rangle$ vanishes for the operator $t_{\mu\nu}=\partial_\mu t_\nu-\partial_\nu t_\mu$ with a conserved current $t_\mu$. This is because  the normalization constant $C_1^{(1)}=0$ for any conserved current, as shown by  \eqref{consbtob}.
\end{enumerate}

  \subsection{Bulk-to-boundary correlators for higher spin operators}
In this subsection, we discuss the application of the bulk-to-boundary correlators to  spin-$n$ operators. The general form is 
\bea 
D_{\mu_1\mu_2\cdots\mu_{2n}}=\sum_{N=1}^{\text{Catalan number}} C^{(n)}_N \mathbb{T}^{(N)}_{\mu_1\cdots \mu_{2n}}\widehat u^{-\Delta-q_N}
\eea where $C^{(n)}_N$ are constants. The superscript $(n)$ denotes the rank of the corresponding operator.  The integer  $q_N$ is equal to the number of the null vectors $n$ in \(\mathbb{T}^{(N)}_{\mu_1\cdots \mu_{2n}}\). We can extrapolate each term to the boundary to obtain the boundary-to-boundary correlators. For a general $q_N>0$, the boundary-to-boundary correlator is 
\bea 
B^{(N)}_{\mu_1\mu_2\cdots\mu_{2n}}&=&\lim_{r'\to\mathscr I^-}r'^\Delta C^{(n)}_N\mathbb{T}^{(N)}_{\mu_1\cdots \mu_{2n}} \widehat u^{-\Delta-q_N}\nn=\left\{\begin{array}{cc}0&0<\Delta<1,\\
\frac{2\pi C^{(n)}_N}{q_N}\frac{\mathbb T^{(N)}_{\mu_1\mu_2\cdots\mu_{2n}}}{(u-v')^{q_N}}\delta^{(2)}(\Omega-\Omega'^{\text{P}}) &\Delta=1,\\
\text{divergent}&\Delta>1.\end{array}\right.
\eea For the tensor structure with $q_N=0$, the corresponding boundary-to-boundary correlators are similar to those of the scalars
\be 
B^{(N)}_{\mu_1\mu_2\cdots\mu_{2n}}=\left\{\begin{array}{cc}C^{(n)}_N\mathbb T^{(N)}_{\mu_1\mu_2\cdots\mu_{2n}}(n\cdot \bar n')^{-\Delta}&0<\Delta<1,\\
\begin{gathered}-2\pi C^{(n)}_N\mathbb T^{(N)}_{\mu_1\mu_2\cdots\mu_{2n}}\ln (u-v')\delta^{(2)}(\Omega-\Omega'^{\text{P}})
+C^{(n)}_N\mathbb T^{(N)}_{\mu_1\mu_2\cdots\mu_{2n}} (n\cdot \bar n')^{-1}\end{gathered} &\Delta=1,\\
\text{divergent}&\Delta>1.\end{array}\right.
\ee 
In previous discussions, our classification of the bulk-to-boundary correlator fits nicely with the KL representation for spin-1 and spin-2 operators. The KL representation of higher spin operators in AdS and dS spacetime is given in \cite{Dusedau:1985ue,Loparco:2023rug}. The parallel analysis in flat spacetime can be found in \cite{Raszillier1967SpectralRF,Oksak:1969ta,Todorov:1969,Mathews:1971zy}. It would be valuable to connect our results to the IR limit of {the} higher spin KL representation. 

 In general, 
a symmetric traceless rank-\(n\) operator transforms in the Lorentz representation
\be 
(j_L,j_R)=\left({n\over2},{n\over2}\right)
\ee where $j_L$ and $j_R$ are half-integers characterizing the irreducible representations of 
SU(2).
Upon restriction to the massive little group \(SO(3)\)\footnote{In {the} KL representation, the two-point correlators are written as the summation of free propagators with {squared mass} $m^2=s$. This is the reason why the massive little group appears here.}, it decomposes as
\begin{equation}
{n\over2}\otimes {n\over2}=\bigoplus_{J=0}^{n}J.\label{sumJ}
\end{equation}

 The non-negativity of the spectral densities multiplying the spin-\(J\) projectors follows from the usual Wightman positivity assumption. Here, if the operator \(t_{\mu_1\cdots\mu_n}\)
 is a real local field acting on a Hilbert
space with a unitary Poincaré representation{, then,} for any test function \(f\),
\be
 \langle 0|t(f)t(f)|0\rangle
 =
 \|t(f)|0\rangle\|^2\ge 0 .
\ee 
This does not impose constraints on gauge-dependent operators in theories with gauge degrees of freedom.
Equivalently, the momentum space spectral tensor $\rho_{\mu_1\cdots\mu_n;\nu_1\cdots\nu_n}(p)$ is a positive-semidefinite matrix. On the mass shell
\(s=-p^2>0\), choose the rest frame \(\tilde{p}=(\sqrt{s},\mathbf 0)\).
The Lorentz transformations preserving \(\tilde{p}\) form the massive
little group \(SO(3)\). Lorentz covariance then implies that the spectral
tensor \(\rho_{\mu_1\cdots\mu_n;\nu_1\cdots\nu_n}(\tilde{p})\) commutes with this \(SO(3)\)
action, i.e.
\be 
D(R)_{\mu_1\cdots\mu_n}{}^{\alpha_1\cdots\alpha_n}
\rho_{\alpha_1\cdots\alpha_n;\nu_1\cdots\nu_n}(\tilde{p})
=
\rho_{\mu_1\cdots\mu_n;\beta_1\cdots\beta_n}(\tilde{p})
{D(R)^{\beta_1\cdots\beta_n}}_{\nu_1\cdots\nu_n},
\qquad R\in SO(3),
\ee 
where \(D(R)\) is the \(SO(3)\) action on the rank-\(n\) STT tensor
space.
Since \eqref{sumJ} is multiplicity-free, Schur's lemma reduces it to one
positive spectral function in each spin-\(J\) channel. With positive
normalization of the fixed-spin projectors, the spectral densities in the
spin-\(J\) channels are non-negative,
\be 
 \rho_J(s)\ge0,\qquad J=0,\ldots,n .
\ee 
   For the STT representation
of rank-$n$, the corresponding KL spectral density is \cite{Todorov:1969}
\begin{equation}
\rho_{\mu_1\cdots\mu_n;\nu_1\cdots\nu_n}(p)
=
\sum_{J=0}^{n}\rho_J(s)\,
\Pi^{(n,J)}_{\mu_1\cdots\mu_n;\nu_1\cdots\nu_n}(p),
\qquad s=-p^2 ,
\end{equation}
where \(\Pi^{(n,J)}\) is the covariant tensor structure for the massive spin-\(J\) component, and
\(\rho_J(s)\) is the corresponding  spectral density. Hence a rank-\(n\) STT operator has
\(n+1\) independent massive KL spectral channels. The number $n+1$ coincides with \eqref{numberSTT}.

Equivalently, in coordinate space,
\begin{equation}
\langle0|T\{t_{\mu_1\cdots\mu_n}(x)t_{\nu_1\cdots\nu_n}(0)\}|0\rangle
 =
\sum_{J=0}^{n}\int_0^\infty ds\,
\rho_J(s)\,
\mathscr P^{(n,J)}_{\mu_1\cdots\mu_n;\nu_1\cdots\nu_n}(\partial;s)\,
G(x;s),\label{klrephs}
\end{equation}
where \(\mathscr P^{(n,J)}\) is the differential operator for the spin-\(J\) covariant tensor structure.
In \cite{Oksak:1969ta,Todorov:1969}, the tensor operator is encoded as a homogeneous polynomial of degree \(n\) in each spinor,
\begin{equation}
{{\tt{t}}}(x,z,\bar z)
=
t_{A_1\cdots A_n;\dot A_1\cdots\dot A_n}(x)
z^{A_1}\cdots z^{A_n}
\bar z^{\dot A_1}\cdots \bar z^{\dot A_n}
\end{equation}
where $t_{A_1\cdots A_n;\dot A_1\cdots\dot A_n}$ is the two-spinor form of the STT tensor \(t_{\mu_1\cdots\mu_n}\).
The auxiliary spinor two-point function is written as
\begin{equation}
\begin{aligned}
&\langle 0|{\tt t}(x,z,\bar z){\tt t}(0,w,\bar w)|0\rangle =\\
&\quad \sum_{J=0}^{n}
\int_0^\infty ds\,\rho_J(s)
\int \frac{d^4p}{(2\pi)^4}\,
e^{{ -}ip\cdot x}\,
2\pi\theta(p^0)\delta(p^2+s)\,
\mathcal K_{n,J}(p;z,w).
\end{aligned}
\end{equation}
Up to a normalization  constant \(N_{n,J}\), the spin-\(J\) block {is~}\cite{Todorov:1969}
\begin{equation}
\mathcal K_{n,J}(p;z,w)
=
N_{n,J}\,[P_z(p)P_w(p)]^n
P_J\!\left(1-\frac{2s\,S_{zw}}{P_z(p)P_w(p)}\right),
\qquad J=0,\ldots,n .
\end{equation}
Here
\begin{equation}
P_z=z^A p_{A\dot A}\bar z^{\dot A},
\qquad
P_w=w^B p_{B\dot B}\bar w^{\dot B},
\end{equation}
and
\begin{equation}
S_{zw}=(zw)(\bar z\bar w)
=
\epsilon_{AB}z^A w^B
\epsilon_{\dot A\dot B}\bar z^{\dot A}\bar w^{\dot B}.
\label{eq:SzwinKL}
\end{equation}
The Legendre polynomial $P_J$ appears because the spin-square equation reduces to the usual \(SO(3)\) angular equation for the  spin-$J$ STT representation.

To pass to ordinary Lorentz indices, we should expand \(\mathcal K_{n,J}\) in
\(z,\bar z,w,\bar w\), extract the coefficient of
\be 
z^{A_1}\bar z^{\dot A_1}\cdots
z^{A_n}\bar z^{\dot A_n}
w^{B_1}\bar w^{\dot B_1}\cdots
w^{B_n}\bar w^{\dot B_n},
\ee
and use the spinor-vector conversion convention
\begin{equation}
t_{A_1\dot A_1\cdots A_n\dot A_n}
=
\sigma^{\mu_1}_{A_1\dot A_1}\cdots
\sigma^{\mu_n}_{A_n\dot A_n}
t_{\mu_1\cdots\mu_n},
\end{equation}
whose inverse is 
\begin{equation}
t_{\mu_1\cdots\mu_n}
 =
(-2)^{-n}
\bar\sigma_{\mu_1}^{\dot A_1 A_1}\cdots
\bar\sigma_{\mu_n}^{\dot A_n A_n}
t_{A_1\cdots A_n;\dot A_1\cdots\dot A_n}.
\end{equation}

The symmetric traceless part of a symmetric tensor $T_{\rho_1\rho_2\cdots\rho_n}$ is obtained by the {formula~}\cite{Liu:2023jnc}
\begin{equation}
T^{\rm STT}_{\rho_1\cdots\rho_n}
=
\sum_{q=0}^{\lfloor n/2\rfloor}
b_{nq}\,
\eta_{(\rho_1\rho_2}\cdots
\eta_{\rho_{2q-1}\rho_{2q}}
T^{\alpha_1}{}_{\alpha_1}\cdots{}^{\alpha_q}{}_{\alpha_q
\rho_{2q+1}\cdots\rho_n)} ,\label{STTpro}
\end{equation}
where parentheses denote unit-weight symmetrization and
\begin{equation}
b_{nq}=(-1)^q\frac{(n-q)!}{4^q q!(n-2q)!}.
\end{equation}
This is the ordinary version of the symmetric subtraction of traces required for a general spin tensor.
Applying this trace subtraction  to both index sets and replacing \(p_\mu\to i\partial_\mu\), one obtains the STT-projected form, which can be formally expressed as follows
\bea 
\mathscr P^{(n,J)}_{\mu_1\cdots\mu_n;\nu_1\cdots\nu_n}(\partial;s)=N_{n,J}\sum_{\varrho=0}^J (-1)^{n-\varrho}d_{J,\varrho}(s)\left(\mathcal R^{(n,\varrho)}_{(\mu_1\cdots\mu_n);(\nu_1\cdots\nu_n)}(\partial)\right)^{\text{STT}}
\label{Prm}
\eea 
where 
\be 
\mathcal R^{(n,\varrho)}_{\mu_1\cdots\mu_n;\nu_1\cdots\nu_n}(\partial)
=
\eta_{\mu_1\nu_1}\cdots\eta_{\mu_\varrho\nu_\varrho}
\partial_{\mu_{\varrho+1}}\cdots\partial_{\mu_n}
\partial_{\nu_{\varrho+1}}\cdots\partial_{\nu_n}\label{mathcalR}
\ee and 
\be 
d_{J,\varrho}(s)=\left(\frac{s}{2}\right)^\varrho \frac{(J+\varrho)!}{(J-\varrho)!(\varrho!)^2}.
\ee The symbol $\left(\mathcal R^{(n,\varrho)}_{(\mu_1\cdots\mu_n);(\nu_1\cdots\nu_n)}(\partial)\right)^{\text{STT}}$ denotes projecting the tensor $\mathcal R^{(n,\varrho)}_{(\mu_1\cdots\mu_n);(\nu_1\cdots\nu_n)}(\partial)$ to the symmetric and traceless part for $\mu$'s and $\nu$'s separately. {Defining a family} of functions 
\bea 
G_q(\delta x;s)=\frac{s^{q/2}(-h)^{-q/2}}{4\pi^2}K_{q}(\sqrt{-sh}),\quad \delta x^\mu=x^\mu-x'^\mu,\quad h=-(\delta x)^2,
\eea then one finds the recursion relation
\be 
\partial_\mu G_q(\delta x;s)=-\delta x_\mu G_{q+1}(\delta x;s).
\ee Then the $m$-th derivative of $G_q$ is 
\begin{equation}
\begin{aligned}
\partial_{\mu_1}\cdots\partial_{\mu_m}G_q(\delta x;s)
&=\sum_{\ell=0}^{\lfloor{\frac{m}{2}\rfloor}}(-1)^{m-\ell}{ \frac{m!}{2^\ell \ell !(m-2\ell )!}}\\
&\quad \eta_{(\mu_1\mu_2}\eta_{\mu_3\mu_4}\cdots\eta_{\mu_{2\ell-1}\mu_{2\ell}}\delta x_{\mu_{2\ell+1}}\cdots \delta x_{\mu_{m})}G_{q+m-\ell}(\delta x;s).\label{partialG}
\end{aligned}
\end{equation} Notice that $G_1(\delta x;s)=G$. Acting on $G$, the operator $\mathcal R^{(n,\varrho)}_{\mu_1\cdots\mu_n;\nu_1\cdots\nu_n}(\partial)$ produces $n-\varrho+1$ structures labeled by $\ell=0,1,\cdots, n-\varrho$. For each fixed $\ell$,  the number of $\delta x$ is $2(n-\varrho-\ell)$ and the number of metric tensors $\eta_{\mu\nu}$ is $\varrho+\ell$. Extrapolating to the null boundary, we conclude that the independent tensor structures are of the form 
\bea 
\mathbb T_{\mu_1\mu_2\cdots\mu_n,\nu_1\cdots\nu_n} \sim \eta \eta \cdots\eta nn\cdots n\eea 
where the number of $\eta$'s is $\varrho+\ell$ and the number of $n$'s is $2n-2(\varrho+\ell)$. Note that one should project it onto the symmetric and traceless part. This is consistent with the bulk-to-boundary correlator of spin-$n$ operators. 
More explicit results can be found in  Appendix \ref{expan}.

\section{Spinning multiplets in Carrollian CFTs}\label{bcca}
In previous sections, we have constructed the spinning bulk-to-boundary correlators. The spin-$s$ boundary operator $\Sigma_{\mu_1\mu_2\cdots\mu_s}$ is related to the bulk field $t_{\mu_1\mu_2\cdots\mu_s}$ via the fall-off condition \eqref{falof}. Unfortunately, the boundary operator $\Sigma_{\mu_1\mu_2\cdots\mu_s}$ is not a 
single primary operator for general $s$. Instead, it contains multiple components. The  local operators in the CCFT have been discussed in \cite{Chen:2021xkw,2023Univ....9..385N}. Here we will clarify the boundary operators $\Sigma_{\mu_1\mu_2\cdots\mu_s}$ in the language of CCFT$_3$. We will consider the {boundary operators of spin 1} and spin 2 firstly and then extend the representation to  higher spin operators.
\subsection{Spin-1 multiplet}
Note that the global symmetry group of CCFT$_3$ is isomorphic to the Poincar\'e group. The latter is generated by the vector fields 
\bs\begin{align}
    \bm\xi_{\text{trans}}&=c^\mu \left(-n_\mu\partial_u+m_\mu \partial_r-\frac{1}{r}Y_\mu^I\partial_I\right),\\
\bm\xi_{\text{Lor}}&=\omega^{\mu\nu}\left(-u n_{\mu\nu}\partial_u+(u+r)n_{\mu\nu}\partial_r+(Y_{\mu\nu}^I+\frac{u}{2r}(Y_{\mu\nu}^I-\bar Y_{\mu\nu}^I))\partial_I\right)
\end{align}\es where $\bm\xi_{\text{trans}}$ and $\bm\xi_{\text{Lor}}$ generate spacetime translation and Lorentz  transformations, respectively. The tensor \(\bar Y_{\mu\nu}^{I}\) is defined as \(\bar Y_{\mu\nu}^I=Y^I_\mu\bar n_\nu-Y^I_\nu\bar n_\mu\). The Lie derivative of the vector $t_\mu$ along $\bm\xi$ is 
\be 
\delta_{\bm\xi}t_\mu=\xi^\nu \partial_\nu t_\mu+\partial_\mu\xi^\nu t_\nu
\ee and it induces the transformation of the boundary operator $\Sigma_\mu$
\be 
\delta_{\bm\xi}\Sigma_\mu=\xi^u\dot\Sigma_\mu+\xi^I \partial_I\Sigma_\mu-\Delta\frac{\xi^r}{r}\Sigma_\mu+\partial_\mu\xi^\nu \Sigma_\nu.\label{lied}
\ee The above equality is valid in the limit $r\to\infty$. To classify local operators at null infinity, it is convenient to study the stabilizer subgroup which is defined to preserve the boundary location
\be 
\xi^u(u_\star,\Omega_\star)=\xi^I(u_\star,\Omega_\star)=0\label{stabilizer}
\ee where $(u_*,\Omega_*)$ is the chosen boundary point.

{
This  stabilizer condition \eqref{stabilizer} is solved by the subgroup $H$ of the Poincar\'e group 
which is used to classify the boundary local operators. In general,  for a boundary multiplet
\(\Sigma_a(\star)\) inserted at the chosen point, the stabilizer subgroup acts only on its component
indices,
\begin{equation}
    U(h)\,\Sigma_a(\star)\,U(h)^{-1}
    =
    \rho_a{}^b(h^{-1})\,\Sigma_b(\star),
    \qquad h\in H.
\end{equation}
Here  $U(h)$ denotes the unitary representation operator associated with  $h$. The subscript $a$  labels the components of the representation and $\rho_a^{\ b}$ is  its matrix realization of the representation. The boundary operator away from the chosen point is obtained by a boundary translation 
\be 
\Sigma_a({\tt x})=e^{{\tt{x}}^\alpha \mathcal P_\alpha}\Sigma_a(\star)e^{-{\tt{x}}^\alpha \mathcal P_\alpha}
\ee where the generators $\mathcal P_\alpha$ move the reference  point to  a general boundary position ${\tt x}$.
}

Without loss of generality \footnote{The stabilizer subgroup  at a general boundary point is discussed in  Appendix \ref{generalpoint}.}, we set 
\be 
(u_\star,\Omega_\star)=(u_\star,w_\star,\bar w_\star)=(0,0,0)
\ee where we have used the stereographic coordinates $(w,\bar w)$ on the unit sphere.
This point is the south pole of the sphere on the $u=0$ slice. The corresponding $n^\mu_\star$ is 
\be 
n^\mu_\star=(1,0,0,-1).
\ee The equations are solved by 
\bs\begin{align} 
&\mathbb K_0\equiv \xi^{(0)}=-\partial_t+\partial_z,\quad \mathbb B_1\equiv \xi^{(1)}=-\partial_x,\quad \mathbb B_2\equiv \xi^{(2)}=-\partial_y,\\
&\mathbb J\equiv\xi^{(3)}=-x\partial_y+y\partial_x,\quad \mathbb D\equiv\xi^{(4)}=-t\partial_z-z\partial_t,\\ &\mathbb K_1\equiv \xi^{(5)}=t\partial_x+x\partial_t-x\partial_z+z\partial_x,\quad \mathbb K_2\equiv\xi^{(6)}=t\partial_y+y\partial_t-y\partial_z+z\partial_y
\end{align}\label{eq:a1-south}\es 
The operator $\mathbb K_0$ generates the light‑front time translation along {the} $v$ direction while $\mathbb B_1/\mathbb B_2$ are transverse translations that generate translations along {the} $x/y$ directions. The operator $\mathbb J$ represents a transverse rotation generator in the $x$-$y$ plane and $\mathbb D$  generates a Lorentz boost along the $z$ direction. Moreover, $\mathbb K_1$ and $\mathbb K_2$ are null rotation generators.
They form a seven-dimensional  subalgebra 
\bs\begin{align}
    &[\mathbb{K}_0,\ \mathbb{D}] = \mathbb{K}_0, \\
&[\mathbb{B}_1,\ \mathbb{J}] = -\mathbb{B}_2, \\
&[\mathbb{B}_1,\ \mathbb{K}_1] = \mathbb{K}_0, \\
&[\mathbb{B}_2,\ \mathbb{J}] = \mathbb{B}_1, \\
&[\mathbb{B}_2,\ \mathbb{K}_2] = \mathbb{K}_0, \\
&[\mathbb{J},\ \mathbb{K}_1] = \mathbb{K}_2, \\
&[\mathbb{J},\ \mathbb{K}_2] = -\mathbb{K}_1, \\
&[\mathbb{D},\ \mathbb{K}_1] = -\mathbb{K}_1, \\
&[\mathbb{D},\ \mathbb{K}_2] = -\mathbb{K}_2.
\end{align}\label{eq:a1-south-brackets}\es The corresponding group is isomorphic to the stabilizer subgroup of CCFT$_3$  at the origin. Note that $\mathbb J, \mathbb K_1$ and $\mathbb K_2$ generate the Wigner little group $\text{ISO}(2)$. Therefore, we may call $\mathbb K_1$ and $\mathbb K_2$ the Wigner translation generators. 

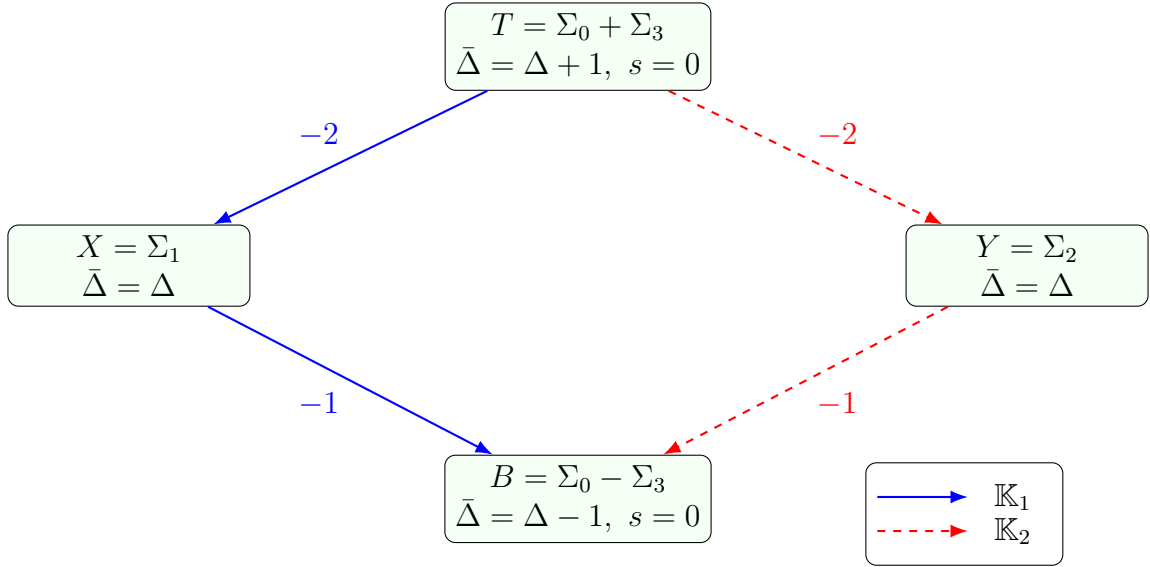
\begin{figure}[htbp]
\centering
\begin{tikzpicture}[node distance=2.5cm,>=Latex]
\tikzset{box/.style={draw,rounded corners,align=center,minimum width=3.2cm,minimum height=0.9cm,fill=green!4}}

\node[box] (top) {$T=\Sigma_0+\Sigma_3$\\$\bar\Delta=\Delta+1,\ s=0$};
\node[box,below left=of top,xshift=-0.8cm] (m1) {$X=\Sigma_1$\\$\bar\Delta=\Delta$};
\node[box,below right=of top,xshift=0.8cm] (m2) {$Y=\Sigma_2$\\$\bar\Delta=\Delta$};
\node[box,below=of $(m1)!0.5!(m2)$] (bot) {$B=\Sigma_0-\Sigma_3$\\$\bar\Delta=\Delta-1,\ s=0$};

\draw[->,thick,blue] (top)--node[above left,text=blue] {$-2$}(m1);
\draw[->,thick,red,dashed] (top)--node[above right,text=red] {$-2$}(m2);
\draw[->,thick,blue] (m1)--node[below left,text=blue] {$-1$}(bot);
\draw[->,thick,red,dashed] (m2)--node[below right,text=red] {$-1$}(bot);

\node[draw,rounded corners,align=left,fill=white,anchor=west] at ($(bot)+(3.8,-0.2)$) {
\begin{tikzpicture}[>=Latex,baseline=(current bounding box.center)]
\draw[->,thick,blue] (0,0.45) -- (1.2,0.45);
\node[anchor=west] at (1.4,0.45) {$\mathbb K_1$};
\draw[->,thick,red,dashed] (0,0) -- (1.2,0);
\node[anchor=west] at (1.4,0) {$\mathbb K_2$};
\end{tikzpicture}
};

\end{tikzpicture}
\caption{$T,X,Y,B$ form a spin-1 multiplet representation. The solid blue line represents the action of $\mathbb K_1$ and the dashed red line represents the action of $\mathbb K_2$. The number on the arrow is the coefficient induced by the action. For instance, $\mathbb K_1 T=-2X$.}
\label{fig:K-real-net}
\end{figure}

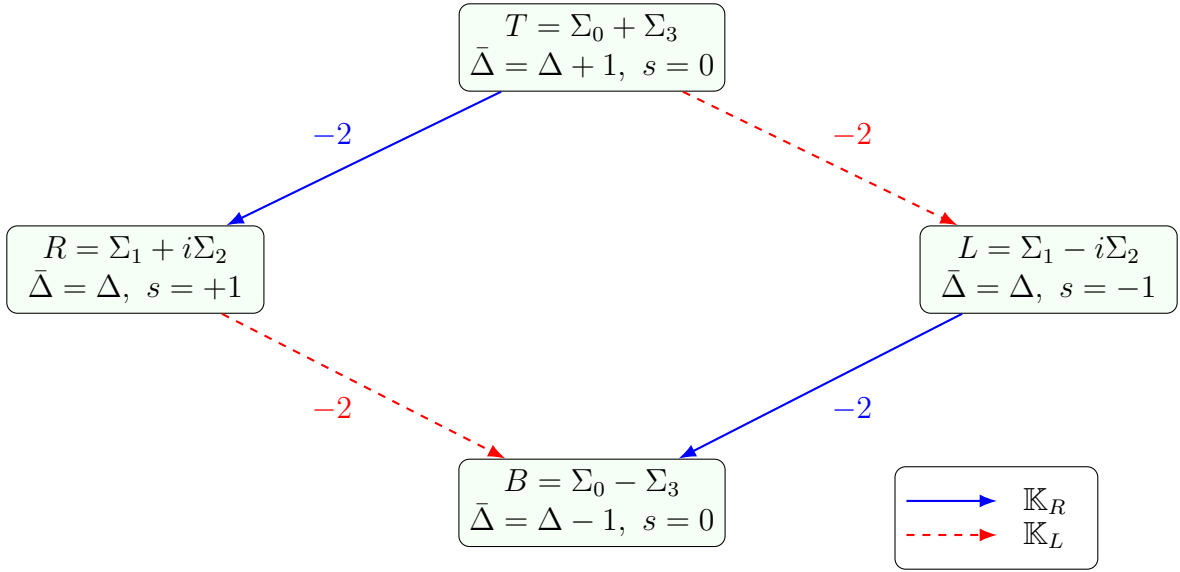
\begin{figure}[htbp]
\centering
\resizebox{0.98\linewidth}{!}{%
\begin{tikzpicture}[node distance=2.5cm,>=Latex]
\tikzset{box/.style={draw,rounded corners,align=center,minimum width=3.4cm,minimum height=0.9cm,fill=green!4}}

\node[box] (top) {$T=\Sigma_0+\Sigma_3$\\$\bar\Delta=\Delta+1,\ s=0$};
\node[box,below left=of top,xshift=-0.8cm] (ar) {$R=\Sigma_1+i\Sigma_2$\\$\bar\Delta=\Delta,\ s=+1$};
\node[box,below right=of top,xshift=0.8cm] (al) {$L=\Sigma_1-i\Sigma_2$\\$\bar\Delta=\Delta,\ s=-1$};
\node[box,below=of $(ar)!0.5!(al)$] (bot) {$B=\Sigma_0-\Sigma_3$\\$\bar\Delta=\Delta-1,\ s=0$};

\draw[->,thick,blue] (top)--node[above left,text=blue] {$-2$}(ar);
\draw[->,thick,red,dashed] (top)--node[above right,text=red] {$-2$}(al);
\draw[->,thick,red,dashed] (ar)--node[below left,text=red] {$-2$}(bot);
\draw[->,thick,blue] (al)--node[below right,text=blue] {$-2$}(bot);

\node[draw,rounded corners,align=left,fill=white,anchor=west] at ($(bot)+(4.0,-0.2)$) {
\begin{tikzpicture}[>=Latex,baseline=(current bounding box.center)]
\draw[->,thick,blue] (0,0.45) -- (1.2,0.45);
\node[anchor=west] at (1.4,0.45) {$\mathbb K_R$};
\draw[->,thick,red,dashed] (0,0) -- (1.2,0);
\node[anchor=west] at (1.4,0) {$\mathbb K_L$};
\end{tikzpicture}
};

\end{tikzpicture}
}
\caption{The spin-1 multiplet representation in the basis $T,R,L,B$. The solid blue line represents the action of $\mathbb K_R$ and the dashed red line represents the action of $\mathbb K_L$. The number on the arrow is the coefficient induced by the action. For instance, $\mathbb K_R T=-2R$. }
\label{fig:K-complex-net}
\end{figure}

We denote $\Sigma^\star_\mu=\Sigma_\mu(u_\star,\Omega_\star)$. Then the variation \eqref{lied} becomes
\be -\delta_{\bm\xi}\Sigma_\mu^\star=\Delta\frac{\xi^r}{r}\Sigma_\mu^\star-\partial_\mu\xi^\nu \Sigma^\star_\nu
\ee where we have chosen the opposite sign to match our convention. For example, 
\be 
\mathbb D\Sigma_0^\star=\Delta\Sigma_0^\star+\Sigma_3^\star.
\ee Note that $\Sigma_0^\star$ is not an eigenvector of the dilatation operator. We redefine the operators as follows 
\bea 
T=\Sigma_0^\star+\Sigma_3^\star,\quad B=\Sigma_0^\star-\Sigma_3^\star,\quad X=\Sigma_1^\star,\quad Y=\Sigma_2^\star
\eea and then 
\bs\begin{align}
   & \mathbb{ D}T=(\Delta+1)T,\quad \mathbb{ D}X=\Delta X,\quad \mathbb{ D}Y=\Delta Y,\quad \mathbb{ D}B=(\Delta-1)B,\\
   &\mathbb{ J}T=0,\quad \mathbb{ J}X=Y,\quad \mathbb{ J}Y=-X,\quad \mathbb{ J}B=0,\\
   &\mathbb{K}_1T=-2X,\quad \mathbb{K}_1X=-B,\quad \mathbb{K}_1Y=0,\quad \mathbb{K}_1B=0,\\
    &\mathbb{K}_2T=-2Y,\quad \mathbb{K}_2X=0,\quad \mathbb{K}_2Y=-B,\quad \mathbb{K}_2B=0,\\
    &\mathbb{K}_0T=\mathbb{K}_0X=\mathbb{K}_0Y=\mathbb{K}_0B=0,\\
    &\mathbb{B}_1T=\mathbb{B}_1X=\mathbb{B}_1Y=\mathbb{B}_1B=0,\\ &\mathbb{B}_2T=\mathbb{B}_2X=\mathbb{B}_2Y=\mathbb{B}_2B=0.
\end{align}\es 
Note that the four components $T,X,Y, B$ form a spin-1 multiplet representation of CCFT$_3$ which is shown in Figure \ref{fig:K-real-net}. In each box, we have labeled the Carrollian conformal weight  $\bar\Delta$ of the corresponding operator.
Notice that $T$ and $B$ have definite spin 0. However, $X$ and $Y$ are not eigenvectors of the rotation operator $\mathbb J$. This  is resolved by introducing the following helicity basis 
\be 
T=\Sigma^\star_0+\Sigma^\star_3,\quad R=\Sigma^\star_1+i\Sigma^\star_2,\quad L=\Sigma^\star_1-i\Sigma^\star_2,\quad B=\Sigma^\star_0-\Sigma^\star_3
\ee and the operators 
\be 
\mathbb K_{R}=\mathbb K_1+i\mathbb K_2,\quad \mathbb K_{L}=\mathbb K_1-i\mathbb K_2.
\ee More precisely, 
\bs\begin{align}
&\mathbb{ D}T=(\Delta+1)T,\quad \mathbb{ D}R=\Delta R,\quad \mathbb{ D}L=\Delta L,\quad \mathbb{ D}B=(\Delta-1)B,\\
&\mathbb{ J}T=0,\quad \mathbb{ J}R=-iR,\quad \mathbb{ J}L=iL,\quad \mathbb{ J}B=0,\\
    &\mathbb{K}_RT=-2R,\quad \mathbb{K}_RR=0,\quad \mathbb{K}_RL=-2B,\quad \mathbb{K}_RB=0,\\
    &\mathbb{K}_LT=-2L,\quad \mathbb{K}_LR=-2B,\quad \mathbb{K}_LL=0,\quad \mathbb{K}_LB=0.
\end{align}\es This spin-1  multiplet representation  is shown in Figure \ref{fig:K-complex-net}. Now the components $R$ and $L$  have definite helicity.

Note that generally $\mathbb{K}_I\mathcal O(0)\not=0$ where $\mathcal O(0)$ is the operator in the spin-1 multiplet representation which is Type Ib in the sense of \cite{Mack:1969rr}. This representation is distinguished from the type Ia highest weight representation of \cite{Chen:2021xkw,2023Univ....9..385N} that obeys the condition   $\mathbb{K}_I\mathcal O(0)=0$. The representation is also different from the  staggered module in CCFT$_3$ considered in \cite{Chen:2023pqf}. Rather interestingly, the constraint $\Delta\ge 1$ leads to the non-negativity of the Carrollian conformal weight $\bar\Delta\ge 0$ for a vector operator.

The above representation is valid for a general vector operator {when a bulk vector field is extrapolated} to the boundary. In what follows, we discuss several examples. 
\begin{enumerate}
\item For the bulk spin-1 operator \eqref{jmuo}, the corresponding boundary operator is 
\be 
\Sigma_\mu=i n_\mu(\Sigma\dot{\overline\Sigma}-\overline\Sigma\dot\Sigma).
\ee At the south pole, we find 
\bea 
T=-2i(\Sigma\dot{\overline\Sigma}-\overline\Sigma\dot\Sigma),\quad X=Y=B=0.
\eea In this case, the second layer and the lowest layer of the representation in Figure \ref{fig:K-real-net} are deleted and the unique non-vanishing operator is $T$ with  conformal weight 
\be 
\bar\Delta=\Delta+1.
\ee This is consistent with \eqref{bardelta}.

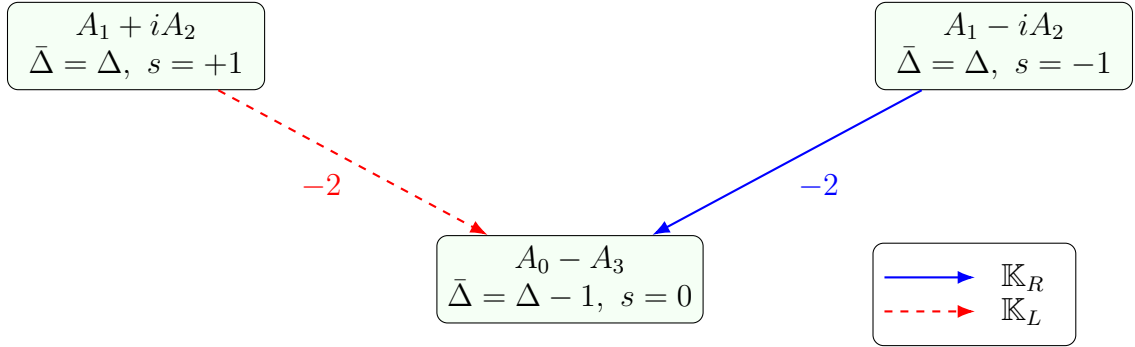
\begin{figure}[htbp]
\centering
\begin{tikzpicture}[node distance=2.5cm,>=Latex]
\tikzset{box/.style={draw,rounded corners,align=center,minimum width=3.4cm,minimum height=0.9cm,fill=green!4}}

\node[box,draw=none,fill=none,text opacity=0] (top) {$T$\\$\bar\Delta=\Delta+1,\ s=0$};
\node[box,below left=of top,xshift=-0.5cm] (ar) {$A_1+iA_2$\\$\bar\Delta=\Delta,\ s=+1$};
\node[box,below right=of top,xshift=0.5cm] (al) {$A_1-iA_2$\\$\bar\Delta=\Delta,\ s=-1$};
\node[box,below=of $(ar)!0.5!(al)$] (bot) {$A_0-A_3$\\$\bar\Delta=\Delta-1,\ s=0$};

\draw[->,thick,red,dashed] (ar)--node[below left,text=red] {$-2$}(bot);
\draw[->,thick,blue] (al)--node[below right,text=blue] {$-2$}(bot);

\node[draw,rounded corners,align=left,fill=white,anchor=west] at ($(bot)+(4.0,-0.2)$) {
\begin{tikzpicture}[>=Latex,baseline=(current bounding box.center)]
\draw[->,thick,blue] (0,0.45) -- (1.2,0.45);
\node[anchor=west] at (1.4,0.45) {$\mathbb K_R$};
\draw[->,thick,red,dashed] (0,0) -- (1.2,0);
\node[anchor=west] at (1.4,0) {$\mathbb K_L$};
\end{tikzpicture}
};

\end{tikzpicture}
\caption{For a spin-1 gauge field, the spin-1 multiplet can be projected onto the gauge-invariant subsector.}
\label{fig:K-complex-net-without-top}
\end{figure}

{%
\item
Consider a gauge field $a_\mu=\frac{A_\mu}{r^\Delta}+\cdots$. Since the potential is gauge
dependent, its fall-off must be specified together with a gauge
representative. In a general covariant
$R_\xi$ gauge, the difference from the Feynman-gauge propagator is
\begin{equation}
G_{\mu\nu}^{(\xi)}(\delta x)-G_{\mu\nu}^{(1)}(\delta x)
=
-\frac{1-\xi}{16\pi^2}
\partial_\mu\partial_\nu\log(\delta x^2+i\epsilon).
\label{coordinate-longitudinal-propagator}
\end{equation} This term is proportional to $p_\mu p_\nu$ in  momentum space and thus corresponds to the longitudinal mode. 
To study the radiative modes, 
we choose the Feynman gauge $\xi=1$, for which
this lower longitudinal sector is absent and the potential correlator
starts with the  fall-off $\Delta=1$. At the south pole,
define
\begin{equation}
B=A_0-A_3,\qquad
T=A_0+A_3,\qquad
R=A_1+iA_2,\qquad
L=A_1-iA_2.
\label{spin-one-components}
\end{equation}
Their Carrollian weights and spins are
\begin{equation}
(\bar\Delta,s)_T=(2,0),\quad
(\bar\Delta,s)_R=(1,+1),\quad
(\bar\Delta,s)_L=(1,-1),\quad
(\bar\Delta,s)_B=(0,0).
\label{spin-one-weights}
\end{equation}
In the middle layer of Figure \ref{fig:K-complex-net}, the modes $R$ and $L$ correspond to the two radiative modes with opposite helicities in the transverse plane.
The choice $\xi=1$ still admits residual gauge transformations
$a_\mu\to a_\mu+\partial_\mu\lambda$ satisfying
$\Box\lambda=0$ \cite{Hirai:2018ijc}. The gauge parameter $\lambda$ may be expanded asymptotically 
\be 
\lambda=\epsilon_0(\Omega)+\frac{\epsilon(u,\Omega)}{r}+\frac{\epsilon_2(u,\Omega)}{r^2}+\cdots
\ee to keep the fall-off conditions. Here $\epsilon_0(\Omega)$ denotes the large gauge transformation\footnote{Terms with fall-off $r^{-1}\log r$ can also be inserted in the expansion. However, such a kind of fall-off can be eliminated by choosing $\epsilon_0(\Omega)$ to be a meromorphic function on the unit sphere. } while $\epsilon(u,\Omega)$ corresponds to a small gauge transformation {and could be any smooth function} on the unit sphere. 
The gauge  transformations of the four modes at the south pole are
\be
\delta T
=2\dot\epsilon,
\qquad 
\delta B=0,\qquad 
\delta R
=-(Y_1^I+iY_2^I)\partial_I\epsilon_0,
\qquad 
\delta L
=-(Y_1^I-iY_2^I)\partial_I\epsilon_0.
\label{spin-one-gauge-transformations}
\ee Therefore, we can eliminate the mode $T$ by choosing  an {appropriate} $\epsilon(u,\Omega)$.  The gauge-invariant subsector ({up to large gauge transformations}) is shown in Figure \ref{fig:K-complex-net-without-top}.

The above reduction does not use the equations of motion. Maxwell's
equation gives the additional {constraint~}\cite{Liu:2023qtr}
\begin{equation}
\dot B=0
\qquad\Longrightarrow\qquad
B=B(\Omega),
\label{B-Maxwell-constraint}
\end{equation}
so the $B$ mode is not dynamical and the $u$-dependent radiative
data are carried by $R$ and $L$ in Maxwell theory.

For comparison, in the infrared analysis of QED by Strominger and {collaborators~}\cite{Strominger:2017zoo}, the soft photon mode is the transverse zero mode

\be
N_z=\int du\,F_{uz}^{(0)}
=A_z^{(0)}\big|_{\mathscr I^+_+}-A_z^{(0)}\big|_{\mathscr I^+_-}
\ee
defined in retarded radial gauge \(A_r=0\), \(A_u|_{\mathscr I^+}=0\), and it enters the soft charge whose Ward identity gives the leading soft photon theorem. The present mode \(B=A_0-A_3=A_r\) is instead a radial/longitudinal component constrained by Maxwell's equations to satisfy \(\partial_u B=0\), so it is only a non-radiative boundary zero mode and should not be identified with the soft photon mode directly.

}

\end{enumerate}
\subsection{Spin-2 multiplet}
Now we extend the previous discussion to  a spin-2 operator $\Sigma_{\mu\nu}$ by adapting it to the language of CCFT$_3$. The Lie-derivative of a rank-2 field is 
\be 
\delta_{\bm\xi}t_{\mu\nu}=\xi^\rho\partial_\rho t_{\mu\nu}+\partial_\mu\xi^\rho t_{\rho\nu}+\partial_\nu\xi^\rho t_{\mu\rho}.
\ee At the south pole, we find the variation of the boundary operator\footnote{The notation $\star$ that represents the reference point has been omitted below.}
\be 
-\delta_{\bm\xi}\Sigma_{\mu\nu}=\Delta \frac{\xi^r}{r}\Sigma_{\mu\nu}-\partial_\mu \xi^\rho \Sigma_{\rho\nu}-\partial_\nu\xi^\rho\Sigma_{\mu\rho}.
\ee At the south pole, we define a set of frame vectors
\be 
e_0^\mu=(1,0,0,0),\quad e_1^\mu=(0,1,0,0),\quad e_2^\mu=(0,0,1,0),\quad e_3^\mu=(0,0,0,1).
\ee Then the boundary operator $\Sigma_{\mu\nu}$ is projected onto the local frame 
\bea 
T_{\alpha\beta}=e_\alpha^\mu e_\beta^\nu \Sigma_{\mu\nu}
\eea where $\alpha,\beta=0,1,2,3$. A convenient basis is $\{e_+^\mu,\ e_1^\mu, \ e_2^\mu,\ e_-^\mu\}$ with
\be 
e_+^\mu=e_0^\mu+e_3^\mu,\quad e_-^\mu=e_0^\mu-e_3^\mu
\ee and then 
\be 
\mathcal T_{\alpha\beta}=e_\alpha^\mu e_\beta^\nu \Sigma_{\mu\nu}
\ee where the indices $\alpha,\beta=+,1,2,-$. The 16 components in the local frame are 
\bs\begin{align}
\mathcal T_{++}&=\Sigma_{00}+\Sigma_{03}+\Sigma_{30}+\Sigma_{33},&
\mathcal T_{+1}&=\Sigma_{01}+\Sigma_{31},\\
\mathcal T_{+2}&=\Sigma_{02}+\Sigma_{32},&
\mathcal T_{+-}&=\Sigma_{00}-\Sigma_{03}+\Sigma_{30}-\Sigma_{33},\\
\mathcal T_{1+}&=\Sigma_{10}+\Sigma_{13},&
\mathcal T_{11}&=\Sigma_{11},\\
\mathcal T_{12}&=\Sigma_{12},&
\mathcal T_{1-}&=\Sigma_{10}-\Sigma_{13},\\
\mathcal T_{2+}&=\Sigma_{20}+\Sigma_{23},&
\mathcal T_{21}&=\Sigma_{21},\\
\mathcal T_{22}&=\Sigma_{22},&
\mathcal T_{2-}&=\Sigma_{20}-\Sigma_{23},\\
\mathcal T_{-+}&=\Sigma_{00}+\Sigma_{03}-\Sigma_{30}-\Sigma_{33},&
\mathcal T_{-1}&=\Sigma_{01}-\Sigma_{31},\\
\mathcal T_{-2}&=\Sigma_{02}-\Sigma_{32},&
\mathcal T_{--}&=\Sigma_{00}-\Sigma_{03}-\Sigma_{30}+\Sigma_{33}.
\end{align}\es

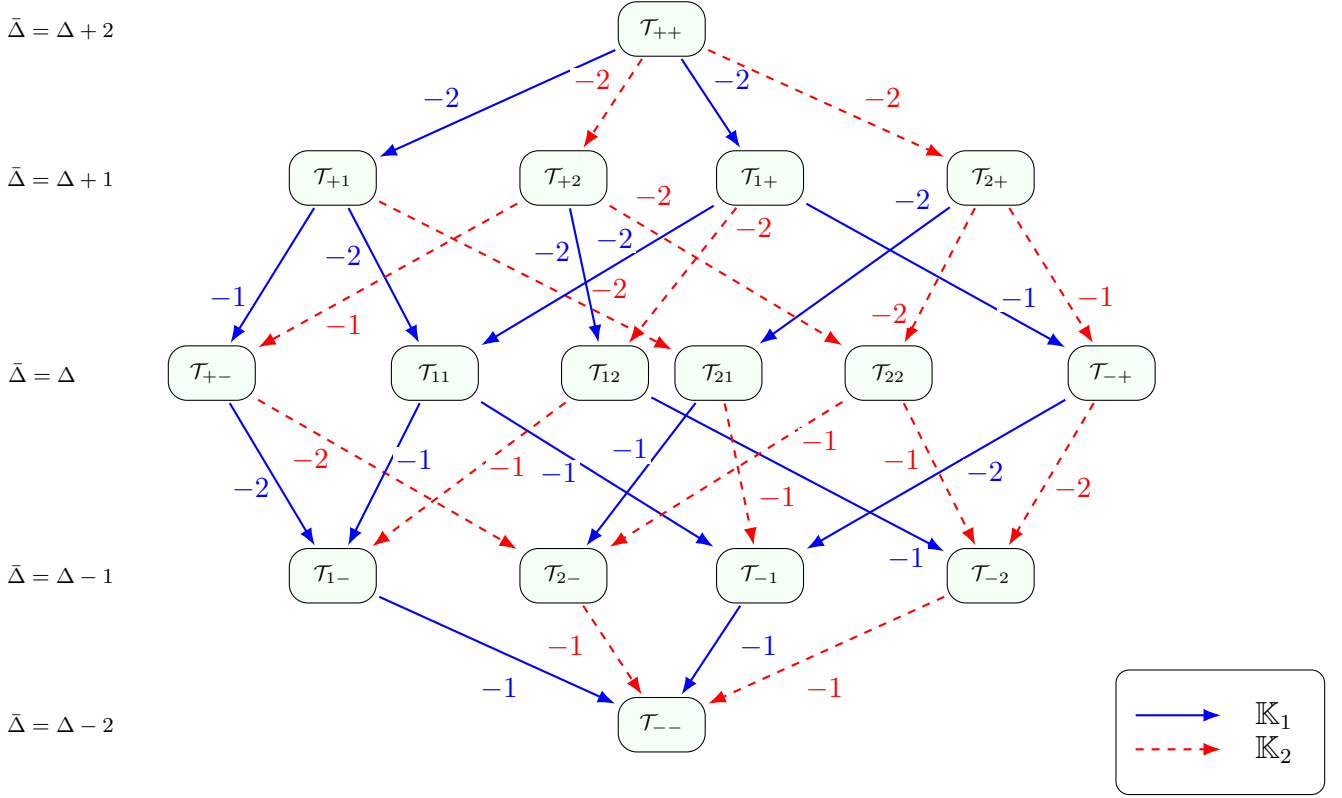
\begin{figure}
    \begin{center}
\resizebox{0.98\linewidth}{!}{%
\begin{tikzpicture}[x=1cm,y=1cm,>=Latex]

\tikzset{
  smallbox/.style={
    draw,
    rounded corners=7pt,
    fill=green!4,
    minimum width=1.15cm,
    minimum height=0.72cm,
    inner xsep=5pt,
    inner ysep=4pt,
    font=\LARGE,
    align=center
  },
  bigbox/.style={
    draw,
    rounded corners=8pt,
    fill=white,
    minimum width=1.55cm,
    minimum height=1.18cm,
    inner xsep=6pt,
    inner ysep=6pt,
    font=\LARGE,
    align=center
  },
  k1/.style={->,blue,thick,shorten <=1pt,shorten >=1pt},
  k2/.style={->,red,thick,dashed,shorten <=1pt,shorten >=1pt},
  lbl1/.style={text=blue,font=\small,inner sep=1pt,fill=white},
  lbl2/.style={text=red,font=\small,inner sep=1pt,fill=white},
  legendbox/.style={draw,rounded corners=5pt,fill=white,inner sep=6pt}
}

\node[smallbox,font=\fontsize{8}{9.6}\selectfont]   (Tpp) at (0.0,10.25) {$\mathcal{T}_{++}$};

\node[smallbox,font=\fontsize{8}{9.6}\selectfont] (Tp1) at (-4.35,8.28) {$\mathcal{T}_{+1}$};
\node[smallbox,font=\fontsize{8}{9.6}\selectfont] (Tp2) at (-1.30,8.28) {$\mathcal{T}_{+2}$};
\node[smallbox,font=\fontsize{8}{9.6}\selectfont] (T1p) at ( 1.30,8.28) {$\mathcal{T}_{1+}$};
\node[smallbox,font=\fontsize{8}{9.6}\selectfont] (T2p) at ( 4.35,8.28) {$\mathcal{T}_{2+}$};

\node[smallbox,font=\fontsize{8}{9.6}\selectfont] (Tpm) at (-5.95,5.70) {$\mathcal{T}_{+-}$};
\node[smallbox,font=\fontsize{8}{9.6}\selectfont] (T11) at (-3.00,5.70) {$\mathcal{T}_{11}$};
\node[smallbox,font=\fontsize{8}{9.6}\selectfont] (T12) at (-0.75,5.70) {$\mathcal{T}_{12}$};
\node[smallbox,font=\fontsize{8}{9.6}\selectfont] (T21) at ( 0.75,5.70) {$\mathcal{T}_{21}$};
\node[smallbox,font=\fontsize{8}{9.6}\selectfont] (T22) at ( 3.00,5.70) {$\mathcal{T}_{22}$};
\node[smallbox,font=\fontsize{8}{9.6}\selectfont] (Tmp) at ( 5.95,5.70) {$\mathcal{T}_{-+}$};

\node[smallbox,font=\fontsize{8}{9.6}\selectfont] (T1m) at (-4.35,3.02) {$\mathcal{T}_{1-}$};
\node[smallbox,font=\fontsize{8}{9.6}\selectfont] (T2m) at (-1.30,3.02) {$\mathcal{T}_{2-}$};
\node[smallbox,font=\fontsize{8}{9.6}\selectfont] (Tm1) at ( 1.30,3.02) {$\mathcal{T}_{-1}$};
\node[smallbox,font=\fontsize{8}{9.6}\selectfont] (Tm2) at ( 4.35,3.02) {$\mathcal{T}_{-2}$};

\node[smallbox,font=\fontsize{8}{9.6}\selectfont]   (Tmm) at (0.0,1.05) {$\mathcal{T}_{--}$};

\node[anchor=east,font=\fontsize{8}{9.6}\selectfont] at (-7.10,10.25) {$\bar\Delta=\Delta+2$};
\node[anchor=east,font=\fontsize{8}{9.6}\selectfont] at (-7.10,8.28)  {$\bar\Delta=\Delta+1$};
\node[anchor=east,font=\fontsize{8}{9.6}\selectfont] at (-7.60,5.70)  {$\bar\Delta=\Delta$};
\node[anchor=east,font=\fontsize{8}{9.6}\selectfont] at (-7.10,3.02)  {$\bar\Delta=\Delta-1$};
\node[anchor=east,font=\fontsize{8}{9.6}\selectfont] at (-7.10,1.05)  {$\bar\Delta=\Delta-2$};

\draw[k1] (Tpp) -- node[pos=0.64,above left=2pt and 0pt,lbl1] {$-2$} (Tp1);
\draw[k2] (Tpp) -- node[pos=0.30,above left=-4pt and 4pt,lbl2] {$-2$} (Tp2);
\draw[k1] (Tpp) -- node[pos=0.30,above right=-4pt and 4pt,lbl1] {$-2$} (T1p);
\draw[k2] (Tpp) -- node[pos=0.64,above right=2pt and 0pt,lbl2] {$-2$} (T2p);

\draw[k1] (Tp1) -- node[pos=0.66,left=3pt,lbl1] {$-1$} (Tpm);
\draw[k1] (Tp1) -- node[pos=0.28,above left=-10pt and 1pt,lbl1] {$-2$} (T11);
\draw[k2] (Tp1) -- node[pos=0.78,above=5pt,lbl2] {$-2$} (T21);

\draw[k2] (Tp2) -- node[pos=0.18,above left=-42pt and 40pt,lbl2] {$-1$} (Tpm);
\draw[k1] (Tp2) -- node[pos=0.78,above left=18pt and 7pt,lbl1] {$-2$} (T12);
\draw[k2] (Tp2) -- node[pos=0.20,above right=8pt and -9pt,lbl2] {$-2$} (T22);

\draw[k1] (T1p) -- node[pos=0.20,above left=-8pt and 12pt,lbl1] {$-2$} (T11);
\draw[k2] (T1p) -- node[pos=0.78,above right=27pt and 30pt,lbl2] {$-2$} (T12);
\draw[k1] (T1p) -- node[pos=0.66,right=7pt,lbl1] {$-1$} (Tmp);

\draw[k1] (T2p) -- node[pos=0.20,above left=7pt and -8pt,lbl1] {$-2$} (T21);
\draw[k2] (T2p) -- node[pos=0.78,left=3pt,lbl2] {$-2$} (T22);
\draw[k2] (T2p) -- node[pos=0.66,right=3pt,lbl2] {$-1$} (Tmp);

\draw[k1] (Tpm) -- node[pos=0.60,left=3pt,lbl1] {$-2$} (T1m);
\draw[k2] (Tpm) -- node[pos=0.22,below left=5pt and -7pt,lbl2] {$-2$} (T2m);

\draw[k1] (T11) -- node[pos=0.22,below left=5pt and -12pt,lbl1] {$-1$} (T1m);
\draw[k1] (T11) -- node[pos=0.22,below right=10pt and 2pt,lbl1] {$-1$} (Tm1);

\draw[k2] (T12) -- node[pos=0.18,below left=10pt and 1pt,lbl2] {$-1$} (T1m);
\draw[k1] (T12) -- node[pos=0.78,below right=10pt and 1pt,lbl1] {$-1$} (Tm2);

\draw[k1] (T21) -- node[pos=0.78,below left=-30pt and -16pt,lbl1] {$-1$} (T2m);
\draw[k2] (T21) -- node[pos=0.66,right=4pt,lbl2] {$-1$} (Tm1);

\draw[k2] (T22) -- node[pos=0.18,below left=0pt and -15pt,lbl2] {$-1$} (T2m);
\draw[k2] (T22) -- node[pos=0.22,below right=5pt and -15pt,lbl2] {$-1$} (Tm2);

\draw[k1] (Tmp) -- node[pos=0.22,below left=10pt and 1pt,lbl1] {$-2$} (Tm1);
\draw[k2] (Tmp) -- node[pos=0.60,right=3pt,lbl2] {$-2$} (Tm2);

\draw[k1] (T1m) -- node[pos=0.60,below left=5pt and 0pt,lbl1] {$-1$} (Tmm);
\draw[k2] (T2m) -- node[pos=0.22,below left=3pt and 3pt,lbl2] {$-1$} (Tmm);
\draw[k1] (Tm1) -- node[pos=0.22,below right=3pt and 3pt,lbl1] {$-1$} (Tmm);
\draw[k2] (Tm2) -- node[pos=0.60,below right=5pt and 0pt,lbl2] {$-1$} (Tmm);

\node[legendbox,anchor=west] at (6,0.95) {
\begin{tikzpicture}[>=Latex,baseline=(current bounding box.center)]
  \draw[k1] (0,0.45) -- (1.2,0.45);
  \node[anchor=west,font=\normalsize] at (1.40,0.45) {$\mathbb K_1$};
  \draw[k2] (0,0.00) -- (1.2,0.00);
  \node[anchor=west,font=\normalsize] at (1.40,0.00) {$\mathbb K_2$};
\end{tikzpicture}
};
\end{tikzpicture}
}
\end{center}\caption{Net representation of spin-2 multiplet. The blue arrow represents the action of $\mathbb K_1$ while the red arrow represents the action of $\mathbb K_2$. Each node in the small box is a component of the boundary operator.}\label{netrank2}
\end{figure}

These operators  are classified according to the Carrollian conformal weights, as shown in Table \ref{spin2mul}. We also transform the table to Figure \ref{netrank2} where operators are placed in layers according to their conformal weights. For example, the operator $\mathcal T_{++}$ has the highest conformal weight and is placed  in the highest layer. The highest layer is called level 2 for the spin-2 multiplet. 
One can use $\mathbb K_I,\ I=1,2$  to act on $\mathcal T_{++}$ and then the conformal dimension of $\mathbb K_I\mathcal T_{++}$ decreases by 1 and the corresponding operator is placed in the next highest layer whose level is 1. The above process can be repeated and each action of the operator $\mathbb K_I$ would reduce the conformal dimension and the level by 1. The operators $\mathbb K_I$  act nilpotently with index five in the sense that
\be 
\mathbb K_I^4\mathcal T_{++}\not=0,\quad \mathbb K_I^5\mathcal T_{++}=0.
\ee  Each vector index follows the chain 
\be 
0\xrightarrow{}+\xrightarrow{\mathbb{K}_I}I\xrightarrow{\mathbb{K}_I}-\xrightarrow{\mathbb{K}_I}0.\label{chains}
\ee 
Therefore, there are only five layers in the diagram. At the next highest layer (level 1), there are 4 operators ($\mathcal T_{+I},\mathcal T_{I+})$. At the middle layer (level 0), there are 6 operators ($\mathcal T_{+-},\mathcal T_{-+},\mathcal T_{IJ})$. At the next lowest layer, there are 4 operators ($\mathcal T_{-I},\mathcal T_{I-}$). 
At the lowest layer (level $-2$), there is a unique operator $\mathcal T_{--}\propto \mathbb K_I^4\mathcal T_{++}$. So the full rank-two tensor product has five layers with multiplicities
\be 
1,\ 4,\ 6,\ 4,\ 1 .
\ee 

The arrows in this net are obtained by acting with $\mathbb K_1,\mathbb K_2$ on the individual tensor components. This full net should not, however, be viewed as generated by $\mathcal T_{++}$ alone. The descendants of $\mathcal T_{++}$ stay in the symmetric sector; the antisymmetric and trace sectors have their own closed sub-nets, as described below.

{
\begin{longtable}{@{}lllll@{}}
\toprule
Components  & Conformal dimension $\bar\Delta$  &  Level &Action of $\mathbb K_1$&Action of $\mathbb K_2$\\ 
\midrule
\endfirsthead
\endhead
$\mathcal T_{++}$  & $\Delta+2$  & 2&$-2\mathcal T_{1+}-2\mathcal T_{+1}$&$-{2}\mathcal T_{2+}-{2}\mathcal T_{+2}$ \\ 
$\mathcal T_{+1}$  & $\Delta+1$  & 1&$-2\mathcal T_{11}-\mathcal T_{+-}$&$-2\mathcal T_{21}$  \\ 
$\mathcal T_{+2}$  & $\Delta+1$  &1&$-2\mathcal T_{12}$&$-2\mathcal T_{22}-\mathcal T_{+-}$  \\ 
$\mathcal T_{+-}$ & $\Delta$  & 0& $-2\mathcal T_{1-}$&$-2\mathcal T_{2-}$ \\ 
$\mathcal T_{1+}$  & $\Delta+1$  & 1 &$-\mathcal T_{-+}-2\mathcal T_{11}$&$-2\mathcal T_{12}$ \\ 
$\mathcal T_{11}$  & $\Delta$ &  0&$-\mathcal T_{-1}-\mathcal T_{1-}$&0 \\ 
$\mathcal T_{12}$  & $\Delta$ & 0&$-\mathcal T_{-2}$&$-\mathcal T_{1-}$\\ 
$\mathcal T_{1-}$  & $\Delta-1$ & -1&$-\mathcal T_{--}$&0 \\ 
$\mathcal T_{2+}$  & $\Delta+1$& 1&$-2\mathcal T_{21}$&$-\mathcal T_{-+}-2\mathcal T_{22}$ \\ 
$\mathcal T_{21}$ & $\Delta$ & 0&$-\mathcal T_{2-}$&$-\mathcal T_{-1}$ \\ 
$\mathcal T_{22}$  & $\Delta$& 0&$0$&$-\mathcal T_{-2}-\mathcal T_{2-}$ \\ 
$\mathcal T_{2-}$  & $\Delta-1$  &-1&$0$&$-\mathcal T_{--}$ \\ 
$\mathcal T_{-+}$ & $\Delta$ & 0&$-2\mathcal T_{-1}$&$-2\mathcal T_{-2}$ \\ 
$\mathcal T_{-1}$  & $\Delta-1$  & -1&$-\mathcal T_{--}$&0 \\ 
$\mathcal T_{-2}$  & $\Delta-1$  & -1&$0$&$-\mathcal T_{--}$ \\ 
$\mathcal T_{--}$ & $\Delta-2$  & -2 &$0$&0\\ 
\bottomrule\caption{\centering The action of $\mathbb K_1$ and $\mathbb K_2$ on general components of a spin-2 multiplet.}\label{spin2mul}
\end{longtable}
} 
Note that one can also use the helicity basis 
\be 
e_+^\mu=e_0^\mu+e_3^\mu,\quad e_R^\mu=e_1^\mu+ie_2^\mu,\quad e_L^\mu=e_1^\mu-ie_2^\mu,\quad e_-^\mu=e_0^\mu-e_3^\mu
\ee to project the spin-2 operator onto the components with definite conformal weights and helicities. We leave the discussion of the spin-2 multiplet representation in the helicity basis to the reader.

Note that the spin-$2$ operator discussed is not irreducible in the bulk. Usually, one can 
decompose the rank-2 field into symmetric and antisymmetric parts. The symmetric part is further decomposed {into a symmetric traceless part} and the trace term. We use a box to denote the spin-1 representation and then the spin-2 representation is in the tensor product space and we can decompose it as follows
\be\vcenter{\hbox{\ydiagram{1}}} \otimes \vcenter{\hbox{\ydiagram{1}}}
= \vcenter{\hbox{\ydiagram{2}}} \oplus \vcenter{\hbox{\ydiagram{1,1}}}
\oplus \bullet.\ee 
This decomposition will be made explicit below, where the corresponding sectors are seen to be closed under the
stabilizer action.
We define  symmetrized and antisymmetrized components by 
\begin{equation}
\cT_{(AB)}=\frac12(\cT_{AB}+\cT_{BA}),
\qquad
\cT_{[AB]}=\frac12(\cT_{AB}-\cT_{BA}).
\end{equation}
Then the symmetric part is
\begin{equation}
\cT_{++},\ \cT_{(+1)},\ \cT_{(+2)},\ \cT_{(+-)},\ \cT_{11},\ \cT_{(12)},\ \cT_{(1-)},\ \cT_{22},\ \cT_{(2-)},\ \cT_{--}.
\end{equation}
The antisymmetric part is 
\begin{equation}
\cT_{[+1]},\ \cT_{[+2]},\ \cT_{[+-]},\ \cT_{[12]},\ \cT_{[1-]},\ \cT_{[2-]}.
\end{equation} The trace term is unique 
\be 
\mathcal T=\mathcal T_{11}+\mathcal T_{22}-\mathcal T_{(+-)}.
\ee 
The above decomposition is compatible with the stabilizer action. The
generators $\mathbb D,\mathbb J,\mathbb K_I$ act on the tensor slots, so they commute with the
symmetrizer and antisymmetrizer. They also preserve the trace sector. Hence the antisymmetric
sector, the trace sector, and the traceless part of the symmetric sector are
closed under the action of $\mathbb K_I$.

We summarize the action of $\mathbb K_I$ on the symmetric sector and the antisymmetric sector in Table \ref{tab:symK} and Table \ref{tab:antiK}.
\begin{longtable}{>{\raggedright\arraybackslash}p{0.30\textwidth}>{\raggedright\arraybackslash}p{0.34\textwidth}>{\raggedright\arraybackslash}p{0.20\textwidth}}
\toprule
Symmetric sector & $\mathbb K_1$ & $\mathbb K_2$\\
\midrule
\endhead
$\cT_{++}$ & $-4\cT_{(+1)}$ & $-4\cT_{(+2)}$\\
$\cT_{(+1)}$ & $-2\cT_{11}-\cT_{(+-)}$ & $-2\cT_{(12)}$\\
$\cT_{(+2)}$ & $-2\cT_{(12)}$ & $-2\cT_{22}-\cT_{(+-)}$\\
$\cT_{(+-)}$ & $-2\cT_{(1-)}$ & $-2\cT_{(2-)}$\\
$\cT_{11}$ & $-2\cT_{(1-)}$ & $0$\\
$\cT_{(12)}$ & $-\cT_{(2-)}$ & $-\cT_{(1-)}$\\
$\cT_{(1-)}$ & $-\cT_{--}$ & $0$\\
$\cT_{22}$ & $0$ & $-2\cT_{(2-)}$\\
$\cT_{(2-)}$ & $0$ & $-\cT_{--}$\\
$\cT_{--}$ & $0$ & $0$\\
\bottomrule \caption{\centering The actions of $\mathbb K_1,\mathbb K_2$  on the symmetric sector.}\label{tab:symK}
\end{longtable}
\begin{longtable}{>{\raggedright\arraybackslash}p{0.30\textwidth}>{\raggedright\arraybackslash}p{0.34\textwidth}>{\raggedright\arraybackslash}p{0.20\textwidth}}
\toprule
\text{Antisymmetric sector}  & $\mathbb K_1$ & $\mathbb K_2$\\
\midrule
\endfirsthead
\endhead
$\cT_{[+1]}$ & $-\cT_{[+-]}$ & $+2\cT_{[12]}$\\
$\cT_{[+2]}$ & $-2\cT_{[12]}$ & $-\cT_{[+-]}$\\
$\cT_{[+-]}$ & $-2\cT_{[1-]}$ & $-2\cT_{[2-]}$\\
$\cT_{[12]}$ & $+\cT_{[2-]}$ & $-\cT_{[1-]}$\\
$\cT_{[1-]}$ & $0$ & $0$\\
$\cT_{[2-]}$ & $0$ & $0$\\
\bottomrule\caption{\centering The actions of $\mathbb K_1,\mathbb K_2$  on the antisymmetric sector.}\label{tab:antiK}
\end{longtable}
For the trace part, we find 
\be 
\mathbb K_1\mathcal T=\mathbb K_2\mathcal T=0,\quad \mathbb J\mathcal T=0.
\ee Therefore, the trace part is closed under the three-dimensional Carrollian conformal algebra (CCA$_3$) generators and it has spin 0. Its Carrollian conformal weight is $\Delta$.

\begin{figure}
\centering
\resizebox{1.0\textwidth}{!}{%
\begin{tikzpicture}[
  >=Latex,
  every node/.style={font=\small,align=center},
  box/.style={draw,rounded corners,minimum width=27mm,minimum height=10mm,inner sep=3pt,fill=green!4},
  trbox/.style={draw,rounded corners,minimum width=29mm,minimum height=12mm,inner sep=3pt,fill=green!4},
  k1/.style={->,thick,blue},
  k2/.style={->,thick,red,dashed}
]
\node[box] (r1) at (0,0) {$R_1=\cT_{++}$\\$\bar\Delta=\Delta+2$};
\node[box] (r2) at (-4,-2.3) {$R_2=\cT_{(+1)}$\\$\bar\Delta=\Delta+1$};
\node[box] (r3) at (4,-2.3) {$R_3=\cT_{(+2)}$\\$\bar\Delta=\Delta+1$};
\node[trbox] (g) at (-8,-5.2) {$\mathcal T=\cT_{11}+\cT_{22}-\cT_{(+-)}$\\$\bar\Delta=\Delta$};
\node[box] (r9) at (-2.7,-5.2) {$R_9=\cT_{11}+\cT_{22}+\cT_{(+-)}$\\$\bar\Delta=\Delta$};
\node[box] (r7) at (2.7,-5.2) {$R_7=\cT_{11}-\cT_{22}$\\$\bar\Delta=\Delta$};
\node[box] (r8) at (8,-5.2) {$R_8=2\cT_{(12)}$\\$\bar\Delta=\Delta$};
\node[box] (r4) at (-4,-8.0) {$R_4=\cT_{(1-)}$\\$\bar\Delta=\Delta-1$};
\node[box] (r5) at (4,-8.0) {$R_5=\cT_{(2-)}$\\$\bar\Delta=\Delta-1$};
\node[box] (r6) at (0,-10.6) {$R_6=\cT_{--}$\\$\bar\Delta=\Delta-2$};

\draw[k1] (r1)--node[left] {$-4$}(r2);
\draw[k2] (r1)--node[right] {$-4$}(r3);
\draw[k1] (r2)--node[left] {$-1$}(r9);
\draw[k1] (r2)--node[above] {$-1$}(r7);
\draw[k2] (r2)--node[above] {$-1$}(r8);
\draw[k1] (r3)--node[above] {$-1$}(r8);
\draw[k2] (r3)--node[right] {$-1$}(r9);
\draw[k2] (r3)--node[above] {$+1$}(r7);
\draw[k1] (r9)--node[left] {$-4$}(r4);
\draw[k2] (r9)--node[right] {$-4$}(r5);
\draw[k1] (r7)--node[left=2pt] {$-2$}(r4);
\draw[k2] (r7)--node[right] {$+2$}(r5);
\draw[k1] (r8)--node[right=2pt] {$-2$}(r5);
\draw[k2] (r8)--node[left] {$-2$}(r4);
\draw[k1] (r4)--node[left] {$-1$}(r6);
\draw[k2] (r5)--node[right] {$-1$}(r6);

\node[
  draw,
  rounded corners,
  fill=white,
  align=left,
  anchor=north east,
  inner sep=4pt
] at (11.5,-9.2) {
\begin{tikzpicture}[>=Latex,baseline=(current bounding box.center)]
  \draw[->,thick,blue] (0,0.45) -- (1.3,0.45);
  \node[anchor=west] at (1.5,0.45) {$\mathbb K_1$};
  \draw[->,thick,red,dashed] (0,0.00) -- (1.3,0.00);
  \node[anchor=west] at (1.5,0.00) {$\mathbb K_2$};
\end{tikzpicture}
};

\end{tikzpicture}%
}
\caption{The trace term forms a singlet and the symmetric traceless sector forms a nine-dimensional Type Ib representation.}
\label{fig:sym-newK-net}
\end{figure}
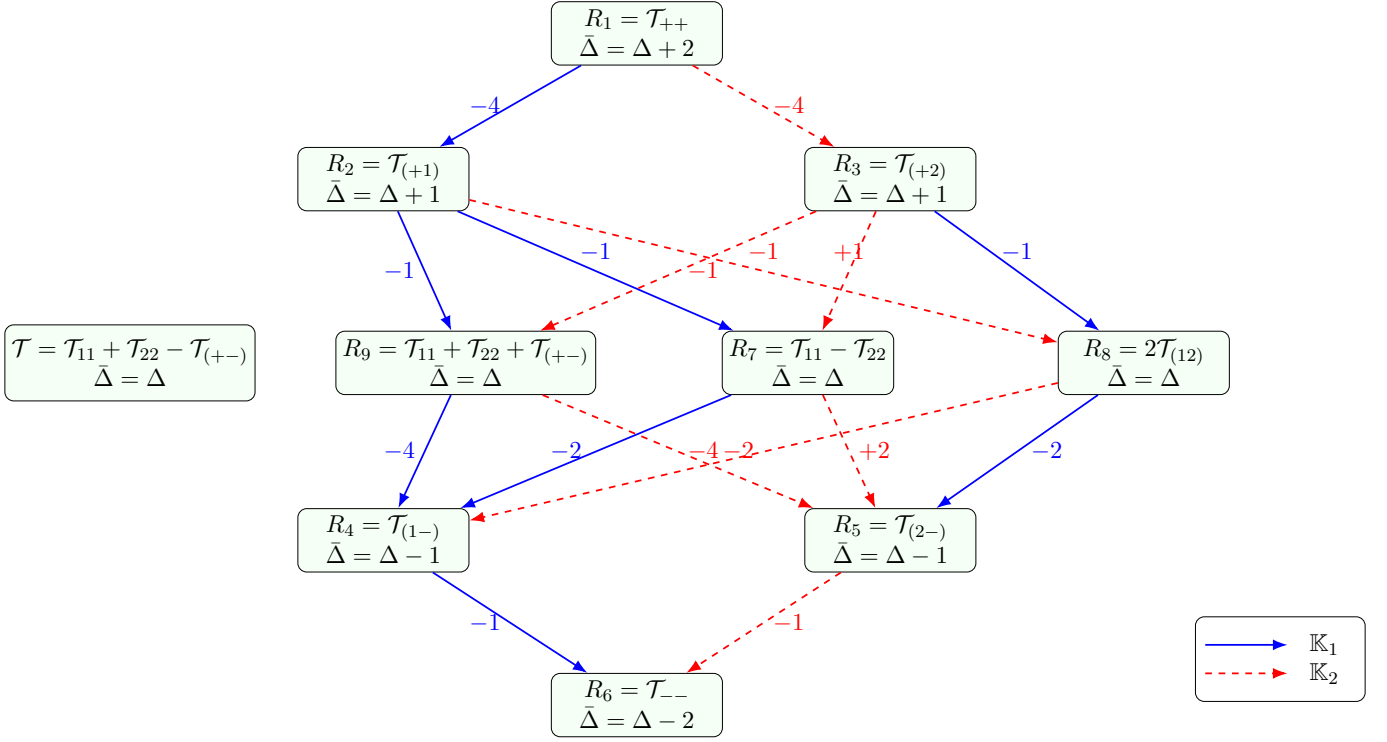

The nine symmetric traceless components are defined as 
\begin{align}
R_1&=\cT_{++},
&R_2&=\cT_{(+1)},
&R_3&=\cT_{(+2)},
&R_4&=\cT_{(1-)},
&R_5&=\cT_{(2-)},\notag\\
R_6&=\cT_{--},
&R_7&=\cT_{11}-\cT_{22},
&R_8&=2\cT_{(12)},
&R_9&=\cT_{11}+\cT_{22}+\cT_{(+-)}.
\label{eq:ST-basis-new}
\end{align}
The actions of $\mathbb K_I$ on $R_7$ and $R_9$ are 
\begin{align}
    \mathbb K_1 R_7&=-2R_4,\quad &\mathbb K_1 R_9&=-4R_4,\quad  
    &\mathbb K_2 R_7&=2R_5,\quad &\mathbb K_2 R_9&=-4R_5.
\end{align}
The action of the spin operator $\mathbb J$ on the basis is 
\begin{align}
\mathbb JR_1&=0,&
\mathbb JR_2&=R_3,&
\mathbb JR_3&=-R_2,&
\mathbb JR_4&=R_5,&
\mathbb JR_5&=-R_4,&\notag\\
\mathbb JR_6&=0,&
\mathbb JR_7&=2R_8,&
\mathbb JR_8&=-2R_7,&
\mathbb JR_9&=0.
\end{align} Thus $R_1,R_6,R_9$ are scalars under SO(2) while $R_2,R_3$ form a spin-1 doublet representation of SO(2). Similarly, $R_4,R_5$ form another spin-1 doublet and $R_7,R_8$ form a spin-2 doublet \footnote{For the gravitational field, $R_7$ and $R_8$ correspond to the $+$ mode and $\times$ mode, respectively.}. In summary, the symmetric traceless part forms a nine-dimensional representation.  
In Figure \ref{fig:sym-newK-net}, we show the 
singlet representation and the nine-dimensional representation. The blue line represents the action of $\mathbb K_1$ and the red line represents the action of $\mathbb K_2$. Note that $R_1=\mathcal T_{++}$ is the highest component of the symmetric
traceless sector. Acting with $\mathbb K_I$ on $R_1$ generates this sector, not
the whole rank-2 tensor product space.

As an illustration, we consider the stress tensor \eqref{stressfreescalar} in a massless free scalar theory. The boundary spin-2 operator is $\Sigma_{\mu\nu}=n_\mu n_\nu \frac{(2\dot{\Sigma}^2-\Sigma \ddot{\Sigma})}{3}$. At the south pole, the non-vanishing components are 
\be 
\Sigma_{00}=\Sigma_{03}=\Sigma_{30}=\Sigma_{33}= \frac{(2\dot{\Sigma}^2-\Sigma \ddot{\Sigma})}{3}.
\ee The unique non-vanishing symmetric traceless component among the $R_i,\ i=1,2,\cdots,9$ is 
\be 
R_1= \frac{4(2\dot{\Sigma}^2-\Sigma \ddot{\Sigma})}{3}.
\ee Its Carrollian conformal weight is $\bar\Delta=\Delta+2=4$ and the spin is zero.

For the antisymmetric sector, we can  find a six-dimensional representation. The operators $\mathcal T_{[+1]},\mathcal T_{[+2]}$ form a spin-1 doublet. Similarly, $\mathcal T_{[1-]} , \mathcal T_{[2-]}$ form another spin-1 doublet. On the other hand, both  $\mathcal T_{[+-]}$ and $\mathcal T_{[12]}$ are scalars. In this basis,  the antisymmetric sector instead has
highest components $\mathcal T_{[+1]}$ and $\mathcal T_{[+2]}$. The representation is shown in Figure \ref{fig:anti-newK-net}. We use the gauge curvature $f_{\mu\nu}=\partial_\mu a_\nu-\partial_\nu a_\mu$ to illustrate the representation. The fall-off index of $f_{\mu\nu}$ is 1 
\be 
f_{\mu\nu}=\frac{F_{\mu\nu}}{r}+\cdots{,}
\ee with 
\be 
F_{\mu\nu}=-n_\mu\dot A_\nu+n_\nu\dot A_\mu.
\ee At the south pole, the non-vanishing components are 
\bea 
F_{01}=\dot A_1,\quad F_{02}=\dot A_2,\quad F_{03}=\dot A_3-\dot A_0,\quad F_{13}=-\dot A_1,\quad F_{23}=-\dot A_2.
\eea Therefore, we find 
\bea 
\begin{gathered}
\mathcal T_{[+1]}=2\dot A_1,\quad \mathcal T_{[+2]}=2\dot A_2,\quad \mathcal T_{[+-]}=-2(\dot A_3-\dot A_0),\\
\mathcal T_{[1-]}=0,\quad \mathcal T_{[2-]}=0,\quad\mathcal T_{[12]}=0
\end{gathered}
\eea 
According to Figure \ref{fig:anti-newK-net}, the conformal weights of $\mathcal T_{[+1]}$  and $\mathcal T_{[+2]}$ are $\bar\Delta=\Delta+1=2$ and they form a spin-1 doublet. On the other hand, the conformal weight of $\mathcal T_{[+-]}$ is ${ \bar\Delta= }\Delta=1$ and it has spin 0. This is consistent with  \eqref{spin-one-weights}. From  equation  \eqref{spin-one-weights}, the conformal weight of $A_1$ and $A_2$ is 1.  Therefore, the conformal weight of $\dot A_1$ and $\dot A_2$ is 2. Similarly, the conformal weight of $B=A_0-A_3$ is 0 and thus the conformal weight of $\dot B=\dot A_0-\dot A_3$
is 1. Note that $B$ is independent of time in Maxwell theory. Equation \eqref{B-Maxwell-constraint} leads to $\dot B=0$ and thus $F_{03}=0$. We conclude that there are only two radiative degrees of freedom $\dot A_1$ and $\dot A_2$ at null infinity  in Maxwell theory. Note that the analysis does not exclude the possibility that $B$ could be dynamical for more general $U(1)$ gauge theories. A similar analysis for the gravitational field can be found in Appendix \ref{gaugeanalysis}. The conclusion is that there are six components in the gauge-invariant subsector. This subsector is shown in Figure \ref{fig:sym-newK-netg}.
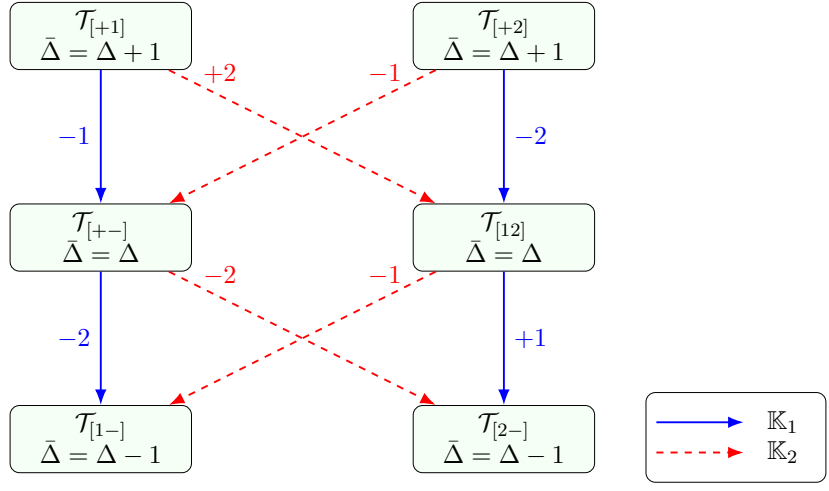
\begin{figure}[p]
\centering
\resizebox{0.6\textwidth}{!}{%
\begin{tikzpicture}[
  >=Latex,
  every node/.style={font=\small,align=center},
  box/.style={draw,rounded corners,minimum width=27mm,minimum height=10mm,inner sep=3pt,fill=green!4},
  k1/.style={->,thick,blue},
  k2/.style={->,thick,red,dashed}
]
\node[box] (a1) at (-3,0) {$\cT_{[+1]}$\\$\bar\Delta=\Delta+1$};
\node[box] (a2) at (3,0) {$\cT_{[+2]}$\\$\bar\Delta=\Delta+1$};
\node[box] (a3) at (-3,-3) {$\cT_{[+-]}$\\$\bar\Delta=\Delta$};
\node[box] (a4) at (3,-3) {$\cT_{[12]}$\\$\bar\Delta=\Delta$};
\node[box] (a5) at (-3,-6) {$\cT_{[1-]}$\\$\bar\Delta=\Delta-1$};
\node[box] (a6) at (3,-6) {$\cT_{[2-]}$\\$\bar\Delta=\Delta-1$};
\draw[k1] (a1)--node[left] {$-1$}(a3);
\draw[k2] (a2)--node[pos=0.30,above right= 7pt and 1pt] {$-1$}(a3);
\draw[k1] (a2)--node[right] {$-2$}(a4);
\draw[k2] (a1)--node[pos=0.30,above left= 7pt and 1pt] {$+2$}(a4);
\draw[k1] (a3)--node[left] {$-2$}(a5);
\draw[k2] (a3)--node[pos=0.30,above left= 7pt and 1pt] {$-2$}(a6);
\draw[k1] (a4)--node[right] {$+1$}(a6);
\draw[k2] (a4)--node[pos=0.30,above right= 7pt and 1pt] {$-1$}(a5);

\node[
  draw,
  rounded corners,
  fill=white,
  align=left,
  anchor=north east,
  inner sep=4pt
] at (7.8,-5.3) {
\begin{tikzpicture}[>=Latex,baseline=(current bounding box.center)]
  \draw[k1] (2,0.45) -- (3.3,0.45);
  \node[anchor=west] at (3.5,0.45) {$\mathbb K_1$};
  \draw[k2] (2,0.00) -- (3.3,0.00);
  \node[anchor=west] at (3.5,0.00) {$\mathbb K_2$};
\end{tikzpicture}
};

\end{tikzpicture}%
}
\caption{\centering The antisymmetric sector forms a six-dimensional Type Ib representation.}
\label{fig:anti-newK-net}
\end{figure}
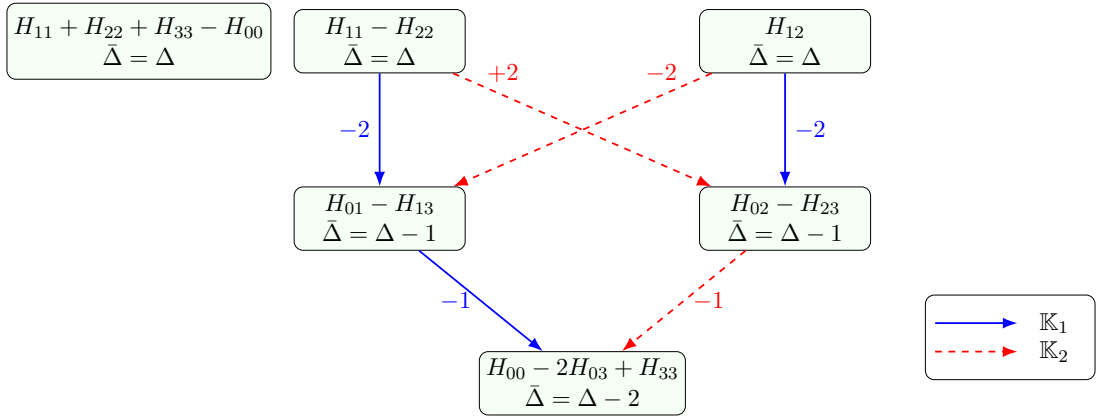
\begin{figure}[p]
\centering
\resizebox{0.8\textwidth}{!}{%
\begin{tikzpicture}[
  >=Latex,
  every node/.style={font=\small,align=center},
  box/.style={draw,rounded corners,minimum width=27mm,minimum height=10mm,inner sep=3pt,fill=green!4},
  trbox/.style={draw,rounded corners,minimum width=29mm,minimum height=12mm,inner sep=3pt,fill=green!4},
  k1/.style={->,thick,blue},
  k2/.style={->,thick,red,dashed}
]
\node[trbox] (g) at (-7,-5.2) {$H_{11}+H_{22}+H_{33}-H_{00}$\\$\bar\Delta=\Delta$};
\node[box] (r7) at (-3.2,-5.2) {$H_{11}-H_{22}$\\$\bar\Delta=\Delta$};
\node[box] (r8) at (3.2,-5.2) {$H_{12}$\\$\bar\Delta=\Delta$};
\node[box] (r4) at (-3.2,-8.0) {$H_{01}-H_{13}$\\$\bar\Delta=\Delta-1$};
\node[box] (r5) at (3.2,-8.0) {$H_{02}-H_{23}$\\$\bar\Delta=\Delta-1$};
\node[box] (r6) at (0,-10.6) {$H_{00}-2H_{03}+H_{33}$\\$\bar\Delta=\Delta-2$};

\draw[k1] (r7)--node[left] {$-2$}(r4);
\draw[k2] (r7)--node[pos=0.30,above left= 7pt and 1pt] {$+2$}(r5);
\draw[k1] (r8)--node[right] {$-1$}(r5);
\draw[k2] (r8)--node[pos=0.30,above right= 7pt and 1pt] {$ -1$}(r4);
\draw[k1] (r4)--node[left] {$-1$}(r6);
\draw[k2] (r5)--node[right] {$-1$}(r6);

\node[
  draw,
  rounded corners,
  fill=white,
  align=left,
  anchor=north east,
  inner sep=4pt
] at (8.1,-9.2) {
\begin{tikzpicture}[>=Latex,baseline=(current bounding box.center)]
  \draw[->,thick,blue] (0,0.45) -- (1.3,0.45);
  \node[anchor=west] at (1.5,0.45) {$\mathbb K_1$};
  \draw[->,thick,red,dashed] (0,0.00) -- (1.3,0.00);
  \node[anchor=west] at (1.5,0.00) {$\mathbb K_2$};
\end{tikzpicture}
};

\end{tikzpicture}%
}
\caption{For a spin-2 gravitational field, the spin-2 multiplet can be projected onto the gauge-invariant subsector by fixing $R_1=R_2=R_3=R_9=0$.}
\label{fig:sym-newK-netg}
\end{figure}

Now we will comment on the spin-$n$ multiplet representation at the end of this section. 
The operator $\Sigma_{\mu_1\mu_2\cdots\mu_n}$ is decomposed into the components
\be 
\mathcal T_{\alpha_1\alpha_2\cdots\alpha_n}=e_{\alpha_1}^{\mu_1}e_{\alpha_2}^{\mu_2}\cdots e_{\alpha_n}^{\mu_n}\Sigma_{\mu_1\mu_2\cdots\mu_n},\quad \alpha_1,\alpha_2,\cdots,\alpha_n=+,1,2,-.
\ee The operator in the highest layer (level $n$) is $\mathcal{T}_{\underbrace{++\cdots+}_{n\text{ times}}}$ whose Carrollian conformal weight is $\Delta+n$. In the symmetric and  traceless sector, the operator
\be 
\mathbb K_1^{l_1}\mathbb K_2^{l_2}\mathcal{T}_{\underbrace{++\cdots+}_{n\text{ times}}}
\ee 
has conformal weight $\Delta+n-l_1-l_2$ and is located at level $n-l_1-l_2$. Due to the chain structure \eqref{chains}, there is a unique operator $\mathcal{T}_{\underbrace{--\cdots-}_{n\text{ times}}}$  in the lowest layer whose level is $-n$. In general, the integers $l_1$ and $l_2$ are constrained by 
\be 
0\le l_1+l_2\le 2n.
\ee 
The other sectors can be obtained in a similar way. In total, 
at level $s-l_1-l_2$, the number of independent components is 
\be 
\left(\begin{array}{c}2n\\ l_1+l_2\end{array}\right).
\ee Then the total number of independent components is equal to the number of components of $\Sigma_{\mu_1\mu_2\cdots\mu_n}$
\be 
\sum_{0\le l\le 2n}\left(\begin{array}{c}2n\\ l\end{array}\right)=2^{2n}=4^n.
\ee

\section{Conclusion and discussion}\label{dis}
In this work, we derived the spinning bulk-to-boundary correlators by solving Ward identities. The structures of the bulk-to-boundary correlators are classified by double-line diagrams using the two-spinor formalism. The total number of independent tensor structures in the solution is given by the Catalan number. Based on this result, we obtain the boundary-to-boundary correlators via extrapolation. For general spinning operators, we discussed the relation between the KL representation and the bulk-to-boundary correlators. We still find a critical fall-off index $\Delta=1$ for general spin-$s$ operators. For a bulk spin-$s$ operator $t_{\mu_1\mu_2\cdots\mu_s}$, the corresponding boundary operator $\Sigma_{\mu_1\mu_2\cdots\mu_s}$ lives in the Type Ib spin-$s$ multiplet representation. As far as we know, these representations have not been discussed in the literature. There are various issues that deserve further study. 

\begin{itemize}
    \item \textbf{Three-point functions.} In this paper, we only considered the bulk-to-boundary correlators. It is known that the boundary three-point correlators are also constrained by Ward identities. Therefore, one would like to explore the three-point correlators that mix bulk and boundary operators. A typical correlator is shown in Figure \ref{fig:example}, where one operator is inserted in the bulk and the other two are placed at null infinity. By extrapolating this correlator to the boundary, we expect to find more interesting results in the future.\footnote{Work in progress.}
    \begin{figure}\centering
        \begin{tikzpicture}[x=1.55cm,y=1.55cm,line cap=round,line join=round]
    \coordinate (ip) at (0,4.5);
    \coordinate (im) at (0,0);
    \coordinate (i0) at (2.25,2.25);
    
    \coordinate (S1J) at (1.35,3.15);
    \coordinate (S2J)  at (1.44,1.44);
    \coordinate (J)  at (1.11,2.58);
    \coordinate (JJ)  at (0.68,2.58);

    \draw[line width=0.45pt] (im) -- (ip) -- (i0) -- (im);

    \draw[line width=0.45pt] (J) -- (S1J);
      \draw[line width=0.45pt] (J) -- (S2J);
      \draw[line width=0.45pt] (J) -- (JJ);

    \node[anchor=south] at (ip) {$i^{+}$};
    \node[anchor=north] at (im) {$i^{-}$};
    \node[anchor=west]  at (i0) {$i^{0}$};

    \node at (1.18,3.76) {$\mathscr{I}^{+}$};
    \node at (1.18,0.83) {$\mathscr{I}^{-}$};
\node at  (JJ) {$\bullet$};\node at  (S1J) {$\bullet$};\node at  (S2J) {$\bullet$};

    \node[anchor=south west] at (S1J) {$\Sigma(u,\Omega_1)$};
    \node[anchor=north west] at (S2J) {$\Sigma(v,\Omega_2)$};
    \node[anchor=east] at (JJ) {$\Phi(x)$};
\end{tikzpicture}\caption{\centering Bulk-boundary-boundary correlator.}\label{fig:example}
    \end{figure}
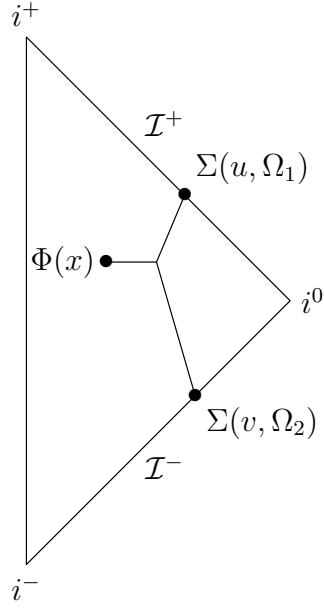
    \item \textbf{Type Ib spin-$s$ multiplet representation.} {When the bulk field is extrapolated to the boundary}, the corresponding boundary operator contains several components. In an appropriate basis, the components can be labeled by the quantum numbers $(\bar\Delta,s)$. Remarkably, they form a Type Ib spin-$s$ multiplet representation that contains multiple components. For spin-1 and spin-2, the representations are shown in Figures~\ref{fig:K-real-net}, \ref{fig:sym-newK-net}, and \ref{fig:anti-newK-net}. It would be nice to explore the properties of the operators in a general spin-$s$ multiplet representation of CCFTs.
    \item \textbf{IR divergences.} In this work, we find the IR divergences for $\Delta>1$ and show that the appearance of divergences when extrapolating bulk correlators to null infinity is not peculiar to \cite{Long:2026cpq}. Note that all the IR divergences are in the form of

    \begin{equation}
\begin{aligned}
B_{\mathrm{bare}}^{(N)}(u,\Omega;v',\Omega')
&=\lim_{r'\to\infty}
\frac{r'^\Delta}
{\left[u-v'+r'\bigl(1+\cos\gamma(\Omega,\Omega')\bigr)-i\epsilon\right]^{\Delta+q_N}},
\\&\qquad \Delta>1,\quad q_N\geq 0.
\end{aligned}
\end{equation}
    
   We have stripped off the irrelevant normalization constant and tensor structures. We set $q_N=0$ to simplify {the} discussion below.
In the study of massless bulk correlators extrapolated to $\mathscr{I}$, 
ref.~\cite{Nguyen:2023miw} pointed out that time-ordered correlators produce 
anomalous $\ln r$ contact terms. In that work, considering $u$-descendant fields 
such as $\partial_u\Sigma$ yields regular Carrollian correlators. However, 
this method is not applicable here, as one can check, the divergent 
$r'^{\Delta-1}$ term persists regardless of the number of derivatives. In ref.~\cite{Liu:2023qtr}, we also introduced a cutoff for the IR modes of the boundary two-point function within the Carrollian Feynman rules, and used analytic continuation to handle divergent integrals in generalized Carrollian amplitudes. Unfortunately, we are not aware of any way to apply the same analytic continuation to the extrapolation limit. In ref.~\cite{Alday:2024yyj}, the AdS to flat/Carrollian limit should also 
be accompanied by a redefinition of the electric Carrollian operator. 
Indeed, by rescaling the boundary operator as 
$\Sigma(v',\Omega')\to \widetilde{\Sigma}(v',\Omega') = r'^{\,1-\Delta}\,\Sigma(v',\Omega')$, 
the electric part of the boundary-to-boundary correlator becomes finite:
\begin{equation}
\langle \Sigma(u,\Omega) \, \widetilde{\Sigma}(v',\Omega') \rangle 
\propto \frac{1}{(u-v')^{\Delta-1}} \, \delta^{(2)}(\Omega-\Omega'^{\text{P}}) .
\end{equation}
However, the reason for renormalizing the operator in this way is not clear to us. 

 Note that the divergence originates from the combination of the large-$r'$ null‑boundary limit and an antipodal angular pinch. Therefore, the problem reduces to handling the large-$r'$ expansion and refining the antipodal contact sector.

 One constructive line is to define the extrapolation limit first as a smeared distribution. 
That is, instead of starting from point insertions, one introduces wavepackets such as
\begin{equation}
B_{r'} \sim r'^{\Delta} \int_{0}^{\infty} d\omega \, \omega^{\Delta-1} \, W_{r'}(\omega) \, e^{-i\omega(u-v' + r'(1+\cos\gamma) - i\epsilon)} .
\end{equation}
This is in the same spirit as LSZ \cite{Lehmann:1957zz}, where asymptotic states or boundary insertions can first be defined with test functions, and the corresponding limits can then be {discussed~}\cite{PhysRev.112.669,Nishijima:1957}{.}  

A second method is to systematically organize the singularity at $r' = \infty$ 
as a distributional extension problem. 
Since $r'^{\Delta}$ cannot be treated as a  locally integrable function 
across the antipodal point when $\Delta > 1$, one can use methods such as the 
Hadamard finite part~\cite{2000JMP....41.7675B}. 
This method naturally yields contact ambiguities:
\begin{equation}
\delta^{(2)}(\Omega-\Omega'^{\text{P}}), \quad \Delta_{S^2} \delta^{(2)}(\Omega-\Omega'^{\text{P}}), \quad\cdots .
\end{equation}
As an example, for an   integer $\Delta\ge 2$,
\begin{align}
B(u,\Omega;v',\Omega') 
&= \text{FP}\left[ \frac{r'^\Delta}{(u-v'+r'(1+\cos\gamma(\Omega,\Omega'))-i\epsilon)^\Delta} \right] 
+ \cdots \nonumber\\
&\quad + a_{\text{log}} \ln(u-v'-i \epsilon) 
\left( \Delta_{S^2} \right)^{\Delta-1} \delta^{(2)}(\Omega-\Omega'^{\text{P}}),
\end{align}
where $\cdots$ contains contact ambiguities that are divergent. 
The logarithmic term is invariant under the scaling transformation $r' \to \lambda r'$, 
which may remind us of the importance of the coefficient $a_{\text{log}}$. 
The naive subtraction is to remove those terms in $\cdots$. 
Unfortunately, the subtracted result does not respect the Carrollian Ward identities 
at the boundary.

A third approach is to use Hamilton-Jacobi renormalization. 
The basic form is given by
\begin{equation}
W_{\text{ren}}[J] = \lim_{r'\to\infty} \bigl( W_{\text{bare}}[J] + S_{\text{ct},r'}[J] \bigr),
\end{equation}
where the electric divergence can be absorbed into a source-sector counterterm such as
\begin{equation}
S_{\text{ct},r'} \sim -\frac{1}{2} \int J \, \alpha_{r'}^{\text{div}} \, \delta^{(2)}(\Omega-\Omega'^{\text{P}}) \, J .
\end{equation}
This logic is consistent with holographic renormalization, where one defines the source functional on a cutoff surface and then subtracts local or quasi-local divergences \cite{Bianchi:2001kw,Skenderis:2002wp,Papadimitriou:2010as}. 
Recent work on scalar Hamilton-Jacobi renormalization in flat holography also provides technical background close to the present problem \cite{Ammon:2025avo}. However, it seems that the renormalization result may depend on the theory, which is not completely fixed by Poincar\'e symmetry.

A fourth line of reasoning starts from the definition of the boundary operators themselves. The leading Ward identities from \cite{Long:2026cpq} fix the leading bulk‑to‑boundary correlator. This is a very strong condition, but it also means that subleading data are not fixed at that level. In the region where $\gamma\to\pi$,
the full asymptotic expansion
\begin{equation}
\Phi \sim r^{-\Delta} \Sigma + r^{-\Delta-1} \Sigma_1 + \dots 
\end{equation}
may contain subleading terms that appear at the same distributional order as the leading electric contact sector. Therefore, if one wishes to define the renormalised electric contact term from the bulk theory itself, subleading Ward identities, operator mixing, and spectral data of the bulk two‑point function could become relevant inputs. The Källén‑Lehmann representation illustrates the connection between fall‑off behavior and long‑distance properties of gapless sectors. This provides a natural entry point for linking the $\Delta>1$ problem to bulk spectral input. 

The IR problems are often not merely about the divergence of an integral, but rather about the definition of observables, external states, boundary conditions, or operators \cite{PhysRev.52.54,Yennie:1961ad,Kulish:1970ut}.
We leave this problem of Carrollian correlators for future work.

   \end{itemize}

\acknowledgments 

The work of J.L. is supported by NSFC Grant No. 12575074. We also acknowledge the High Energy Physics Base and its AI Club at the School of Physics, Huazhong University of Science and Technology for providing a collaborative discussion platform and computing/research resources.

\appendix

\section{Asymptotic expansion of the bulk-to-bulk correlator} \label{nextL}
{%

}

{In this appendix, we use the Ward identities to fix the subleading bulk-to-boundary correlators. We still expand the bulk-to-bulk correlator asymptotically near $\mathscr I^+$
\be
G_{\mu_1\mu_2\cdots\mu_n}(x;x')
=\sum_{\Delta}\sum_{k=0}^\infty
\frac{D^{(\Delta),(k)}_{\mu_1\mu_2\cdots\mu_n}(u,\Omega;x')}{r^{\Delta+k}}.
\label{expansionGd}
\ee
The power indices are real and the summation is not necessarily a unit-spaced power expansion. We have collected the terms whose power indices differ by an integer into the same $\Delta$-series, so any two distinct $\Delta_1,\ \Delta_2$ in the summation obey $\Delta_1-\Delta_2\not\in\mathbb Z$. More general expansions containing logarithmic terms are possible. However, we will ignore these cases in this work.

The differential operators of the Poincar\'e generators do not mix different $\Delta$-series in the Ward identities. We will omit the superscript $\Delta$ in the following discussion and define
\be
a_\mu=u\bar m_\mu-x'_\mu.
\ee

For $k\ge 1$, we can find the $\mathrm N^k\mathrm{LO}$ Ward identities for translation by reading off the coefficient of $r^{-\Delta-k}$ in the bulk Ward identities, yielding
\begin{equation}
\big(\partial_\rho^{x'}-n_\rho\partial_u\big)
D^{(k)}_{\mu_1\mu_2\cdots\mu_n}
=Y_\rho^I\partial_I D^{(k-1)}_{\mu_1\mu_2\cdots\mu_n}
+(\Delta+k-1)m_\rho D^{(k-1)}_{\mu_1\mu_2\cdots\mu_n}.
\label{eq:general-Delta-order-k-translation-Ward}
\end{equation}
Ward identities for Lorentz transformations can be simplified using Ward identities for translation,
\begin{align}
&\left[
-\widehat u n_{\alpha\beta}\partial_u
-(\Delta+k)n_{\alpha\beta}
+Y_{\alpha\beta}^I\delta_I
+\mathcal S_{\alpha\beta}^{(n)}
\right]D^{(k)}_{\mu_1\mu_2\cdots\mu_n}
\nonumber\\
&\qquad=(a_\alpha Y_\beta^I-a_\beta Y_\alpha^I)
\partial_I D^{(k-1)}_{\mu_1\mu_2\cdots\mu_n}
+(\Delta+k-1)(a_\alpha m_\beta-a_\beta m_\alpha)
D^{(k-1)}_{\mu_1\mu_2\cdots\mu_n}.
\label{eq:general-Delta-order-k-reduced-Lorentz-Ward}
\end{align}
The LO Ward identities are obtained by setting $k=0$ and dropping the right-hand sides. The NLO and NNLO Ward identities are the cases $k=1$ and $k=2$, respectively.
Here the spin generator $\mathcal S_{\alpha\beta}^{(n)}$ acting on all the $n$ covariant Lorentz indices is
\begin{equation}
(\mathcal S_{\alpha\beta}^{(n)}D)_{\mu_1\cdots\mu_n}
=\sum_{j=1}^{n}\bigg[
\eta_{\mu_j\alpha}
D_{\mu_1\cdots\mu_{j-1}\,\beta\,\mu_{j+1}\cdots\mu_n}
-\eta_{\mu_j\beta}
D_{\mu_1\cdots\mu_{j-1}\,\alpha\,\mu_{j+1}\cdots\mu_n}
\bigg].
\end{equation}
Thus, the source of the $\mathrm N^k\mathrm{LO}$ Ward identities is determined by the complete result at order $\mathrm N^{k-1}\mathrm{LO}$. At leading order, the bulk-to-boundary correlator has been discussed in the text. In general, it has the form
\be
D^{(0)}_{\mu_1\mu_2\cdots\mu_n}
=\sum_{N=1}^{\#_n}C_N^{(n/2)}
\frac{\mathbb T^{(N)}_{\mu_1\mu_2\cdots\mu_n}}
{\widehat u^{\Delta+q_N}},
\label{eq:general-Delta-LO-solution}
\ee
where $\mathbb T^{(N)}$ is constructed from $n$, $\eta$ and $\epsilon$, and $q_N$ is the number of $n$'s in $\mathbb T^{(N)}$. We borrow the solutions for spin 0, 1 and 2 in the following,
\bs\label{leadingD}
\begin{align}
D(u,\Omega;x')={}&\frac{C^{(0)}}{\widehat u^\Delta},\\
D_{\mu\nu}(u,\Omega;x')={}&
\frac{C^{(1)}_1\eta_{\mu\nu}}{\widehat u^\Delta}
+\frac{C^{(1)}_2n_\mu n_\nu}{\widehat u^{\Delta+2}},\\
D_{\mu\nu\rho\sigma}(u,\Omega;x')={}&
\frac{1}{\widehat u^\Delta}
\Big(C_1^{(2)}\eta_{\mu\nu}\eta_{\rho\sigma}
+C_2^{(2)}\eta_{\mu\rho}\eta_{\nu\sigma}
+C_3^{(2)}\eta_{\mu\sigma}\eta_{\nu\rho}
+C_{14}^{(2)}\epsilon_{\mu\nu\rho\sigma}\Big)
\nonumber\\
&+\frac{1}{\widehat u^{\Delta+2}}
\Big(C_4^{(2)}\eta_{\mu\nu}n_\rho n_\sigma
+C_5^{(2)}\eta_{\mu\rho}n_\nu n_\sigma
+C_6^{(2)}\eta_{\mu\sigma}n_\nu n_\rho\\
&\qquad+C_7^{(2)}\eta_{\nu\rho}n_\mu n_\sigma
+C_8^{(2)}\eta_{\nu\sigma}n_\mu n_\rho
+C_9^{(2)}\eta_{\rho\sigma}n_\mu n_\nu
\nonumber\\
&\qquad+C_{11}^{(2)}\epsilon_{\mu\nu\rho\lambda}n^\lambda n_\sigma
+C_{12}^{(2)}\epsilon_{\mu\nu\sigma\lambda}n^\lambda n_\rho
+C_{13}^{(2)}\epsilon_{\mu\rho\sigma\lambda}n^\lambda n_\nu\Big)\nonumber\\
&+\frac{C_{10}^{(2)}}{\widehat u^{\Delta+4}}
n_\mu n_\nu n_\rho n_\sigma.\nn
\end{align}
\es

Consider the Poincar\'e-covariant structure in the bulk
\begin{equation}
\mathcal G^{(k,N)}(x;x')
=C^{(n/2)}_{k,N}
\frac{\mathbb T_{\mu_1\cdots\mu_n}^{(N)}(\delta x,\eta,\epsilon)}
{\left(-\frac{1}{2}\delta x^2\right)^{\Delta+k+q_N}}
\equiv\sum_{k'=0}^\infty
\frac{D^{(k';k,N)}_{\mu_1\mu_2\cdots\mu_n}}{r^{\Delta+k+k'}}.
\label{eq:general-Delta-order-s-Cartesian-lift}
\end{equation}
This structure satisfies the complete bulk Ward identities. The tensor $\mathbb T^{(N)}(\delta x,\eta,\epsilon)$ has the same form as $\mathbb T^{(N)}(n,\eta,\epsilon)$, with each $n_\mu$ replaced by $\delta x_\mu$. Using
\be
\delta x_\mu=rn_\mu+a_\mu,\qquad
-\frac{1}{2}\delta x^2=r\widehat u-\frac{a^2}{2},
\ee
we can write
\be
\mathcal G^{(k,N)}(x;x')
=\frac{C^{(n/2)}_{k,N}}{r^{\Delta+k}}
\frac{\mathbb T^{(N)}_{\mu_1\cdots\mu_n}(n+a/r,\eta,\epsilon)}
{\left(\widehat u-\frac{a^2}{2r}\right)^{\Delta+k+q_N}}.
\ee
Its coefficients can therefore be obtained by expanding the numerator and denominator in $1/r$. In particular, its leading coefficient
\begin{equation}
D^{(0;k,N)}_{\mu_1\mu_2\cdots\mu_n}
=C^{(n/2)}_{k,N}
\frac{\mathbb T^{(N)}_{\mu_1\mu_2\cdots\mu_n}}
{\widehat u^{\Delta+k+q_N}}
\label{eq:general-Delta-order-s-leading}
\end{equation}
satisfies the LO bulk-to-boundary Ward identities with effective fall-off index $\Delta+k$, hence is part of the homogeneous solution for the $\mathrm N^k\mathrm{LO}$ Ward identities. More generally, for $k'\ge1$, the coefficient $D^{(k';k,N)}$ satisfies the $\mathrm N^{k+k'}\mathrm{LO}$ Ward identities sourced by $D^{(k'-1;k,N)}$.

For $k=1$, this starts from the LO solution $D^{(0)}=\sum_N D^{(0;0,N)}$. Suppose that the complete result at order $k-1$ is
\be
D^{(k-1)}_{\mu_1\cdots\mu_n}
=\sum_{k'=0}^{k-1}\sum_{N=1}^{\#_n}
D^{(k-k'-1;k',N)}_{\mu_1\cdots\mu_n}.
\label{eq:general-Delta-order-k-minus-one}
\ee
Since the Ward identities are linear, a particular solution sourced by this complete result is
\be
D^{(k),p}_{\mu_1\cdots\mu_n}
=\sum_{k'=0}^{k-1}\sum_{N=1}^{\#_n}
D^{(k-k';k',N)}_{\mu_1\cdots\mu_n}.
\label{eq:general-Delta-order-k-complete-particular}
\ee
The difference between the complete solution and this particular solution satisfies the homogeneous Ward identities with effective fall-off index $\Delta+k$,
\be
D^{(k),\mathrm{hom}}_{\mu_1\cdots\mu_n}
=\sum_{N=1}^{\#_n}C^{(n/2)}_{k,N}
\frac{\mathbb T^{(N)}_{\mu_1\cdots\mu_n}}
{\widehat u^{\Delta+k+q_N}}
=\sum_{N=1}^{\#_n}D^{(0;k,N)}_{\mu_1\cdots\mu_n}.
\label{eq:general-Delta-order-k-homogeneous}
\ee
Hence the complete result of the $\mathrm N^k\mathrm{LO}$ Ward identities is
\begin{align}
D^{(k)}_{\mu_1\cdots\mu_n}
={}&D^{(k),p}_{\mu_1\cdots\mu_n}
+D^{(k),\mathrm{hom}}_{\mu_1\cdots\mu_n}=\sum_{k'=0}^{k}\sum_{N=1}^{\#_n}
D^{(k-k';k',N)}_{\mu_1\cdots\mu_n}.
\label{eq:general-Delta-order-k-complete}
\end{align}
The constants $C^{(n/2)}_{k,N}$ in the new homogeneous solution are independent of those at lower orders. Note that
\bea
C_{0,N}^{(n/2)}=C_N^{(n/2)},\quad
C_{1,N}^{(n/2)}=\widetilde C_N^{(n/2)},\quad
C_{2,N}^{(n/2)}=\widehat C_N^{(n/2)}.
\eea

The NLO solution is given by a particular and a homogeneous solution,
\be
D^{(1)}_{\mu_1\cdots\mu_n}
=D^{(1),p}_{\mu_1\cdots\mu_n}
+D^{(1),\mathrm{hom}}_{\mu_1\cdots\mu_n}.
\label{eq:general-Delta-D1-complete}
\ee
The particular solution $D^{(1),p}$ is sourced by the LO bulk-to-boundary correlator $D^{(0)}$. It is the sum of the $r^{-\Delta-1}$ coefficients of $\mathcal G^{(0,N)}$,
\begin{align}
D^{(1),p}_{\mu_1\mu_2\cdots\mu_n}
={}&\sum_{N=1}^{\#_n}
\frac{C^{(n/2)}_N}{\widehat u^{\Delta+q_N}}
\mathbb Q^{(N)}_{\mu_1\mu_2\cdots\mu_n}+\sum_{N=1}^{\#_n}
\frac{C^{(n/2)}_N(\Delta+q_N)a^2}{2\widehat u^{\Delta+q_N+1}}
\mathbb T^{(N)}_{\mu_1\mu_2\cdots\mu_n}.
\label{eq:general-Delta-D1-particular}
\end{align}
The tensor structure $\mathbb Q^{(N)}$ is given by
\be
\mathbb Q_{\mu_1\mu_2\cdots\mu_n}^{(N)}=a_\alpha\frac{\partial\mathbb T^{(N)}_{\mu_1\mu_2\cdots\mu_n}}
{\partial n_\alpha}\Big|_{\eta,\epsilon}.
\ee
The derivatives with respect to $n$ act on the tensor polynomials before imposing $n^2=0$.
For example,
\bs\begin{align}
\mathbb T^{(1)}_{\mu\nu}=\eta_{\mu\nu}
\quad&\Rightarrow\quad\mathbb Q^{(1)}_{\mu\nu}=0,\\
\mathbb T^{(2)}_{\mu\nu}=n_\mu n_\nu
\quad&\Rightarrow\quad\mathbb Q^{(2)}_{\mu\nu}=a_\mu n_\nu+a_\nu n_\mu.
\end{align}\es
The homogeneous solution $D^{(1),\mathrm{hom}}$ has the same form as $D^{(0)}$ but with the effective fall-off index shifted upward by $1$,
\begin{equation}
D^{(1),\mathrm{hom}}_{\mu_1\mu_2\cdots\mu_n}
=\sum_{N=1}^{\#_n}\widetilde C^{(n/2)}_N
\frac{\mathbb T^{(N)}_{\mu_1\mu_2\cdots\mu_n}}
{\widehat u^{\Delta+1+q_N}}.
\label{eq:general-Delta-D1-homogeneous}
\end{equation}
For spin 0 and 1, the particular solutions read
\bs\label{particularD}
\begin{align}
D^{(1),p}={}&\frac{\Delta C^{(0)}a^2}{2\widehat u^{\Delta+1}},\\
D_{\mu\nu}^{(1),p}={}&
\frac{\Delta C_1^{(1)}a^2}{2\widehat u^{\Delta+1}}\eta_{\mu\nu}
+C_2^{(1)}\frac{a_\mu n_\nu+n_\mu a_\nu}{\widehat u^{\Delta+2}}+\frac{(\Delta+2)C_2^{(1)}a^2}{2\widehat u^{\Delta+3}}n_\mu n_\nu.
\end{align}
\es
The spin-2 particular solution follows from \eqref{eq:general-Delta-D1-particular} using the tensor structures in \eqref{leadingD}. For example,
\bs
\begin{align}
\mathbb Q^{(4)}_{\mu\nu\rho\sigma}
={}&\eta_{\mu\nu}(a_\rho n_\sigma+n_\rho a_\sigma),\\
\mathbb Q^{(10)}_{\mu\nu\rho\sigma}
={}&a_\mu n_\nu n_\rho n_\sigma+n_\mu a_\nu n_\rho n_\sigma+n_\mu n_\nu a_\rho n_\sigma+n_\mu n_\nu n_\rho a_\sigma,\\
\mathbb Q^{(11)}_{\mu\nu\rho\sigma}
={}&\epsilon_{\mu\nu\rho\lambda}(a^\lambda n_\sigma+n^\lambda a_\sigma).
\end{align}
\es

Since the NNLO Ward identities are linear, their source can be decomposed according to the two terms in the complete NLO solution. The complete NNLO solution therefore takes the form
\be
D^{(2)}_{\mu_1\cdots\mu_n}
=D^{(2),pp}_{\mu_1\cdots\mu_n}
+D^{(2),p}_{\mu_1\cdots\mu_n}
+D^{(2),\mathrm{hom}}_{\mu_1\cdots\mu_n}.
\ee
Here the iterated particular solution $D^{(2),pp}$ is sourced by $D^{(1),p}$, while $D^{(2),p}$ is sourced by $D^{(1),\mathrm{hom}}$. They are the sum of $D^{(2;0,N)}$ and $D^{(1;1,N)}$, respectively. The particular solutions $D^{(2),p}$ are given by \eqref{eq:general-Delta-D1-particular} with $C_N^{(n/2)}\to\widetilde C_N^{(n/2)}$ and $\Delta\to\Delta+1$. The new homogeneous solutions at NNLO are obtained by replacing $C_N^{(n/2)}\to\widehat C_N^{(n/2)}$ and $\Delta\to\Delta+2$ in \eqref{eq:general-Delta-LO-solution},
\be
D^{(2),\mathrm{hom}}_{\mu_1\cdots\mu_n}
=\sum_{N=1}^{\#_n}\widehat C_N^{(n/2)}
\frac{\mathbb T^{(N)}_{\mu_1\cdots\mu_n}}
{\widehat u^{\Delta+2+q_N}}.
\ee
Finally, expanding $\mathcal G^{(0,N)}$ to order $r^{-\Delta-2}$ gives the iterated particular solutions,
\begin{align}
D^{(2),pp}_{\mu_1\cdots\mu_n}
={}&\sum_{N=1}^{\#_n}
\frac{C_N^{(n/2)}}{\widehat u^{\Delta+q_N}}
\Bigg[
\frac{a_\alpha a_\beta}{2}
\frac{\partial^2\mathbb T^{(N)}_{\mu_1\cdots\mu_n}}
{\partial n_\alpha\partial n_\beta}\Big|_{\eta,\epsilon}\nonumber\\
&\qquad+\frac{(\Delta+q_N)a^2}{2\widehat u}
\mathbb Q^{(N)}_{\mu_1\cdots\mu_n}+\frac{(\Delta+q_N)(\Delta+q_N+1)a^4}{8\widehat u^2}
\mathbb T^{(N)}_{\mu_1\cdots\mu_n}\Bigg].
\label{eq:general-Delta-D2-iterated-particular}
\end{align}
The three terms come from the second-order expansion of the numerator, the product of the first-order numerator and denominator terms, and the second-order expansion of the denominator. For spin 0 and 1, this gives
\bs
\begin{align}
D^{(2),pp}={}&
\frac{\Delta(\Delta+1)C^{(0)}a^4}{8\widehat u^{\Delta+2}},\\
D_{\mu\nu}^{(2),pp}={}&
\frac{\Delta(\Delta+1)C_1^{(1)}a^4}{8\widehat u^{\Delta+2}}\eta_{\mu\nu}
+\frac{C_2^{(1)}}{\widehat u^{\Delta+2}}a_\mu a_\nu+\frac{(\Delta+2)C_2^{(1)}a^2}{2\widehat u^{\Delta+3}}
(a_\mu n_\nu+n_\mu a_\nu)
\nonumber\\
&+\frac{(\Delta+2)(\Delta+3)C_2^{(1)}a^4}{8\widehat u^{\Delta+4}}
n_\mu n_\nu.
\end{align}
\es
For spin 2, \eqref{eq:general-Delta-D2-iterated-particular} applies to all the tensor structures in \eqref{leadingD}, including the $\epsilon$ terms. For the structure with four $n$'s, its second-derivative term is
\begin{align}
\frac{a_\alpha a_\beta}{2}
\frac{\partial^2\mathbb T^{(10)}_{\mu\nu\rho\sigma}}
{\partial n_\alpha\partial n_\beta}
={}&a_\mu a_\nu n_\rho n_\sigma
+a_\mu n_\nu a_\rho n_\sigma
+a_\mu n_\nu n_\rho a_\sigma+n_\mu a_\nu a_\rho n_\sigma
+n_\mu a_\nu n_\rho a_\sigma
+n_\mu n_\nu a_\rho a_\sigma.
\end{align}

Examples:
\begin{enumerate}
\item Comparing the above results with \eqref{Dmunu1}, we should identify
\begin{align}
\Delta=0,\quad C_1^{(1)}=0,\quad
C_2^{(1)}=\frac{1-\xi}{16\pi^2},\quad
\widetilde C_1^{(1)}=-\frac{1+\xi}{16\pi^2},\quad
\widetilde C_2^{(1)}=0.
\end{align}
The coefficients are consistent with \eqref{514}.

\item To compare with the spin-2 propagator, we keep the gauge parameters arbitrary. The complete propagator has the large-$r$ expansion
\begin{equation}
G_{\mu\nu\rho\sigma}(x;x')
=rD_{\mu\nu\rho\sigma}^{(0)}(u,\Omega;x')
+D_{\mu\nu\rho\sigma}^{(1)}(u,\Omega;x')
+\frac{1}{r}D_{\mu\nu\rho\sigma}^{(2)}(u,\Omega;x')
+O(r^{-2}).
\end{equation}
Therefore, we find $\Delta=-1$ for general gauge parameters and the non-vanishing coefficients up to NNLO are
\bs
\begin{align}
C_{10}^{(2)}
={}&-\frac{\gamma_5}{32\pi^2},\\
\widetilde C_4^{(2)}=\widetilde C_9^{(2)}
={}&-\frac{\gamma_3+\frac{\gamma_5}{4}}{16\pi^2},
\nonumber\\
\widetilde C_5^{(2)}=\widetilde C_6^{(2)}
=\widetilde C_7^{(2)}=\widetilde C_8^{(2)}
={}&-\frac{\gamma_4+\frac{\gamma_5}{4}}{16\pi^2},\\
\widehat C_1^{(2)}
={}\frac{1-\gamma_3-\frac{\gamma_5}{8}}{8\pi^2},\qquad
\widehat C_2^{(2)}=\widehat C_3^{(2)}
={}&-\frac{1+\gamma_4+\frac{\gamma_5}{8}}{8\pi^2}.
\end{align}
\es
For a special choice of the gauge parameters, the fall-off index could be $\Delta=0$ or $\Delta=1$. We still find the gauge dependence of the fall-off index
\bea
\Delta=\left\{\begin{array}{cl}
-1&\alpha\not=1\ \text{and}\ (\alpha-3)\xi+\alpha+1\not=0,\\
0&\xi\not=1\ \text{and}\ [\alpha=1\ \text{or}\ (\alpha-3)\xi+\alpha+1=0],\\
1&\xi=1\ \text{and}\ \alpha=1.
\end{array}\right.
\eea
\end{enumerate}

}

\section{The lower bound on the fall-off index \texorpdfstring{\(\Delta\)}{Delta}}
\label{spin-1KLrepproof}

{%
In this appendix we analyze how spectral densities satisfying the UV/IR conditions \eqref{UVIR} and local integrability conditions lead to a lower bound  on $\Delta$. We focus on the spin-1 case and then extend to the other spin fields.

We start with \eqref{eq:KL-forspin-1}. The terms to be extrapolated to null infinity
can be written as
\begin{equation}
\begin{aligned}
G_{\mu\nu}(x;x')={}&
\frac{\eta_{\mu\nu}}{4\pi^2(-h)}
\int_0^\infty ds\,
\sqrt{-hs}K_1(\sqrt{-hs})\,s\rho_v^1(s)\\
&+
\frac{\eta_{\mu\nu}}{4\pi^2(-h)^2}
\int_0^\infty ds\,
(-hs)K_2(\sqrt{-hs})\,\rho_v(s)\\
&-
\frac{(x-x')_\mu(x-x')_\nu}{4\pi^2(-h)^3}
\int_0^\infty ds\,
(-hs)^{3/2}K_3(\sqrt{-hs})\,\rho_v(s).
\end{aligned}
\label{spin-1GinKLspec}
\end{equation}

Denote the three spectral
integrals in \eqref{spin-1GinKLspec} by $I_1(h)$, $I_2(h)$ and
$I_3(h)$, in the order in which they appear. 
For $\nu=1,2,3$, the Bessel identity and its small-argument
limit are
\begin{equation}
\frac{d}{dz}\bigl[z^\nu K_\nu(z)\bigr]
=-z^\nu K_{\nu-1}(z)<0,
\qquad
\lim_{z\to0^+}z^\nu K_\nu(z)
=2^{\nu-1}\Gamma(\nu).
\end{equation}
Thus the Bessel factors are positive and bounded uniformly in
$h<0$. We split each spectral integral at a fixed small $s_0>0$.
The IR parts obey
\begin{equation}
\begin{aligned}
    \left|
\int_0^{s_0}ds\,
(-hs)^{\nu/2}K_\nu(\sqrt{-hs})s\rho_v^1
\right|
\leq
2^{\nu-1}\Gamma(\nu)
\int_0^{s_0}ds\,|s\rho_v^1|&=o(s_0^{\epsilon_1}),\\
 \left|
\int_0^{s_0}ds\,
(-hs)^{\nu/2}K_\nu(\sqrt{-hs})\rho_v
\right|
\leq
2^{\nu-1}\Gamma(\nu)
\int_0^{s_0}ds\,|\rho_v|&=o(s_0^{\epsilon}),\\
\end{aligned}
\label{spin-1-KL-IR-estimate}
\end{equation}
Hence the IR
contribution tends to zero as $s_0\to0^+$.

For the UV part, choose a fixed $h_0<0$ such that
$\sqrt{-h_0}>\max\{\alpha,\alpha_1\}$. At fixed $h_0$,
\begin{equation}
(-h_0s)^{\nu/2}K_\nu(\sqrt{-h_0s})
=\mathcal{O}\!\left(
s^{\nu/2-1/4}e^{-\sqrt{-h_0s}}
\right),
\qquad s\to\infty.
\end{equation}
The above large-$s$ behavior, combined with the  local integrability of spectral densities and the UV conditions
\eqref{KL-UV-convergence}, therefore gives
\begin{equation}
\int_{s_0}^{\infty}ds\,
(-h_0s)^{\nu/2}K_\nu(\sqrt{-h_0s})|s\rho_v^1(s)|
<\infty,\qquad \int_{s_0}^{\infty}ds\,
(-h_0s)^{\nu/2}K_\nu(\sqrt{-h_0s})|\rho_v(s)|
<\infty{.}
\label{spin-1-KL-UV-dominator}
\end{equation}
For $h\leq h_0$, monotonicity gives
\begin{equation}
0\leq(-hs)^{\nu/2}K_\nu(\sqrt{-hs})
\leq(-h_0s)^{\nu/2}K_\nu(\sqrt{-h_0s}).
\end{equation}
For each fixed $s>0$, the left-hand side tends to zero as
$h\to-\infty$. To evaluate the UV  contribution explicitly, keep $s_0>0$
and $h_0$ fixed as above. For $h<h_0$, define the ratio
\begin{equation}
R_\nu(s;h,h_0)
=
\left(\frac{-h}{-h_0}\right)^{\nu/2}
\frac{K_\nu(\sqrt{-hs})}
     {K_\nu(\sqrt{-h_0s})}.
     \label{R-wronski-def}
\end{equation}
The modified Bessel equation gives
\begin{equation}
\frac{d}{ds}
\left[
s\frac{d}{ds}K_\nu(\sqrt{-hs})
\right]
=
\left(
\frac{-h}{4}+\frac{\nu^2}{4s}
\right)K_\nu(\sqrt{-hs}).
\end{equation}
We obtain the Wronskian identity
\begin{equation}
\begin{aligned}
&\frac{d}{ds}\Bigg[
s\Bigg(
K_\nu(\sqrt{-h_0s})
\frac{d}{ds}K_\nu(\sqrt{-hs})-
K_\nu(\sqrt{-hs})
\frac{d}{ds}K_\nu(\sqrt{-h_0s})
\Bigg)
\Bigg]\\=&
\frac{h_0-h}{4}
K_\nu(\sqrt{-hs})K_\nu(\sqrt{-h_0s}).
\end{aligned}
\end{equation}
Integrating from $s$ to infinity therefore gives
\begin{equation}
\begin{aligned}
\partial_s R_\nu(s;h,h_0)
={}&
\left(\frac{-h}{-h_0}\right)^{\nu/2}
\frac{h-h_0}
     {4sK_\nu(\sqrt{-h_0s})^2}
\int_s^\infty dv\,
K_\nu(\sqrt{-hv})K_\nu(\sqrt{-h_0v})
<0.
\end{aligned}
\end{equation}
Thus $R_\nu$ is positive and decreasing. Moreover,
$R_\nu(s;h,h_0)\to0$ as $s\to\infty$, so
\begin{equation}
\int_{s_0}^\infty ds\,
|\partial_s R_\nu(s;h,h_0)|
=
R_\nu(s_0;h,h_0).
\end{equation}

On a finite interval $[s_0,S]$, 
\begin{equation}
\begin{aligned}
&\int_{s_0}^{S}ds\,
(-hs)^{\nu/2}K_\nu(\sqrt{-hs})
\rho^{\mathrm{eff}}(s)\\
=&\left[
R_\nu(s;h,h_0)
\int_{s_0}^{s}dv\,
(-h_0v)^{\nu/2}K_\nu(\sqrt{-h_0v})
\rho^{\mathrm{eff}}(v)
\right]_{s_0}^{S}
\\
\quad&-
\int_{s_0}^{S}ds\,
\partial_s R_\nu(s;h,h_0)
\int_{s_0}^{s}dv\,
(-h_0v)^{\nu/2}K_\nu(\sqrt{-h_0v})
\rho^{\mathrm{eff}}(v).
\end{aligned}
\end{equation}
Here $\rho^{\mathrm{eff}}$ denotes either $s\rho_v^1$ or
$\rho_v$, restricted to its density function part.
The lower boundary term vanishes because the inner integral
is zero at $s=s_0$. By \eqref{spin-1-KL-UV-dominator},
the inner integral is absolutely convergent and hence uniformly
bounded in its upper endpoint. For fixed $h<h_0<0$, the
large-argument asymptotics of $K_\nu$ give
$R_\nu(S;h,h_0)\to0$ as $S\to\infty$.
Therefore, the upper boundary term also vanishes. Consequently,
\begin{equation}
\begin{aligned}
&\left|
\int_{s_0}^\infty ds\,
(-hs)^{\nu/2}K_\nu(\sqrt{-hs})
\rho^{\mathrm{eff}}(s)
\right|\\
\leq&
R_\nu(s_0;h,h_0)\times
\int_{s_0}^\infty ds\,
(-h_0s)^{\nu/2}K_\nu(\sqrt{-h_0s})
|\rho^{\mathrm{eff}}(s)|.
\end{aligned}
\end{equation}
The last integral is finite and independent of $h$.
For fixed $s_0$ and $h_0$, the large-argument asymptotics
of $K_\nu$ give
\begin{equation}
R_\nu(s_0;h,h_0)
=
O\!\left(
(-h)^{\nu/2-1/4}e^{-\sqrt{-hs_0}}
\right)
\longrightarrow0,
\qquad h\to-\infty.
\end{equation}
Hence
\begin{equation}
\lim_{h\to-\infty}
\int_{s_0}^{\infty}ds\,
(-hs)^{\nu/2}K_\nu(\sqrt{-hs})
\rho^{\mathrm{eff}}(s)=0
\qquad(s_0>0\text{ fixed}).
\label{STT-KL-UV-estimate}
\end{equation}

Combining this result with the uniform IR estimate
\eqref{spin-1-KL-IR-estimate} and then taking $s_0\to0^+$ proves that the complete integrals tend to zero. Therefore,

\begin{equation}
    \lim_{h\to -\infty} I_1(h)=0,\qquad \lim_{h\to -\infty} I_2(h)=0,\qquad \lim_{h\to -\infty} I_3(h)=0{.}
\end{equation}
For the density function contribution, the preceding estimates,
together with \eqref{spin-1GinKLspec}, imply
\begin{equation}
rG_{\mu\nu}(x;x')\longrightarrow0
\end{equation}
as $r\to\infty$ at fixed $(u,\Omega,x')$ with $\widehat u<0$.
Assume, in addition, that the correlator admits a nonzero
leading pure-power asymptotic behavior,
\begin{equation}
G_{\mu\nu}(x;x')
=
r^{-\Delta}
\left[D_{\mu\nu}(u,\Omega;x')+o(1)\right],
\qquad \Delta\in\mathbb{R},
\end{equation}
where $D_{\mu\nu}$ is nonzero somewhere in the spacelike
extrapolation region. Comparing this behavior with
$rG_{\mu\nu}\to0$ yields
\begin{equation}
\Delta>1.
\end{equation}
The existence of such a pure-power leading term is an additional
assumption and does not follow from the spectral conditions alone.

We next present the consequence of Wightman positivity for the
leading independent $\Delta$-series. For a  physical
spectrum,
\begin{equation}
\rho_v^0(s)\geq0,\qquad \rho_v^1(s)\geq0,
\qquad 0\leq\rho_v^1(s)\leq\rho_v(s)
\end{equation}
almost everywhere. Using
$K_3(z)=K_1(z)+4K_2(z)/z$ and the definitions of $I_i$, we have\footnote{This positivity argument does not apply to the covariantly
gauge-fixed potential, whose Wightman two-point function
is not positive on the full gauge-field state space.
In particular, in Feynman gauge one has $I_3=0$ but
$(-h)I_1+I_2=-h>0$, so the spectral positivity inequality fails.}
\begin{equation}
0\leq(-h)I_1+I_2\leq I_3.
\label{spin-1-KL-positivity-ratio}
\end{equation}
If $I_3=0$, both coefficients vanish.
Otherwise, the coefficient of
$\eta_{\mu\nu}$ is suppressed by one power of $r$ relative to the
$n_\mu n_\nu$ coefficient:
\begin{equation}
\frac{\left|[(-h)I_1+I_2]/(-h)^2\right|}
{r^2I_3/(-h)^3}
\leq\frac{-h}{r^2}=O(r^{-1}).
\label{spin-1-KL-leading-suppression}
\end{equation}
Thus, for the globally first nonzero pure-power series, the leading
coefficient multiplying the $\eta_{\mu\nu}$ structure vanishes,
which is precisely $C_1^{(1)}=0$.

The same estimate also constrains subsequent independent
$\Delta$-series. Suppose that the coefficient of $n_\mu n_\nu$
has a nonzero leading term proportional to
$r^{-\Delta_{\min}}$. Then
\eqref{spin-1-KL-leading-suppression} implies that the full
coefficient of $\eta_{\mu\nu}$ is
$O(r^{-\Delta_{\min}-1})$.
Within the assumed ordered pure-power expansion, distinct
independent series have disjoint sets of exponents.
A nonzero term in one series therefore cannot be cancelled
by terms from another. Consequently, any subsequent independent
series whose leading $\eta_{\mu\nu}$ coefficient is nonzero
must satisfy
\begin{equation}
\Delta\geq\Delta_{\min}+1.
\end{equation}
By our convention, terms whose exponents differ by an integer
are grouped into the same series. The case
$\Delta=\Delta_{\min}+1$ thus belongs to the first series
rather than to a distinct independent series. Hence
\begin{equation}
\Delta>\Delta_{\min}+1>2.
\label{spin-1-KL-series-separation}
\end{equation}
The final inequality uses $\Delta_{\min}>1$, established above
in the absence of massless-pole contributions.

Finally, allow $\delta(s)$ terms in the effective combinations, but no
derivatives of $\delta(s)$:
\begin{equation}
s\rho_v^1(s)=a\delta(s)+\cdots,
\qquad
\rho_v(s)=b\delta(s)+\cdots,
\end{equation}
where the omitted ordinary parts obey the conditions in \eqref{UVIR} and are locally integrable. Since
$zK_1(z)\to1$, $z^2K_2(z)\to2$ and $z^3K_3(z)\to8$ as $z\to0^+$,
the pole contribution to \eqref{spin-1GinKLspec} is
\begin{equation}
G_{\mu\nu}(x;x')
=\frac{a}{4\pi^2(-h)}\eta_{\mu\nu}
+\frac{2b}{4\pi^2(-h)^2}\eta_{\mu\nu}
-\frac{8b}{4\pi^2(-h)^3}(x-x')_\mu(x-x')_\nu .
\end{equation}

After extrapolation, we get
\begin{equation}
    D_{\mu\nu}=-\frac{a\eta_{\mu\nu}}{8\pi^2\widehat u}+\frac{bn_{\mu}n_{\nu}}{4\pi^2\widehat u^3}.
\end{equation}
Thus $\Delta=1$ when $a$ and $b$ are not both zero. 

In summary,  we have proved 
\begin{equation}
    \Delta\geq 1.
\end{equation}

}

\section{Bulk-to-boundary correlator from the higher spin KL representation}
\label{expan}

In this appendix, we collect all pieces of \eqref{klrephs}, \eqref{STTpro}, \eqref{Prm}, \eqref{mathcalR} and \eqref{partialG} to obtain the explicit  expanded bulk STT two-point function

{ \begin{equation}
\begin{aligned}
&\left\langle0\left|T\{
t_{\mu_1\cdots\mu_n}(x)
t_{\nu_1\cdots\nu_n}(x')
\}\right|0\right\rangle
\\
=&
\sum_{J=0}^{n}
\sum_{\varrho=0}^{J}
\sum_{c=0}^{n-\varrho}
\widetilde{\mathcal B}_{nJ\varrho c}
\left[
  \eta_{\mu_1\nu_1}\cdots\eta_{\mu_{\varrho+c}\nu_{\varrho+c}}
   \delta x_{\mu_{\varrho+c+1}}\cdots \delta x_{\mu_n}
   \delta x_{\nu_{\varrho+c+1}}\cdots  \delta x_{\nu_n}
\right]^{\text{STT}}
\\
&\quad\times
\int_0^\infty \mathrm d s\,\rho_J(s)\,
\frac{s^{\varrho+L/2}(-h)^{-L/2}}{4\pi^2}
K_L\!\left(\sqrt{-s h}\right),
\qquad
L=1+2(n-\varrho)-c .
\end{aligned}
\label{eq:STT-KL-physical-en}
\end{equation}
Here the bracket denotes the STT projection on the indices $\{\mu_i\}_{i=1}^n$ and $\{\nu_i\}_{i=1}^n$ separately. For instance, the \text{STT} projection on a tensor
\(Y_{\mu_1\cdots\mu_n;\nu_1\cdots\nu_n}\) is defined as below
\begin{equation}
\begin{aligned}
\left[
Y_{\mu_1\cdots\mu_n;\nu_1\cdots\nu_n}
\right]^{\text{STT}}
&=
\sum_{q=0}^{\lfloor n/2\rfloor}
\sum_{t=0}^{\lfloor n/2\rfloor}
b_{nq}b_{nt}\,
\eta_{(\mu_1\mu_2}\cdots\eta_{\mu_{2q-1}\mu_{2q}}
\eta_{(\nu_1\nu_2}\cdots\eta_{\nu_{2t-1}\nu_{2t}}
\\
&\quad\times
\eta^{\alpha_1\alpha_2}\cdots\eta^{\alpha_{2q-1}\alpha_{2q}}
\eta^{\beta_1\beta_2}\cdots\eta^{\beta_{2t-1}\beta_{2t}}\\
&\quad\times
Y^{\rm s}_{\alpha_1\cdots \alpha_{2q}\mu_{2q+1}\cdots\mu_n);\,
  \beta_1\cdots \beta_{2t}\nu_{2t+1}\cdots\nu_n)} ,
\end{aligned}
\label{eq:STT-proj-def-en}
\end{equation}
where 
\be
Y^{\rm s}_{\mu_1\cdots\mu_n;\nu_1\cdots\nu_n}=Y_{(\mu_1\cdots\mu_n);(\nu_1\cdots\nu_n)}.
\ee
The parentheses on the two sides of the semicolon denote unit-weight
symmetrization in $\mu$'s and $\nu$'s separately.

The summations in \eqref{eq:STT-KL-physical-en} are organized as follows.
\begin{itemize}
  \item \(\sum_{J=0}^{n}\) decomposes the rank-\(n\) STT two-point function
  into the \(n+1\) massive little-group spin blocks.
  \item \(\sum_{\varrho=0}^{J}\) comes from the Legendre expansion of the
  spin-J block.  The integer \(\varrho\) counts the number of \(\mu\)-\(\nu\) pairs in metrics when using \eqref{eq:SzwinKL}.
  \item \(\sum_{c=0}^{n-\varrho}\) is the compact expression of the remaining
  derivative-pairing sector.  After the \(\varrho\) pairs of indices for metrics are fixed, there remain \(2(n-\varrho)\) derivative slots, evenly split between $\mu$'s and $\nu$'s.  The integer $c$ denotes the number of additional $\mu-\nu$ pairs in the metrics produced when these derivatives act on $G(\delta x;s)$.  Same-side metric contractions are trace terms and are removed
  by the STT projection.
\end{itemize}

The coefficient is
\begin{equation}
\begin{aligned}
\widetilde{\mathcal B}_{nJ\varrho c}
:=
N_{nJ}
(-1)^{n-\varrho+c}
\frac{(J+\varrho)!}{2^\varrho(J-\varrho)!(\varrho!)^2}
\frac{[(n-\varrho)!]^2}{c!(n-\varrho-c)!^2}.
\end{aligned}
\label{eq:B-coeff-en}
\end{equation}

{%
Now we employ this representation and prove that the fall-off index still obeys $\Delta\ge 1$ under certain conditions.

We first consider a spectrum described by  spectral
densities, without $\delta(s)$ or its derivatives in the effective
combinations below. To collect all contributions to the same tensor
structure, we define
\begin{equation}
\rho^{\mathrm{eff}}_\varrho(s)
=
s^\varrho\sum_{J=\varrho}^{n}
N_{n,J}
\frac{(J+\varrho)!}
{2^\varrho(J-\varrho)!(\varrho!)^2}\,
\rho_J(s),
\qquad \varrho=0,\ldots,n.
\label{STT-KL-effective}
\end{equation}
By collecting terms with $k=\varrho+c$ in
\eqref{eq:STT-KL-physical-en}, we obtain
\begin{equation}
\begin{aligned}
&\left\langle0\left|T\{
t_{\mu_1\cdots\mu_n}(x)
t_{\nu_1\cdots\nu_n}(x')
\}\right|0\right\rangle\\
=&\sum_{k=0}^{n}
\Big[
\eta_{\mu_1\nu_1}\cdots\eta_{\mu_k\nu_k}
\delta x_{\mu_{k+1}}\cdots\delta x_{\mu_n}
\delta x_{\nu_{k+1}}\cdots\delta x_{\nu_n}
\Big]^{\mathrm{STT}}
\int_0^\infty ds\,\widetilde{G}_k(s,h).
\end{aligned}
\label{STT-KL-regrouped}
\end{equation}
The coefficients are
\begin{equation}
\begin{aligned}
\widetilde{G}_k(s,h)
={}&\frac{(-1)^{n-k}}
{4\pi^2(-h)^{1+2(n-k)}}
\sum_{\varrho=0}^{k}
\frac{[(n-\varrho)!]^2}
{(k-\varrho)![(n-k)!]^2}\frac{
(-hs)^{(1+2n-k-\varrho)/2}
K_{1+2n-k-\varrho}(\sqrt{-hs})
}{(-h)^{k-\varrho}}\,
\rho^{\mathrm{eff}}_\varrho(s).
\end{aligned}
\label{STT-KL-rhotilde}
\end{equation}

The spectral integrals depend on $r$ through $h$, so we
retain all terms. Since
$\delta x_\mu\delta x_\nu=r^2n_\mu n_\nu+O(r)$, the leading
tensors in \eqref{STT-KL-regrouped} are
$r^{2(n-k)}\mathcal T^{(k)}(\Omega)$, where
\begin{equation}
\mathcal T^{(k)}_{\mu_1\cdots\mu_n;\nu_1\cdots\nu_n}(\Omega)
=
\Big[
\eta_{\mu_1\nu_1}\cdots\eta_{\mu_k\nu_k}
n_{\mu_{k+1}}\cdots n_{\mu_n}
n_{\nu_{k+1}}\cdots n_{\nu_n}
\Big]^{\mathrm{STT}}.
\label{STT-boundary-tensors}
\end{equation}
These $n+1$ tensors are linearly independent.

We impose conditions on the effective spectral densities 
at the two ends of the mass spectrum. For the UV
part, require
\begin{equation}
\exists\,\alpha_\varrho>0,\qquad
|\rho^{\mathrm{eff}}_\varrho(s)|
=o\!\left(e^{\alpha_\varrho\sqrt{s}}\right),
\qquad s\to\infty.
\label{STT-KL-UV}
\end{equation}
For the IR part, require
\begin{equation}
\exists\,\epsilon_\varrho>0,\qquad
|\rho^{\mathrm{eff}}_\varrho(s)|
=o\!\left(s^{-1+\epsilon_\varrho}\right),
\qquad s\to0^+.
\label{STT-KL-IR}
\end{equation}
We also require all the effective combinations to be locally
integrable on $(0,\infty)$.

We now show that the Bessel integrals entering
\eqref{STT-KL-rhotilde} tend to zero as $h\to-\infty$.
Fix $0\leq\varrho\leq k\leq n$ and put
$\nu=1+2n-k-\varrho\geq1$. The Bessel identity and its
small-argument limit give
\begin{equation}
\frac{d}{dz}\bigl[z^\nu K_\nu(z)\bigr]
=-z^\nu K_{\nu-1}(z)<0,
\qquad
\lim_{z\to0^+}z^\nu K_\nu(z)
=2^{\nu-1}\Gamma(\nu).
\label{STT-KL-Bessel-properties}
\end{equation}
Thus $(-hs)^{\nu/2}K_\nu(\sqrt{-hs})$ is positive,
bounded by $2^{\nu-1}\Gamma(\nu)$, and decreases as $-h$
increases.

We split each integral at a fixed small $s_0>0$.
On the infrared interval, \eqref{STT-KL-IR} gives
\begin{equation}
\begin{aligned}
\left|
\int_0^{s_0}ds\,
(-hs)^{\nu/2}K_\nu(\sqrt{-hs})
\rho^{\mathrm{eff}}_\varrho(s)
\right|
&\leq
2^{\nu-1}\Gamma(\nu)
\int_0^{s_0}ds\,
|\rho^{\mathrm{eff}}_\varrho(s)|
=o\!\left(s_0^{\epsilon_\varrho}\right)
,\qquad s_0\to0^+.
\end{aligned}
\label{STT-KL-IR-estimate}
\end{equation}

For the ultraviolet part, choose a fixed $h_0<0$ with
$\sqrt{-h_0}$ larger than  $\text{max}\{\alpha_\varrho\}$.
At fixed $h_0$, the large-$s$ behavior is
\begin{equation}
(-h_0s)^{\nu/2}K_\nu(\sqrt{-h_0s})
=O\!\left(
s^{\nu/2-1/4}e^{-\sqrt{-h_0}\sqrt{s}}
\right),
\qquad s\to\infty.
\label{STT-KL-UV-kernel}
\end{equation}
The local integrability of $\rho^{\mathrm{eff}}_\varrho$
on finite intervals $[s_0,S]$, together with
\eqref{STT-KL-UV} and \eqref{STT-KL-UV-kernel}, therefore gives
\begin{equation}
\int_{s_0}^{\infty}ds\,
(-h_0s)^{\nu/2}K_\nu(\sqrt{-h_0s})
|\rho^{\mathrm{eff}}_\varrho(s)|<\infty.
\label{STT-KL-UV-dominator}
\end{equation}
For $h\leq h_0$, monotonicity in
\eqref{STT-KL-Bessel-properties} gives
\begin{equation}
0\leq(-hs)^{\nu/2}K_\nu(\sqrt{-hs})
\leq(-h_0s)^{\nu/2}K_\nu(\sqrt{-h_0s}).
\end{equation}

To make the ultraviolet decay explicit, use the ratio
introduced in \eqref{R-wronski-def}
\begin{equation}
R_\nu(s;h,h_0)
=
\left(\frac{-h}{-h_0}\right)^{\nu/2}
\frac{K_\nu(\sqrt{-hs})}
     {K_\nu(\sqrt{-h_0s})},
\qquad h<h_0.
\end{equation}
The argument using the Wronskian in the spin-1 case now applies to every
$\nu=1+2n-k-\varrho\geq1$ and gives
\begin{equation}
\lim_{h\to-\infty}
\int_0^\infty ds\,
(-hs)^{\nu/2}K_\nu(\sqrt{-hs})
\rho^{\mathrm{eff}}_\varrho(s)
=0,
\qquad
\nu=1+2n-k-\varrho.
\label{STT-KL-integrals-vanish}
\end{equation}

Along the extrapolation, $-h/r\to-2\widehat u>0$.
For each $k$, multiplying \eqref{STT-KL-rhotilde} by
$r^{1+2(n-k)}$ leaves a finite sum of the integrals in
\eqref{STT-KL-integrals-vanish}, multiplied by the bounded
factors \((r/(-h))^{1+2(n-k)}\allowbreak(-h)^{-(k-\varrho)}\).
Consequently,
\begin{equation}
\lim_{r\to\infty}
r^{1+2(n-k)}
\int_0^\infty ds\,\widetilde{G}_k(s,h)=0,
\qquad k=0,\ldots,n.
\label{STT-KL-vanishing}
\end{equation}
After multiplication by the corresponding tensors in
\eqref{STT-KL-regrouped}, this shows that the  full correlator is $o(r^{-1})$.
Denoting the correlator in \eqref{STT-KL-regrouped} by
$G_{\mu_1\cdots\mu_n;\nu_1\cdots\nu_n}(x;x')$, we have
\begin{equation}
rG_{\mu_1\cdots\mu_n;\nu_1\cdots\nu_n}(x;x')
\longrightarrow0\label{rGo}
\end{equation}
as $r\to\infty$ at fixed $(u,\Omega,x')$ with $\widehat u<0$.

Assume, in addition, that the correlator admits a nonzero
leading pure-power asymptotic behavior,
\begin{equation}
G_{\mu_1\cdots\mu_n;\nu_1\cdots\nu_n}(x;x')
=
r^{-\Delta}
\left[
D_{\mu_1\cdots\mu_n;\nu_1\cdots\nu_n}(u,\Omega;x')
+o(1)
\right],
\qquad \Delta\in\mathbb R,
\end{equation}
where $D_{\mu_1\cdots\mu_n;\nu_1\cdots\nu_n}$ is nonzero
somewhere in the spacelike extrapolation region.
Comparing this behavior with \eqref{rGo} yields
\begin{equation}
\Delta>1.
\label{STT-KL-continuous-bound}
\end{equation}

Taking the leading pure-power extrapolation of
\eqref{STT-KL-regrouped} and using $-h/r\to-2\widehat u$, we obtain
\begin{equation}
\begin{aligned}
D_{\mu_1\cdots\mu_n;\nu_1\cdots\nu_n}(u,\Omega;x')
&=
\sum_{k=0}^{n}
\widetilde{\mathcal A}_{\Delta;n,k}^{\rm bb}
\mathcal T^{(k)}_{\mu_1\cdots\mu_n;\nu_1\cdots\nu_n}(\Omega)
\frac{1}{[-\widehat u]^{\Delta+2(n-k)}} .
\end{aligned}
\label{eq:final-D-en}
\end{equation}
Here
\begin{equation}
\widetilde{\mathcal A}_{\Delta;n,k}^{\rm bb}
=
\lim_{h\to-\infty}
\left(\frac{-h}{2}\right)^{\Delta+2(n-k)}
\int_0^\infty ds\,\widetilde{G}_k(s,h),
\end{equation}
with $\widetilde{G}_k$ given in \eqref{STT-KL-rhotilde}.
Thus the leading correlator has the same form as the LO Ward solution.

We now derive the consequence of Wightman positivity for these
coefficients. As in the spin-1 case, for a gauge-invariant local
operator in a unitary QFT, the physical spin-channel measures
are non-negative. When they are expressed in terms of spectral densities, we have
$\rho_J(s)\geq0$ almost everywhere.
Since $N_{nJ}$ and the combinatorial coefficients in
\eqref{STT-KL-effective} are non-negative, we have 
\begin{equation}
\rho^{\mathrm{eff}}_\varrho(s)\geq0,
\qquad \varrho=0,\ldots,n.
\label{STT-KL-effective-positivity}
\end{equation}
The same finite sum also gives
\begin{equation}
0\leq\rho^{\mathrm{eff}}_\varrho(s)
\leq
\frac{(n+\varrho)!}
{2^\varrho(n-\varrho)!(\varrho!)^2}
s^\varrho\rho^{\mathrm{eff}}_0(s).
\label{STT-KL-effective-comparison}
\end{equation}
Channels with $N_{nJ}=0$ do not contribute.

The Bessel recurrence relation
$K_{m+1}(x)=K_{m-1}(x)+2mK_m(x)/x$ gives, for
integer $m\geq1$ and $x>0$,
\begin{equation}
x^{m+1}K_{m+1}(x)\geq x^mK_m(x),
\qquad
K_{m+2}(x)\geq K_m(x).
\end{equation}
Since
$1+2n=(1+2n-k-\varrho)+2\varrho+(k-\varrho)$,
repeated application of  the above inequalities  yields
\begin{equation}
(-hs)^\varrho(-hs)^{\nu/2}K_\nu(\sqrt{-hs})
\leq
(-hs)^{(1+2n)/2}K_{1+2n}(\sqrt{-hs}),
\qquad
\nu=1+2n-k-\varrho.
\label{STT-KL-Bessel-comparison}
\end{equation}
Combining this with \eqref{STT-KL-effective-comparison},
we obtain
\begin{equation}
\begin{aligned}
&\int_0^\infty ds\,
\frac{(-hs)^{\nu/2}K_\nu(\sqrt{-hs})}
     {(-h)^{k-\varrho}}
\rho^{\mathrm{eff}}_\varrho(s)\\
&\quad\leq\frac{(n+\varrho)!}
{2^\varrho(n-\varrho)!(\varrho!)^2(-h)^k}
\int_0^\infty ds\,
(-hs)^{(1+2n)/2}K_{1+2n}(\sqrt{-hs})
\rho^{\mathrm{eff}}_0(s).
\end{aligned}
\label{STT-KL-effective-comparison-integral}
\end{equation}
All terms in \eqref{STT-KL-rhotilde} satisfy the same inequality. Hence there exists a finite constant $C_{n,k}$
such that
\begin{equation}
r^{2(n-k)}
\left|\int_0^\infty ds\,\widetilde{G}_k(s,h)\right|
\leq
C_{n,k}\left(\frac{-h}{r^2}\right)^k
r^{2n}
\left|\int_0^\infty ds\,\widetilde{G}_0(s,h)\right|.
\label{STT-KL-leading-suppression}
\end{equation}
Since $-h\sim r$, the $k$-th
structure is suppressed relative to $k=0$ by at least
$r^{-k}$.

Let $\Delta_{\min}$ denote the leading fall-off index among
the nonzero independent $\Delta$-series.
For this globally first nonzero pure-power series,
\eqref{STT-KL-leading-suppression} gives
\begin{equation}
\widetilde{\mathcal A}_{\Delta_{\min};n,k}^{\rm bb}=0,
\qquad k=1,\ldots,n.
\label{STT-KL-general-leading-vanish}
\end{equation}
For $n=1$, this is precisely $C_1^{(1)}=0$.

The same estimate also constrains subsequent independent
$\Delta$-series. We use
$\widetilde{\mathcal A}_{\Delta;n,k}^{\rm bb}$ for the leading
Ward coefficient of each such series.
If the $k=0$ term starts at $r^{-\Delta_{\min}}$, then
\eqref{STT-KL-leading-suppression} implies that the full
coefficient associated with the $k$-th structure is
$O(r^{-\Delta_{\min}-k})$ after including its tensor factor.

We next allow $\delta(s)$ terms in the effective combinations,
but no derivatives of $\delta(s)$:
\begin{equation}
\rho^{\mathrm{eff}}_\varrho(s)
=
s^\varrho\sum_{J=\varrho}^{n}
N_{n,J}
\frac{(J+\varrho)!}
{2^\varrho(J-\varrho)!(\varrho!)^2}\,
\rho_J(s)
=
a_\varrho\delta(s)+\cdots,
\qquad \varrho=0,\ldots,n.
\label{STT-KL-massless}
\end{equation}
The omitted  parts satisfy \eqref{STT-KL-IR} and
\eqref{STT-KL-UV}, as well as local integrability on
$(0,\infty)$.

Substituting \eqref{STT-KL-massless} into \eqref{STT-KL-rhotilde} and noticing \eqref{STT-KL-Bessel-properties}, we obtain

\begin{equation}
D_{\mu_1\cdots\mu_n;\nu_1\cdots\nu_n}(u,\Omega;x')
=
\sum_{k=0}^{n}
\frac{(-1)^{n-k}a_k
\Gamma\!\bigl(1+2(n-k)\bigr)}{8\pi^2}
\frac{
\mathcal T^{(k)}_{\mu_1\cdots\mu_n;\nu_1\cdots\nu_n}(\Omega)
}{
[-\widehat u]^{1+2(n-k)}
}.
\label{STT-KL-massless-match}
\end{equation}
This has the form \eqref{eq:final-D-en} with $\Delta=1$.
Since the tensors $\mathcal T^{(k)}$ are independent,
the leading coefficient is nonzero when the $a_k$ are
not all zero. Assuming Wightman positivity, the individual physical-channel measures are non-negative and contain only ordinary \(\delta(s)\) terms. Since \(s^{\varrho}\delta(s)=0\) for \(\varrho>0\), it follows that \(a_{\varrho}=0\).

Combining the density function result with these massless
contributions, we find that any nonzero leading pure-power extrapolation
within the stated class satisfies
\begin{equation}
\Delta\geq1.
\end{equation}
The equality holds when there is a nonzero massless-pole
contribution in the effective combinations. Otherwise the
inequality is strict.

Finally, we illustrate an explicit realization of the spectral densities, which is the finite power-law sum
\begin{equation}
\rho_J(s)
=
\sum_{\kappa=0}^{n}
F_{J,\kappa}s^{\Delta-\kappa-2},
\qquad s>0,\qquad \Delta>1.
\label{STT-KL-power}
\end{equation}
For fixed $\kappa$ and $\varrho$, the spin channels combine into
\begin{equation}
\mathfrak H_{\kappa\varrho}
=
\sum_{J=\varrho}^{n}
N_{n,J}
\frac{(J+\varrho)!}
{2^\varrho(J-\varrho)!(\varrho!)^2}
F_{J,\kappa}.
\label{STT-KL-H}
\end{equation}
We first group equal powers in each effective combination.
Any non-integrable power must cancel there before the KL
integral is taken.
For an integrable power, the Bessel integral is
\begin{equation}
\begin{aligned}
\int_0^\infty ds\,
s^a(-hs)^{\nu/2}K_\nu(\sqrt{-hs})
&=
2^{2a+\nu+1}
\Gamma(a+\nu+1)\Gamma(a+1)
(-h)^{-a-1},
\qquad a>-1,
\end{aligned}
\label{STT-KL-power-integral-general}
\end{equation}
where $a=\Delta+\varrho-\kappa-2$ and
$\nu=1+2n-k-\varrho$.

We require this example to have a finite extrapolation after
multiplication by $r^\Delta$.
 With the tensor factor included, a term with indices
$(\kappa,k)$ scales as $r^{\kappa-k}$.
At $k=0$, convergence of the spectral integral or the absence
of positive powers of $r$ therefore requires
$\mathfrak H_{\kappa0}=0$ for $\kappa>0$.
Proceeding inductively in $k$, we find that the terms with $\varrho<k$
have already vanished when $\kappa>k$.
The remaining term then requires $\mathfrak H_{\kappa k}=0$.
Thus
\begin{equation}
\mathfrak H_{\kappa\varrho}=0
\qquad\text{for }\varrho<\kappa,
\label{STT-KL-triangular}
\end{equation}
and
\begin{equation}
\rho^{\mathrm{eff}}_\varrho(s)
=
\sum_{\kappa=0}^{\varrho}
\mathfrak H_{\kappa\varrho}
s^{\Delta+\varrho-\kappa-2}.
\label{STT-KL-effective-power}
\end{equation}
These effective combinations satisfy the IR,
UV and local-integrability conditions for $\Delta>1$.

At leading order, the scaling $r^{\kappa-k}$ requires
$\kappa=k$. Since $\varrho\leq k$ in
\eqref{STT-KL-rhotilde}, the triangular condition
\eqref{STT-KL-triangular} then leaves only
$\kappa=\varrho=k$.
Substituting the Bessel integral into
\eqref{STT-KL-rhotilde} and comparing with
\eqref{eq:final-D-en}, we obtain
\begin{equation}
\widetilde{\mathcal A}_{\Delta;n,k}^{\rm bb}
=
\mathfrak H_{kk}(-1)^{n-k}
\frac{2^{\Delta-4}}{\pi^2}
\Gamma\!\bigl(\Delta+2(n-k)\bigr)
\Gamma(\Delta-1).
\label{STT-KL-power-match}
\end{equation}
For $n=0,1,2$, setting $N_{0,0}=1$,
$N_{1,0}=N_{1,1}=1$, and
$N_{2,0}=\frac43,\ N_{2,1}=2,\ N_{2,2}=\frac23$, we obtain the result in the main text.

}
}

\section{Stabilizer group}

 This appendix fixes the stabilizer group used in the present context and  connects it to the stabilizer group in the literature. In what follows, the seven-dimensional algebra spanned by
\(\{\mathbb{D},\mathbb{K}_0,\mathbb{K}_1,\mathbb{K}_2,\mathbb{B}_1,\mathbb{B}_2,\mathbb{J}\}\)
in \eqref{eq:a1-south} and \eqref{eq:a1-south-brackets} will be denoted by \(\mathfrak h_S\). 

 \subsection{From the south pole to a general point}\label{generalpoint}
In this appendix, the south pole is denoted by $p_S=(0,\bm S)$ and the orthonormal basis is $\bm e_I,\ I=1,2$, with $\bm e_1\times \bm e_2=-\bm S=(0,0,1)$.
Let the point be
\begin{equation}
  p_\star=(u_\star,\Omega_\star),\qquad \bm N=\bm n(\Omega_\star),\qquad \bm N^2=1.
\end{equation}
We choose an orthonormal basis \(\bm E_1,\bm E_2\) of the tangent plane at \(\bm N\) whose orientation is fixed by
\begin{equation}
 \bm  E_I\cdot \bm N=0,\qquad \bm E_I\cdot \bm E_J=\delta_{IJ},\qquad
  \bm E_1\times \bm E_2=-\bm N .
  \label{eq:tangent-frame-orientation}
\end{equation}
Choose a rotation \(R\in SO(3)\) such that
\begin{equation}
  R\bm S=\bm N,\qquad R\bm e_I=\bm E_I.
\end{equation}
This rotation sends the south-pole direction $\bm S$ to \(\bm N\) and the orthonormal basis $\bm e_I$ to $\bm E_I$. Then applying the translation
\begin{equation}
  A=\frac{u_\star}{2}\bigl(P_0-N^iP_i\bigr)=a^\mu P_\mu,\qquad
  {a^\mu=\frac{u_\star}{2}(1,-\bm N)}
\end{equation}
to $u$ induces
\begin{equation}
  \delta u=a^0-\mathbf a\cdot \bm n 
  \label{eq:translation-i-plus}
\end{equation}  on \(\mathscr{I}^+\).
Therefore, at \(\bm n=\bm N\), the above translation gives
\begin{equation}
  \delta u
  =\frac{u_\star}{2}-\left(-\frac{u_\star}{2}\right)
  =u_\star .
\end{equation}
Hence the group element
\begin{equation}
  g_\star
  =\exp\!\left[\frac{u_\star}{2}(P_0-N^iP_i)\right]\circ R
  \label{eq:g-star}
\end{equation}
{satisfies} \(g_\star p_S=p_\star\), and the stabilizer group and algebra obey
\begin{equation}
  H_{p_\star}=g_\star H_{p_S}g_\star^{-1},\qquad
  \mathfrak h_{p_\star}=(g_\star)_*\mathfrak h_S .
  \label{eq:stabilizer-conjugation}
\end{equation}
Thus the generators of the stabilizer group at \(p_\star\) are written explicitly as 
\bs  \label{eq:a1-star-basis} \begin{align}
  D^\star&=-N^iJ_{0i}
  +\frac{u_\star}{2}\bigl(P_0-N^iP_i\bigr),\\
  J^\star&=N^1J_{23}+N^2J_{31}+N^3J_{12},\\
  K^\star_0&=-P_0-N^iP_i,\\
  B^\star_I&=-E_I^iP_i,\\
  K^\star_I&=-E_I^iJ_{0i}
  +E_I^iN^jJ_{ij}
  -u_\star E_I^iP_i .
\end{align}\es

To verify directly that the vector fields in \eqref{eq:a1-star-basis}
preserve \(p_\star\), write their induced action on \(\mathscr I^+\) as
\begin{equation}
  \xi\big|_{\mathscr I^+}
  =f(u,\Omega)\,\partial_u
  +Y^\theta(\Omega)\,\partial_\theta
  +Y^\phi(\Omega)\,\partial_\phi .
\end{equation}
The stabilizer condition at \(p_\star=(u_\star,\Omega_\star)\) is
\begin{equation}
  f(u_\star,\Omega_\star)=0,\qquad
  Y^\theta(\Omega_\star)=Y^\phi(\Omega_\star)=0 .
  \label{eq:direct-stabilizer-condition}
\end{equation}
Let
\begin{equation}
  \alpha=\bm N\cdot \bm n,\qquad
  \beta_I=\bm E_I\cdot \bm n,
\end{equation}
and
\begin{equation}
  N_\theta=\bm N\cdot \bm e_\theta,\qquad
  N_\phi=\bm N\cdot \bm e_\phi,\qquad
  E_{I\theta}=\bm E_I\cdot \bm e_\theta,\qquad
  E_{I\phi}=\bm E_I\cdot \bm e_\phi .
\end{equation}
Here \(\bm e_\theta,\bm e_\phi\) are the unit tangent vectors on the sphere.  From the bulk expressions in \eqref{eq:a1-star-basis}, the induced data on \(\mathscr I^+\) are\bs
  \label{eq:Iplus-star-action}\begin{align}
  K^\star_0:\quad&
  \begin{alignedat}[t]{3}
    f_{K^\star_0} &= \alpha-1,\qquad&
    Y_{K^\star_0} &= 0.
  \end{alignedat}
  \label{eq:B14a}
  \\[0.45em]
  B^\star_I:\quad&
  \begin{alignedat}[t]{3}
    f_{B^\star_I} &= \beta_I,\qquad&
    Y_{B^\star_I} &= 0.
  \end{alignedat}
  \label{eq:B14b}
  \\[0.45em]
  D^\star:\quad&
  \begin{alignedat}[t]{3}
    f_{D^\star}
      &= -u\alpha+\frac{u_\star}{2}(1+\alpha),\qquad&
    Y_{D^\star}
      &= N_\theta\,\partial_\theta
       + \frac{N_\phi}{\sin\theta}\,\partial_\phi.\end{alignedat}
  \label{eq:B14c}
  \\[0.45em]
  J^\star:\quad&
  \begin{alignedat}[t]{3}
    f_{J^\star}
      &= 0,\qquad&
    Y_{J^\star}
      &= N_\phi\,\partial_\theta
       - \frac{N_\theta}{\sin\theta}\,\partial_\phi.\end{alignedat}
  \label{eq:B14d}
  \\[0.45em]
   K^\star_I:\quad&
  \begin{alignedat}[t]{2}
    f_{K^\star_I}
      &= -(u-u_\star)\beta_I,
  \end{alignedat}
  \label{eq:B14e}
  \\[-0.1em]
  &
  \begin{aligned}
    Y_{K^\star_I}
      &=
      -\bigl[(\alpha-1)E_{I\theta}-\beta_I N_\theta\bigr]\partial_\theta
      -\frac{(\alpha-1)E_{I\phi}-\beta_I N_\phi}{\sin\theta}\,\partial_\phi .
  \end{aligned}
  \label{eq:B14f}
\end{align}\es 
 At the target angle \(\Omega=\Omega_\star\), one has
\begin{equation}
  \alpha=1,\qquad
  \beta_I=0,\qquad
  N_\theta=N_\phi=0 .
  \label{eq:target-identities}
\end{equation}
Substituting \eqref{eq:target-identities} into
\eqref{eq:Iplus-star-action} gives
\begin{equation}
  f_{K^\star_0}\big|_{p_\star}
  =f_{B^\star_I}\big|_{p_\star}
  =f_{D^\star}\big|_{p_\star}
  =f_{J^\star}\big|_{p_\star}
  =f_{K^\star_I}\big|_{p_\star}=0,
\end{equation}
and
\begin{equation}
  Y_{K^\star_0}\big|_{p_\star}
  =Y_{B^\star_I}\big|_{p_\star}
  =Y_{D^\star}\big|_{p_\star}
  =Y_{J^\star}\big|_{p_\star}
  =Y_{K^\star_I}\big|_{p_\star}=0 {.}
\end{equation}

Therefore all generators in \eqref{eq:a1-star-basis} satisfy the
stabilizer condition \eqref{eq:direct-stabilizer-condition} directly.
For \(u_\star=0\), \(\bm N=\bm S=(0,0,-1)\), \(\bm E_1=\bm e_1\), and \(\bm E_2=\bm e_2\), the generators reduce precisely to  \eqref{eq:a1-south}.

\subsection{Isomorphism}
Since the algebra $\mathfrak h_{p_\star}$ is the pushforward of $\mathfrak h_S$, we get the isomorphism between these two algebras immediately 
\be \mathfrak h_{p_\star}\simeq \mathfrak h_S.
\nn
\ee 
More explicitly, the algebra $\mathfrak h_{p_\star}$  would be 
\bs\begin{align}
[K^\star_0,D^\star] &= K^\star_0,&
[B^\star_1,J^\star] &= -B^\star_2,&
[B^\star_2,J^\star] &= B^\star_1,\\
[B^\star_1,K^\star_1] &= K^\star_0,&
[B^\star_2,K^\star_2] &= K^\star_0,\\
[J^\star,K^\star_1] &= K^\star_2,&
[J^\star,K^\star_2] &= -K^\star_1,&
[D^\star,K^\star_1] &= -K^\star_1,&
[D^\star,K^\star_2] &= -K^\star_2,
\end{align}  \label{eq:star-brackets}\es 
with all other Lie brackets vanishing.

In \cite{Chen:2021xkw}, the stabilizer group of a CCFT$_3$ is generated by  
the generators
\begin{equation}
  \{D,K_0,K_1,K_2,B_1,B_2,J_{12}\},
\end{equation}
with  brackets which are isomorphic to \(\mathfrak h_{p_\star}\).
 These generators are selected by preserving the boundary point $(u=0,z=\bar z=0)$ {on the Carrollian manifold} with topology $\mathbb R\times\mathbb R^2$. 
Interestingly, there is another seven-dimensional group that preserves the Rindler horizon \cite{Li:2024kbo}. It is  
generated by 
\begin{equation}
  \{\xi_{TZ},\xi_+,\xi_{+X},\xi_{+Y},\xi_X,\xi_Y,\xi_{XY}\}.
\end{equation}
We will denote this algebra by $\mathfrak{h}_{Rindler}$.  
After a straightforward calculation, 
we find \begin{equation}
\mathfrak{h}_S\simeq\mathfrak{h}_{p_\star}\simeq \mathfrak{h}\simeq\mathfrak h_{Rindler}.
\end{equation}
 In table \ref{four-algebra}, we list the one-to-one correspondence for these four seven-dimensional algebras.
\begingroup
\small
\setlength{\tabcolsep}{4pt}
\renewcommand{\arraystretch}{1.35}
\setlength{\LTleft}{\fill}
\setlength{\LTright}{\fill}

\begin{longtable}{@{}
>{\centering\arraybackslash}p{0.12\linewidth}
>{\centering\arraybackslash}p{0.12\linewidth}
>{\centering\arraybackslash}p{0.16\linewidth}
>{\centering\arraybackslash}p{0.20\linewidth}
@{}}
\toprule
\(\mathfrak{h}_S\) & \(\mathfrak{h}\) & \(\mathfrak{h}_{Rindler}\) & \(\mathfrak h_{p_\star}\)\\
\midrule
\endfirsthead

\toprule
\(\mathfrak{h}_S\) & \(\mathfrak{h}\) & \(\mathfrak{h}_{Rindler}\) & \(\mathfrak h_{p_\star}\)\\
\midrule
\endhead

\(\mathbb{D}\)   & \(D\)      & \(\xi_{TZ}\)  & \(D^\star\)\\
\(\mathbb{K}_0\) & \(K_0\)    & \(\xi_+\)     & \(K^\star_0\)\\
\(\mathbb{K}_1\) & \(K_1\)    & \(\xi_{+X}\)  & \(K^\star_1\)\\
\(\mathbb{K}_2\) & \(K_2\)    & \(\xi_{+Y}\)  & \(K^\star_2\)\\
\(\mathbb{B}_1\) & \(B_1\)    & \(\xi_X\)     & \(B^\star_1\)\\
\(\mathbb{B}_2\) & \(B_2\)    & \(\xi_Y\)     & \(B^\star_2\)\\
\(\mathbb{J}\)   & \(J_{12}\) & \(-\xi_{XY}\) & \(J^\star\)\\

\bottomrule
\caption{\centering One-to-one correspondence among the four seven-dimensional algebras.}
\label{four-algebra}
\end{longtable}

\endgroup

\section{Gravitational operator at null infinity}\label{gaugeanalysis}
In this appendix, we consider the rank-2 symmetric gravitational field $h_{\mu\nu}$ with a linearized gauge transformation 
\be 
h_{\mu\nu}\to h_{\mu\nu}-\partial_\mu\xi_\nu-\partial_\nu\xi_\mu.
\ee We impose the following fall-off condition for the gravitational field and the gauge parameter\footnote{The gauge parameter may also allow large gauge transformations that have been turned off. We expect that these large gauge transformations do not affect the main conclusion.}
\be 
h_{\mu\nu}=\frac{H_{\mu\nu}}{r}+\cdots,\quad \xi_\mu=\frac{\epsilon_\mu}{r}+\cdots.
\ee 
Now the gauge transformation of the boundary spin-2 operator at the south pole is 
\bs\begin{align}
    &H_{00}\to H_{00}-2\dot\epsilon_0,\quad H_{01}\to H_{01}-\dot\epsilon_1,\nonumber\\
    &H_{02}\to H_{02}-\dot\epsilon_2,\quad H_{03}\to H_{03}-\dot \epsilon_{0}-\dot\epsilon_3,\\
    &H_{11}\to H_{11},\quad H_{12}\to H_{12},\quad H_{13}\to H_{13}-\dot\epsilon_1,\nonumber\\
    &H_{22}\to H_{22},\quad H_{23}\to H_{23}-\dot\epsilon_2,\quad H_{33}\to H_{33}-2\dot\epsilon_3.
\end{align}\es There are six gauge-invariant modes 
\begin{align}
    H_{11}, H_{22},H_{12},H_{00}-H_{03}-H_{30}+H_{33},H_{01}-H_{13}, H_{02}-H_{23}.
\end{align}
Equivalently, the gauge-invariant modes are 
\bea 
R_4, R_5,R_6,R_7,R_8,\mathcal T+R_9.
\eea The gauge transformations of the other four modes are 
\bea 
\begin{gathered}
R_1\to R_1-4\dot\epsilon_0-4\dot\epsilon_3,\quad R_2\to R_2-2\dot\epsilon_1,\\
R_3\to R_3-2\dot\epsilon_2,\quad R_{9}\to R_9-2\dot\epsilon_0+2\dot\epsilon_3.
\end{gathered}
\eea One can fix the gauge $R_1=R_2=R_3=R_9=0$ at null infinity. Further analysis of the gauge-invariant modes should impose the equation of motion that depends on the theory. In Einstein gravity, there are only two dynamical gauge-invariant modes at null infinity. In a gauge-invariant spin-2 theory, one may delete the gauge-dependent modes and thus we obtain Figure \ref{fig:sym-newK-netg}.

\bibliography{refs}

\end{document}